\documentclass[lettersize,journal]{IEEEtran}
\usepackage{amsmath,amsfonts}
\usepackage{algorithmic}
\usepackage{algorithm}
\usepackage{array}
\usepackage[caption=false,font=footnotesize,labelfont=rm,textfont=rm]{subfig}

\usepackage{textcomp}
\usepackage{stfloats}
\usepackage{url}
\usepackage{verbatim}
\usepackage{graphicx}
\usepackage{cite}
\usepackage{stfloats}
\usepackage{makecell}
\usepackage{multirow}
\usepackage{amsthm,amsmath,amssymb}
\usepackage{mathrsfs}
\usepackage{epsfig}
\usepackage{epstopdf}
\usepackage{diagbox}
\usepackage{soul}
\usepackage{ulem}
\usepackage{caption}
\usepackage{etoolbox}
\AtBeginEnvironment{tabular}{\normalsize}

\usepackage{color, xcolor}
\usepackage{booktabs}
\begin{document}


\title{A Comprehensive Survey of 3GPP Release 19 ISAC Channel Modeling: From Empirical Features to \\ Unified Methodology and Standardized Simulator}

\author{Yameng Liu, Yuxiang Zhang, Jianhua Zhang, Yuanpeng Pei, Changsheng Zhao, Shilin Luo, Lei Tian, \\ Yingyang Li, Wei Hong, Jianming Wu, Guangyi Liu, Yan Li, Tao Jiang, Chuangxin Jiang, Junchen Liu, \\ Yongqiang Fei, Woo-Suk Ko, Jing Xu, Bin Liang, Takahiro Tomie    

\thanks{This research is supported in part by National Natural Science Foundation of China under Grant 62525101, Grant 62201087, in part by the National Key R\&D Program of China under Grant 2023YFB2904803, in part by the Guangdong Major Project of Basic and Applied Basic Research under Grant 2023B0303000001, in part by the Natural Science Foundation of Beijing-Xiaomi Innovation Joint Foundation under Grant L243002, and in part by the Beijing University of Posts and Telecommunications-China Mobile Research Institute Joint Institute. \textit{(Corresponding author: Yuxiang Zhang; Jianhua Zhang.)}}
\thanks{Yameng Liu, Yuxiang Zhang, Jianhua Zhang, Yuanpeng Pei, Changsheng, Shilin Luo, and Lei Tian are with the State Key Laboratory of Networking and Switching Technology, Beijing University of Posts and Telecommunications, Beijing, China (email: liuym@bupt.edu.cn; zhangyx@bupt.edu.cn; jhzhang@bupt.edu.cn; Peiyp@bupt.edu.cn; zhaochangsheng@bupt.edu.cn; luoshilin@bupt.edu.cn).} 
\thanks{Yingyang Li and Wei Hong are with the Xiaomi Industry Standardization and Research Department, Beijing, China (e-mail: liyingyang@xiaomi.com, hongwei@xiaomi.com).} 
\thanks{Jianming Wu is with the vivo Communications Research Institute, vivo Mobile Communication, Beijing, China (e-mail: jianming.wu@vivo.com).}
\thanks{Tao Jiang and Guangyi Liu is with the Future Research Laboratory, China Mobile Research Institution, Beijing, China, (email: jiangtao@chinamobile.com; liuguangyi@chinamobile.com).}
\thanks{Yan li is with the Department of Wireless and Device Technology Research, 
China Mobile Research Institute, Beijing, China (email: liyanwx@chinamobile.com).} 
\thanks{Chuangxin Jiang and Junchen Liu are with the algorithm department, ZTE corporation, Xi'an, China (email: jiang.chuangxin1@zte.com.cn, liu.junchen@zte.com.cn).} 
\thanks{Yongqiang Fei is with the CICT Mobile Communications Technology  Co., Ltd, also with the State Key Laboratory of Wireless Mobile Communications Technology, CATT, Beijing, China (e-mail: feiyongqiang@cictmobile.com).}
\thanks{Woo-Suk Ko is with the Communication \& Media Standard Laboratory, LG Electronics Inc., Seoul, Republic of Korea (e-mail: woosuk.ko@lge.com).}
\thanks{Jing Xu and Bin Liang are with the Standards Research Center, Beijing OPPO Telecommunications Corp., Ltd., Beijing, China (e-mail: xj@oppo.com; liangbin@oppo.com).}
\thanks{Takahiro Tomie is with the 6G-Tech Department, NTT DOCOMO, INC., Kanagawa, Japan (e-mail: ngochao.tran.fr@nttdocomo.com).}} 

\markboth{Journal of \LaTeX\ Class Files,~Vol.~14, No.~8, August~2021}%
{Shell \MakeLowercase{\textit{et al.}}: A Sample Article Using IEEEtran.cls for IEEE Journals}


\maketitle

\begin{abstract}
Integrated Sensing and Communication (ISAC) has been identified as a key 6G application by International Telecommunication Union (ITU) and 3rd Generation Partnership Project (3GPP). 
Channel measurement and modeling is a prerequisite for ISAC system design and has attracted widespread attention from both academia and industry. 3GPP Release 19 initiated the ISAC channel study item in December 2023 and finalized its modeling specification in May 2025 after extensive technical discussions. However, a comprehensive survey that provides a systematic overview—from empirical channel features to modeling methodologies and standardized simulators—remains unavailable. 
In this paper, the key requirements and challenges in ISAC channel research are first analyzed, followed by a structured overview of the standardization workflow throughout the 3GPP Release 19 process.
Then, critical aspects of ISAC channels, including physical objects, target channels, and background channels, are examined in depth, together with additional features such as spatial consistency, environment objects, Doppler characteristics, and shared clusters, supported by measurement-based analysis. To establish a unified ISAC channel modeling framework, an Extended Geometry-based Stochastic Model (E-GBSM) is proposed, incorporating all the aforementioned ISAC channel characteristics. Finally, a standardized simulator is developed based on E-GBSM, and a two-phase calibration procedure aligned with 3GPP Release 19 is conducted to validate both the model and the simulator, demonstrating close agreement with industrial reference results.
Overall, this paper provides a systematic survey of 3GPP Release 19 ISAC channel standardization and offers insights into best practices for new feature characterization, unified modeling methodology, and standardized simulator implementation, which can effectively supporting ISAC technology evaluation and future 6G standardization.
\end{abstract}


\section{Introduction}\label{section1}
\subsection{Overview of 6G ISAC}


Radar and communications, as key utilities of electromagnetic waves, have traditionally developed separately. Communication is primarily concerned with information transmission, while radar focuses on target sensing and identification \cite{zhang2023integrated,liu2023shared,zhang2024latest}. However, the expansion of frequency bands and bandwidth in communications, along with the evolution from single-antenna systems to large-scale antenna arrays \cite{zhang20173} in both radar and communication systems, has led to significant similarities in their hardware architectures and channel characteristics \cite{liu2023seventy,wen2025survey}. These convergences make the integration and co-design of radar and communication systems increasingly feasible. This emerging research area is commonly referred to as Integrated Sensing and Communication (ISAC). As research progresses, ISAC technology has been more broadly defined as the integration of sensing and communication functionalities within a unified system, enabling Base Stations (BS) or User Terminals (UT) to simultaneously exchange information and perceive their surrounding environment \cite{kumari2017ieee,nie2022predictive,zhang2018multibeam,liu2025coupling}. By sharing a majority of the software, hardware, and information resources, ISAC systems bring tremendous advantages to improving spectrum utilization and reducing costs \cite{liu2022survey,cui2021integrating,zhang2021overview,pucci2022system}. 

In June 2023, the International Telecommunication Union (ITU) finalized the Recommendation on the \textit{Framework and Overall Objectives of the Future Development of IMT for 2030 and Beyond} \cite{itu6G,liu2024cooperative}. Building upon the three primary use cases of 5G, the document expands and enhances the framework by identifying six key usage scenarios—including ISAC—and defining 15 capability indicators such as higher-accuracy positioning and sensing-related metrics. In the same month, the 3rd Generation Partnership Project (3GPP) released TR 22.837 standardization \cite{3gpp22837}, which specifies 32 potential use cases for wireless sensing, including target detection, tracking, and environment reconstruction. In parallel, the IMT-2030 Promotion Group, a national-level research initiative, has published several reports on ISAC, further underscoring the central role of ISAC in the evolution of future communication technologies.

\textbf{Basic Definitions:}
1)	\textit{Sensing transmitter (STx)}: the Transmission Reception Point (TRP) or a UT that sends out the sensing signal which the sensing service will use in its operation. A sensing Tx can be located in the same or different TRP or a UT as the sensing receiver.
2)	\textit{Sensing receiver (SRx)}: the TRP or a UT that receives the sensing signal which the sensing service will use in its operation. A sensing Rx can be located in the same or different TRP or a UT as the sensing Tx.
3)	\textit{Sensing Target (ST)}: target that need to be sensed by deriving characteristics of the objects within the environment from the sensing signal.
4)	\textit{Background environment}: background clutter and/or Environment Objects (EOs) that are not the sensing target(s).
5)	\textit{Monostatic sensing}: sensing where the sensing transmitter and sensing receiver are co-located in the same TRP or UT.  
6)	\textit{Bistatic sensing}: sensing where the sensing transmitter and sensing receiver are in different TRPs or UTs. 

\subsection{Requirements and Challenges of ISAC Channel Research}

\begin{figure*}[t]
\centering
\includegraphics[width=6.5in]{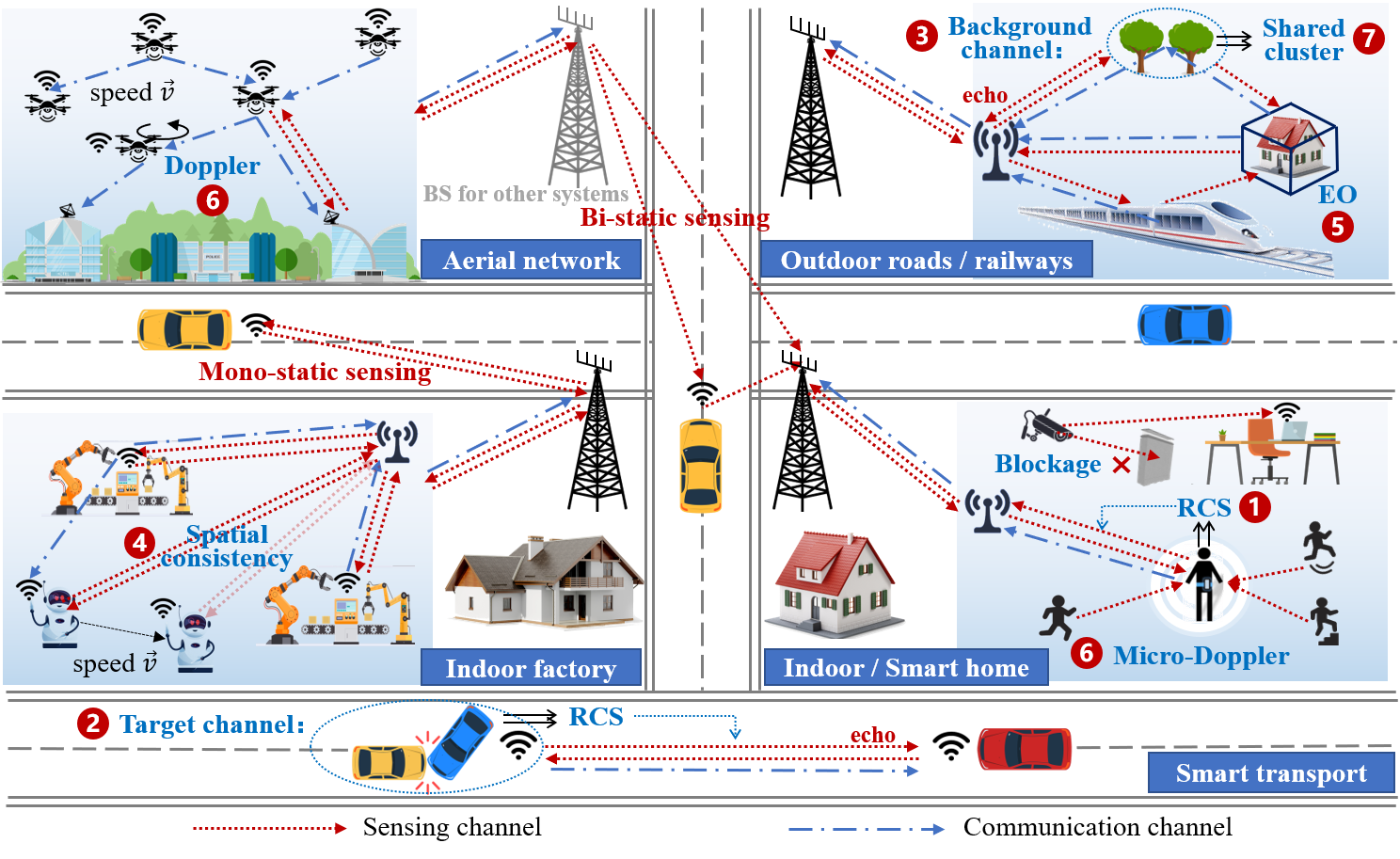}
\caption{Illustration of the ISAC channel with new features, including 1) RCS, 2) target channel, 3) background channel, 4) spatial consistency, 5) EO, 6) Doppler effect, and 7) shared cluster. The red and blue dotted lines denote the sensing and communication channels, respectively.}
\label{fig_1_scenario}
\end{figure*}

The radio channel is the transmission medium between a Tx and a Rx. It plays a crucial role in determining the ultimate performance limit of mobile communications systems \cite{molisch2012wireless,zhang2020channel,tang2021channel,yuan2022spatial}. Channel models can be categorized as the deterministic model, stochastic model, and deterministic-stochastic hybrid model. Among them, Geometry-Based Stochastic Model (GBSM) is mainstream due to its low complexity and scalability, and has been widely adopted in standardized communication channel modeling by ITU \cite{itu2412} and 3GPP \cite{3gpp38901}. It adopts a clustered structure that groups multipaths into stochastic clusters, leveraging both environment-specific features and stochastic modeling \cite{zhang2016interdisciplinary,yu2022implementation}. GBSM is two-dimensions (2D) in 4G systems and expands to three-dimensions (3D) in 5G systems by introducing the zenith angles \cite{zhang20173}.

In contrast to 5G communication systems that primarily focus on overall channel propagation characteristics, ISAC systems emphasize extracting effective target information from environmental clutter to support sensing tasks such as localization, tracking, and identification \cite{wang2020small,xu2023spatial}. This necessitates the inclusion of additional deterministic components in the channel model to characterize the specific propagation behavior of sensing targets \cite{ref_刘亚萌WCL,yang2024integrated}. Moreover, ISAC introduces new sensing use cases and monostatic sensing modes, which are not covered by existing communication configurations \cite{3gppRan116}. Therefore, conventional 5G channel models are insufficient for ISAC performance evaluation, and an Extended-GBSM (E-GBSM) for 6G ISAC is urgently needed \cite{gong2024new}.

Fig. \ref{fig_1_scenario} illustrates the ISAC channels with new features observed in typical application scenarios. Accurately and realistically capturing these characteristics poses significant challenges for ISAC channel modeling. 
Firstly, in order to realize the sensing tasks, a necessary prior step in sensing channel modeling is to determine the ST’s position, velocity, and scattering propagation characteristics. This modeling process has been widely recognized and adopted in the 3GPP standardization work \cite{3gppRan116} and academic research \cite{zhang2016interdisciplinary,yu2022implementation}. As for the scattering characteristic, the radar community has introduced the concept of Radar Cross Section (RCS), which is defined as the effective area of a target that captures and reflects the electromagnetic wave energy \cite{Youssef1989Radar}. To address the challenges of RCS modeling, computational electromagnetic (CEM) methods, including finite difference time domain (FDTD) \cite{Cakir2014FDTD}, the moment method (MoM) \cite{Lucente2008Iteration}, and the finite element method (FEM) \cite{Dunn2006Numerical} are applied to solve Maxwell's equations to emulate electromagnetic interactions with targets, offering high accuracy in RCS calculations. However, as for the typical STs such as Unmanned Aerial Vehicles (UAVs), humans, and vehicles, the high computational complexity of electromagnetic simulations limits their applicability in standardized models, making it a key challenge to develop a generalized and scalable RCS modeling approach for ISAC systems.

ISAC Tx and Rx can be mainly categorized into four types: Transmission Reception Point (TRP), normal UT, aerial UT, and vehicle UT. Correspondingly, nine propagation links of \textit{1) TRP-TRP, 2) TRP-normal UT, 3) TRP-vehicle UT, 4) TRP-aerial UT, 5) normal UT-normal UT, 6) normal UT-vehicle UT, 7) normal UT-aerial UT, 8) vehicle UT-vehicle UT, 9) aerial UT-aerial UT} are defined, which covers all potential propagation situations in ISAC \cite{3gppRan121FL}. Because the ISAC channel incorporates the ST, the ST-related channel is divided into Tx-target and target-Rx links, each corresponding to one of the nine propagation types described above. However, accurately characterizing the full Tx-target-Rx channel, which is a component of the ISAC channel, remains an open question. During the standardization discussions, some companies \cite{3gpp117oppo,3gpp117nokia} have proposed using a statistical cluster from communications to describe the ST-related channel, while others \cite{3gpp117bupt,3gpp117vivo} have theorized concatenation channel models. More empirical research is needed to explore the channel characteristics of typical scenarios and further validate the relevant models.

\begin{figure}[t]
\centering
\includegraphics[width=3.4in]{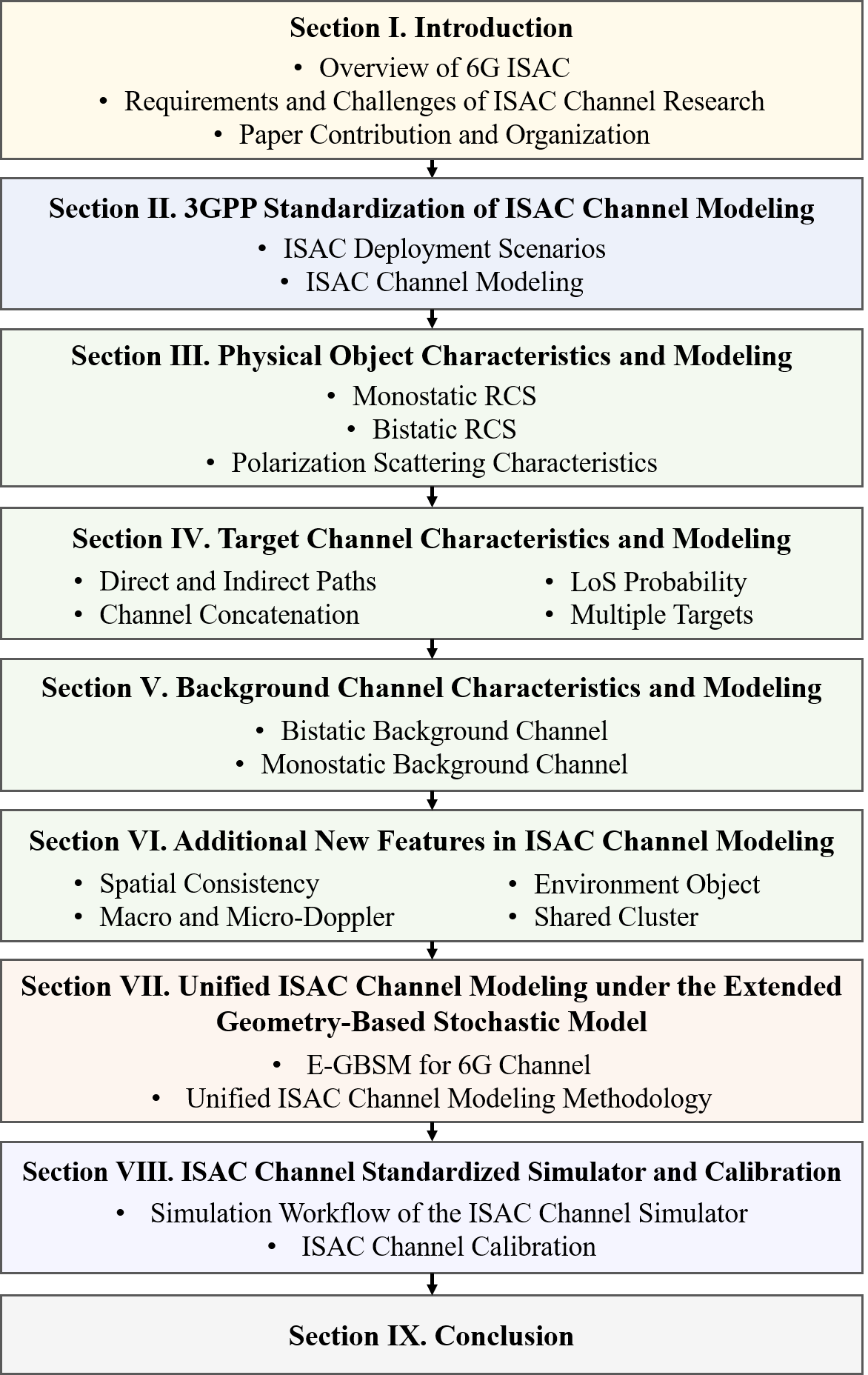}
\caption{Paper organization and content flow.}
\label{fig_1_sections}
\end{figure}

As for the ST-unrelated channel component (multipaths contributed by the environment, such as the trees labeled by the upper-right circle of Fig. \ref{fig_1_scenario} and the corresponding red sensing paths), although they do not carry useful information about the target of interest, they actually constitute a larger fraction of the ISAC channel, as practically validated in \cite{zhang2025research}. Accurately characterizing these environmental / background multipaths will significantly enhance the performance of target information extraction, such as localization and identification \cite{wang2020small}. The ST-unrelated background channelS, similar to communication channels, focus on the overall propagation characteristics without considering intermediate nodes. Therefore, in the separated Tx-Rx (defined as bistatic) sensing mode, the corresponding models and parameters from 5G communication standardization \cite{3gpp38901} can be directly reused as the starting point for research. These 5G models and parameters depend on the distance and relative position between separated Tx and Rx, making them unsuitable for the new Tx-Rx co-located echo (defined as monostatic) sensing mode, which includes propagation situations \textit{1)}, \textit{5)}, \textit{8)}, and \textit{9)}. Developing a general approach for modeling monostatic background channels requires further empirical analysis and research.


Channels generated by nodes within the correlation distance exhibit similar characteristics, reflecting the spatial continuity of the propagation environment \cite{ademaj2018ray}. Accurately characterizing this spatial consistency is essential for preserving and evaluating the continuity of channel variations. In the context of 5G communication, spatial consistency has been addressed for both channel large-scale and small-scale parameters \cite{3gpp38901}, across different propagation links, constrained by the correlation distance. In ISAC systems, spatial consistency becomes even more crucial for sensing performance evaluation, particularly in multi-node cooperation sensing and under mobility scenarios \cite{liu2024cooperative}. Given the dual role of the channel in supporting both communication and sensing functions, spatial consistency models should incorporate the unique characteristics introduced by STs, without compromising for existing communication channel correlations. 

In propagation channels, a Line-of-Sight (LoS) path may exist between the Tx and Rx (or between Tx / Rx and ST). Meanwhile, various types of scatterers contribute to the non-LoS (NLoS) propagation paths through mechanisms such as reflection, scattering, and diffraction, resulting in NLoS clusters. The LoS path provides the most direct and effective information for enabling sensing functions such as localization in ISAC systems \cite{kuutti2018survey}. However, to enhance target information acquisition, especially under NLoS conditions, it is crucial to characterize the large, regular EO, such as walls and ground surfaces, along with the NLoS paths they generate. These NLoS paths have a significant impact on the channel and can be effectively utilized for improved sensing performance. Exploring how to accurately model EOs, particularly their propagation characteristics of single-hop mirror reflections, is one of the key challenges \cite{jiang2025novel,3gppRan121FL}.

Doppler effect plays a critical role in communication systems, causing frequency shifts in the received signal due to the relative motion between the Tx and Rx. In traditional communication systems, these shifts can lead to signal degradation, such as inter-symbol interference and reduced capacity, especially in high-speed scenarios \cite{wang2023doppler}. Accurate modeling and compensation of Doppler shifts are essential for maintaining link reliability. In ISAC systems, characterizing the Doppler effect enables effective velocity sensing and identification of STs. Unlike communication channels, ISAC channels require considering the relative motion between the Tx, Rx, and STs, as all three act as mobile nodes. Additionally, ISAC applications, like human posture recognition in smart homes and UAV state detection in low-altitude networks (as shown in Fig. \ref{fig_1_scenario}), demand more advanced micro-Doppler studies. 

In the existing research, ISAC channel models are widely utilized in ISAC systems to evaluate upper-level technology development and system design \cite{nguyen2022access,rahman2019framework,zhang2022integrated,yuan2020spatio}, where communication and
sensing channels are generated independently.  Communication and sensing channels are intrinsically different because of different propagation configurations, e.g.,
Tx-Rx relative positions. However, due to the multiplexing of hardware resources (e.g., antennas) and the same environment, some objects might serve as shared propagation scatterers for both the communication and sensing channels. Accurately characterizing this feature will help assess the mutual assistance between communication and sensing. However, whether this ISAC channel is truly shared, to what extent it is shared, and how to model and evaluate remain challenges that require empirical validation and analysis.

In 5G communication channels, blockage modeling has been studied as an add-on feature to the channel model. In 3GPP TR 38.901 \cite{3gpp38901}, Blockage \textit{A} and \textit{B} are defined, using a stochastic method and a geometric method, respectively, to capture e.g., human and vehicular blocking. In ISAC channels, STs with known location, orientation, and size can also potentially block the background channel or the paths of other target channels. Investigating the blocking effects caused by different typical STs and their coupling with the ISAC environment, as well as how to extend the modeling based on 3GPP, requires further empirical validation.



\begin{figure*}
\centering
\includegraphics[width=7.2in]{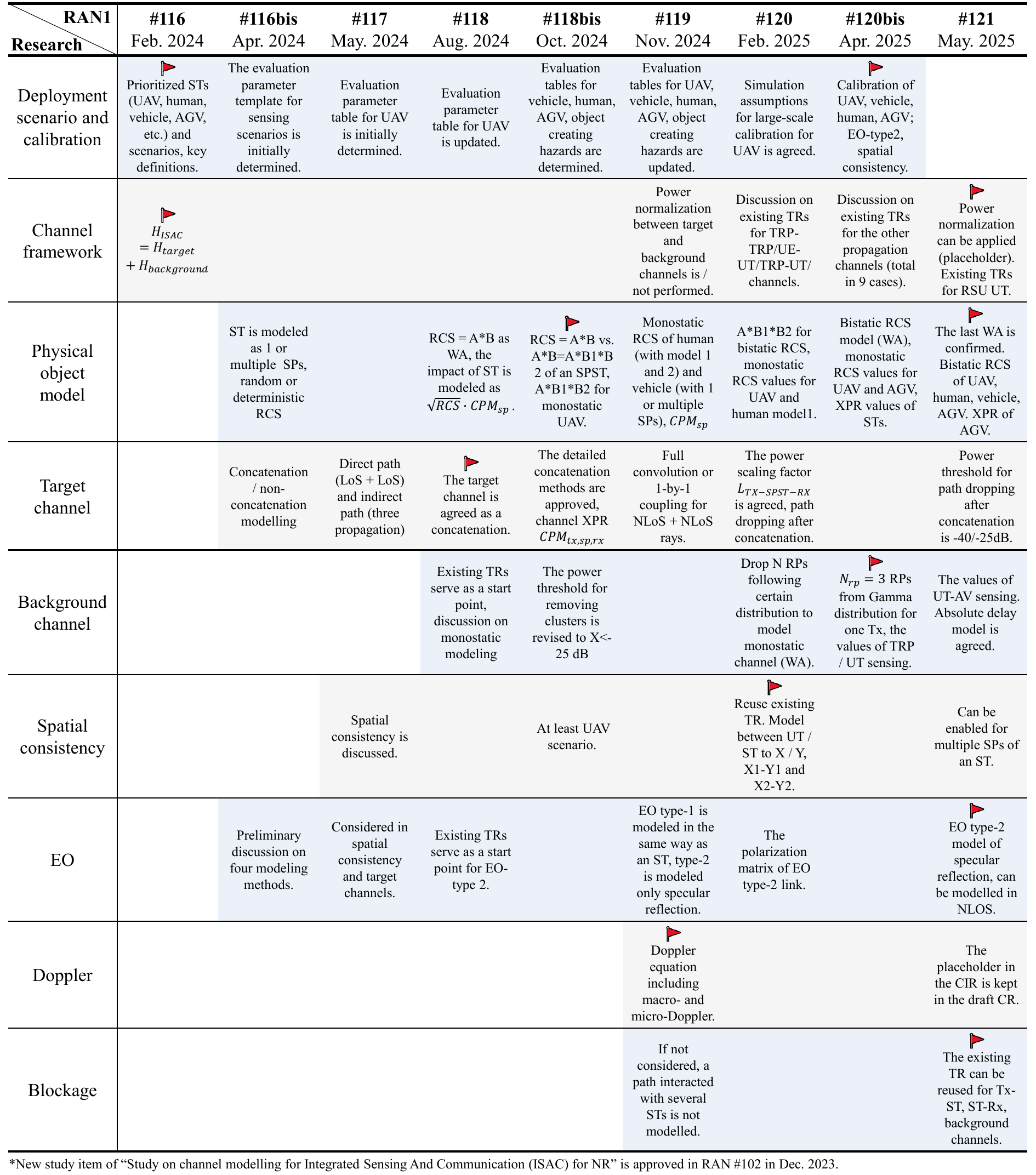}
\caption{3GPP ISAC channel standardization progress and timeline, where the abbreviations include Working Assumption (WA), Technical Report (TR), Scattering Point (SP), Sensing Target (ST), Reference Point (RP), Transmission Reception Point (TRP), User Terminal (UT), and Channel Impulse Response (CIR).}
\label{fig_2}
\end{figure*}

\subsection{Paper Contribution and Organization}


ISAC channel measurement and modeling plays a pivotal role in the design, standardization, and optimization of ISAC systems. 
However, there has not been a comprehensive survey yet that provides a systematic overview that spans from empirical channel measurements to modeling theories and to standardized simulations. The main contributions of this paper are summarized as follows:

\begin{itemize}
\item{A systematic examination is conducted on the requirements and challenges of ISAC channel research, and an overview of the standardization workflow and timeline in 3GPP Release 19 is provided. The discussion focuses on key ISAC scenarios and target types, the evolution of the ISAC channel modeling framework, and the standardization efforts related to critical ISAC channel features.}
\item{The latest standardization progress and global research on the emerging characteristics of ISAC channels and their modeling are comprehensively reviewed. The review provides an in-depth discussion covering critical channel components, including physical objects, target channels, background channels, along with additional features such as spatial consistency, EO, Doppler effect, and shared clusters. All aspects are supported by extensive empirical measurements and analysis.}
\item{Representative 6G channel and system-level simulators are surveyed and compared. A unified and standardization-oriented ISAC channel model is then established by evolving the 3D GBSM into an extended framework (E-GBSM). The E-GBSM characterizes the concatenated Tx-Target and Target-Rx sub-channels, and incorporates all key ISAC channel features, such as RCS, monostatic background paths, EO, shared clusters, and micro-Doppler effect).}
\item{Based on the proposed unified ISAC channel model, the BUPTCMCCCMG-IMT2030 channel simulator is developed, along with the ISAC channel simulation procedure and calibration methodology aligned with 3GPP TR 38.901. Calibration results demonstrate high consistency with reference implementations at 6 GHz and 30 GHz under both TRP-monostatic and bistatic sensing configurations, thereby verifying the engineering applicability of the standardized channel model and simulator.}
\end{itemize}

\section{3GPP Standardization of ISAC Channel Modeling} \label{section2}


To address the gap in current standardized communication channel models that prevents effective sensing evaluation, the 3GPP Technical Speciﬁcation Group (TSG) of Radio Access Network (RAN) approved a new study item \textit{Study on channel modelling for Integrated Sensing And Communication (ISAC) for NR} for the development of ISAC channel in meeting RAN \#102 \cite{3gppRan102} in Dec. 2023. The work was conducted by RAN working group 1 (RAN1) that is responsible for the speciﬁcation of the physical layer within RAN. This proposal aims to develop a channel model to support studies on typical ISAC scenario and use cases, and including two sub-agenda items of \textit{ISAC deployment scenarios and calibrations} and \textit{ISAC channel modelling}. The detailed discussions starts at RAN1 \#116 meeting in Feb. 2024 and ended at RAN1 \#121 meeting in May 2025. The timeline and important reached consensuses of these meetings are presented in Fig. \ref{fig_2} \cite{3gppRan121sum}. After nine formal meetings (\#116, \#116bis, \#117, \#118, \#118bis, \#119, \#120, \#120bis, and \#121) with online and offline discussions, and some email discussions, the channel model for ISAC is established and added into the 3GPP TR 38.901 V19.0.0 standard wireless channel \cite{3gpp38901}. 


\subsection{ISAC Deployment Scenarios}

The typical existing communication scenarios and STs of UAV, humans indoors, humans outdoors, automotive vehicles (at least outdoors), automated guided vehicles (e.g. in indoor factories), and objects creating hazards on roads/railways (examples defined in TR 22.837) are considered as a starting point, which are approved at the RAN1 \#116 meeting in Feb. 2024. Any TRP and / or UT location in the corresponding communication scenario can be selected as sensing Tx and Rx locations. During \#116bis to \#119 meeting, RAN1 has successively finalized the evaluation param-eters for the agreed typical STs. The defined parameter categories include:
\begin{enumerate}
    \item Applicable communication scenarios.
    \item Sensing Txs and Rxs properties.
    \item ST including LoS/NLoS, outdoor/indoor, 2D or 3D mobility, 2D or 3D distribution, orientation, and physical characteristics (e.g., size).
    \item Minimum 3D distances between pairs of Tx/Rx and ST.
    \item Minimum 3D distance between STs.
    \item Optionally, EO, e.g., types, characteristics, mobility, distribution, etc.
\end{enumerate}
The determination and updates of these parameter values are based on TR 38.901 \cite{3gpp38901}, while the parameter values defined in TR 36.777 \cite{3gppTR36777} for aerial UT, defined in TR 38.808 \cite{3gppTR38808} for indoor room scenario, defined in TR 37.885 \cite{3gppTR37885}, 38.859 \cite{3gppTR38859} for automotive Urban grid / highway scenarios are also taken into account in the corresponding ISAC evaluation scenarios. These settings can be used directly for the system simulation and evaluation, and certain percentage of TRPs / UTs that have sensing capabilities should be configured in the channel calibration.

\begin{table*}[t]
    \caption{Target RCS measurement and simulation campaigns in 3GPP Release 19}
    \centering
    \begin{tabular}{m{2cm}<{\centering}|m{3cm}<{\centering}|m{8cm}<{\centering}|m{3cm}<{\centering}}
        \toprule\hline
        \textbf{Sensing modes} & \textbf{Targets} & \textbf{Participating organizations} & \textbf{Reference} \\
        \hline
        \multirow{10}{*}{Monostatic}
         & Vehicle & \makecell{BUPT, Huawei, NIST, Xiaomi, ZTE}  & \cite{ref_北邮119提案} \cite{ref_华为119提案} \cite{ref_NIST120提案} \cite{ref_小米120提案} \cite{ref_中兴118提案} \\
        \cline{2-4}
         & Human & \makecell{BUPT, NIST, ZTE, Xiaomi, vivo, Apple,\\ OPPO, CATT, LGE } & \cite{ref_北邮119提案} \cite{ref_NIST119提案} \cite{ref_中兴118提案} \cite{ref_小米120提案} \cite{ref_维沃119提案} \cite{ref_苹果120b提案} \cite{ref_Oppo120提案} \cite{ref_大唐119提案} \cite{ref_LGE118b提案} \\
        \cline{2-4}
         & UAV with small size& \makecell{BUPT, Huawei, vivo, ZTE, \\Samsung, CATT, LGE, QC } & \cite{ref_北邮118提案} \cite{ref_华为118提案} \cite{ref_维沃118提案} \cite{ref_中兴118提案} \cite{ref_三星118提案}\cite{ref_大唐119提案} \cite{ref_LGE118b提案} \cite{ref_高通120提案} \\
        \cline{2-4}
         & UAV with large size& \makecell{BUPT, Xiaomi, NIST, Samsung }& \cite{ref_北邮119提案} \cite{ref_NIST119提案} \cite{ref_小米120b提案} \cite{ref_三星120b提案}\\
        \cline{2-4}
         & AGV & \makecell{BUPT, Huawei, LGE }& \cite{ref_北邮120b提案} \cite{ref_华为119提案} \cite{ref_LGE120提案}\\
        \hline
        \multirow{5}{*}{Bistatic}
         & Vehicle & BUPT, Huawei, ZTE, Xiaomi & \cite{ref_北邮120b提案} \cite{ref_华为120b提案} \cite{ref_中兴118提案} \cite{ref_小米120提案} \\
        \cline{2-4}
         & Human & BUPT, Apple, Xiaomi, DOCOMO  & \cite{ref_北邮120b提案} \cite{ref_苹果120b提案} \cite{ref_小米120b提案} \cite{ref_docomo120b提案}\\
        \cline{2-4}
         & UAV with small size & BUPT, vivo & \cite{ref_北邮118提案} \cite{ref_维沃118提案} \\
        \cline{2-4}
         & UAV with large size & BUPT & \cite{ref_北邮120b提案} \\
        \cline{2-4}
         & AGV & BUPT& \cite{ref_北邮120b提案}\\
        \hline\bottomrule
    \end{tabular}
    \label{company contribution}
\end{table*}

\subsection{ISAC Channel Modeling}

Different from the communication standard channel model in 3GPP TR 38.901 V18.0.0 version, the common framework of the ISAC channel model is agreed at the RAN1 \#116 meeting in Feb. 2024, which is composed of a component of target channel and a component of background channel:
\begin{equation}
\label{equ_tar-sma}
H_{ISAC}=H_{target} + H_{background},
\end{equation}
where the target channel $H_{target}$ includes all multipath components impacted by the ST(s). Background channel $H_{background}$ includes other multipath components not belonging to target channel. A pair of sensing Tx and Rx can sense one or multiple STs within the shared environment, and the same ST can be modeled in the ISAC channels of multiple pairs of sensing Tx and Rx. Moreover, power normalization between target and background channels was discussed and a placeholder was left for further study.

Several key new features of the ISAC channel have been discussed during the 3GPP standardization process, as shown in Fig. \ref{fig_2}. Firstly, the physical model of sensing targets was agreed to be modeled as single or multiple Scattering Points (SP) in the \#116bis meeting. In the \#118bis meeting, the RCS of STSP was defined according to models $A\cdot B$ or $A\cdot B_1\cdot B_2$, thus establishing the framework for RCS research. By the final \#121 meeting in May 2025, the monostatic and bistatic RCS models and parameters for typical STs were agreed upon. Secondly, in the Aug. 2024 \#118 meeting, a key proposal for the target channel was approved, equating the ST to Tx or Rx, and the target channel was modeled through the concatenated sub-channels of Tx-TAR and TAR-Rx. Following meetings, the implementation and simplification methods, as well as power thresholds, were determined. Thirdly, for the background channel, the bistatic mode reused existing communication TRs, while the monostatic background channel underwent prolonged discussion. In the Apr. 2025 \#120bis meeting, a 3-Reference-Point (RP) based modeling method was approved, and the relevant parameters for typical ISAC scenarios and targets were quickly converged and determined. Additional features such as spatial consistency, EO, Doppler, and blockage were discussed intermittently during meetings, leading to certain consensus.

In the following sections, this paper focuses on summarizing the research on these new features of ISAC channel models, proposes a unified channel modeling methodology extending the GBSM, and introduces a standardization-compatible channel simulator with the corresponding calibration parameters.


\section{Physical Object Characteristics and Modeling} \label{section3}

Accurate representation of the influence of physical objects on the channel is essential for modeling the target channel. Physical objects are classified into two categories: STs and EOs. STs are objects of interest for sensing . According to 3GPP, the main types of STs include UAVs of large and small sizes, vehicles, humans, and Automated Guided Vehicles (AGVs). EOs are non-target objects with known locations. The ISAC channel model supports two types of EOs. The first type (type-1 EO) shares similar characteristics with STs and is modeled in the same manner. The second type (type-2 EO) is larger in size and is modeled differently from STs. This section focuses on the details related to STs, which are also applicable to type-1 EOs.

In order to better sensing the ST, an ST  is modeled based on its size through either a single or multiple scattering points. Each Scattering Point of an ST (SPST) is used to model the total scattering effect of an ST in the channel. The 3GPP believes that STs with larger sizes, such as vehicles and AGVs, need to define both single and multiple scattering point models, while targets with smaller sizes, such as humans and UAVs of large and small sizes, only need to define a single scattering point model. 
After being scattered by physical objects, the scattering direction energy distribution, polarization state, and phase of the electromagnetic wave will all change. The influence of the scattering point on these aspects needs to be modeled at least in two aspects, i.e., RCS and cross-polarization matrix (CPM).The RCS of the SPST is a scalar value and is defined as the hypothetical area required to intercept the incident power at the SPST such that if the total intercepted power are re-radiated, the power density actually observed at the receiver would be produced. The polarization matrix of the SPST includes the effects of the phase and amplitude of copolarization and cross-polarization at the SPST. The polarization matrix of the SPST is separately modeled from other polarization effects introduced by stochastic clusters and/or EOs in the target channel.

\begin{figure}[!h]
  \centering
    \begin{minipage}[c]{\linewidth}
    \centering
      \subfloat[]{
        \includegraphics[width=0.45\linewidth]{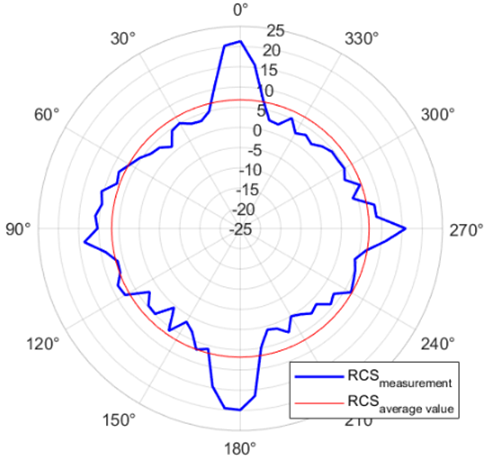}
      }
      \subfloat[]{
        \includegraphics[width=0.45\linewidth]{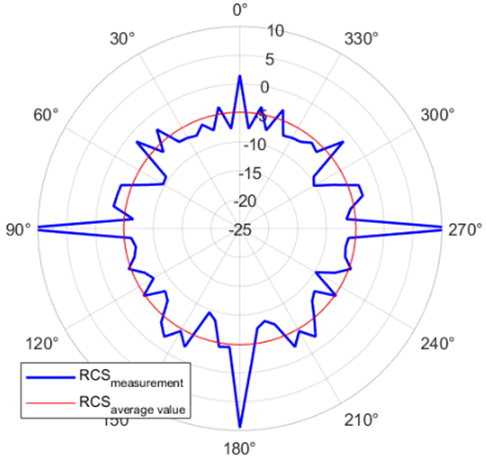}
      }
    \end{minipage}

    \begin{minipage}[c]{\linewidth}
    \centering
    \subfloat[]{
    \includegraphics[width=0.45\linewidth]{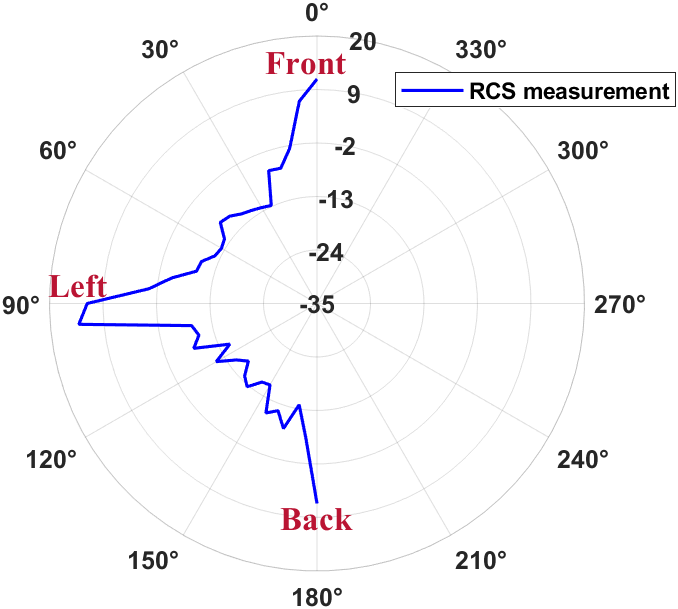}
      }
       \subfloat[]{
        \includegraphics[width=0.45\linewidth]{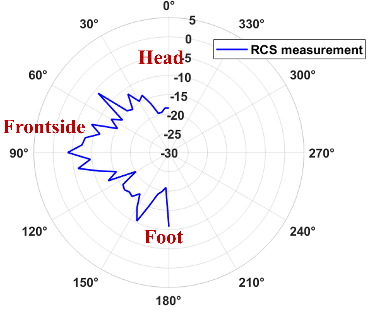}
      }
     \end{minipage}
    \caption{Monostatic RCS distribution of different STs (a) vehicle, (b) UAV with large size, (c) AGV, and (d) human under RCS model 2\cite{ref_北邮119提案}, \cite{ref_北邮121提案}.}
    \label{Different ST RCS}
\end{figure}

In 3GPP Release 19, several companies put forward insightful contributions to monostatic RCS modeling. For example, LG Electronics (LGE) Inc. emphasize the need for dedicated research into forward scattering mechanisms\cite{ref_LGE118b提案}, and vivo Mobile Communication Co., Ltd. propose representing the human body's RCS using multiple STs approach\cite{ref_维沃118提案}. In parallel, extensive measurement campaigns and simulations were conducted by various participants. Table \ref{company contribution} summarizes the RCS measurements and simulations conducted by various companies for several typical sensing targets during the 3GPP Release 19 period. These campaigns are categorized into monostatic sensing and bistatic sensing. 

\subsection{Monostatic RCS} 

To support subsequent research, 3GPP specifies the orientation of an ST as follows: The front of a vehicle, the face of a human, a UAV with large size or an AGV facing the direction with azimuth angle \(\phi=0^\circ\) and zenith angle \(\theta=90^\circ\) in the local coordinate system. For AGVs, the ‘front’ refers to the short edge along the horizontal direction. The top side of the ST is defined as facing the direction with zenith angle \(\theta=0^\circ\).

\begin{figure}[!htbp]
    \centering
    \subfloat[]{\includegraphics[width=0.5\linewidth]{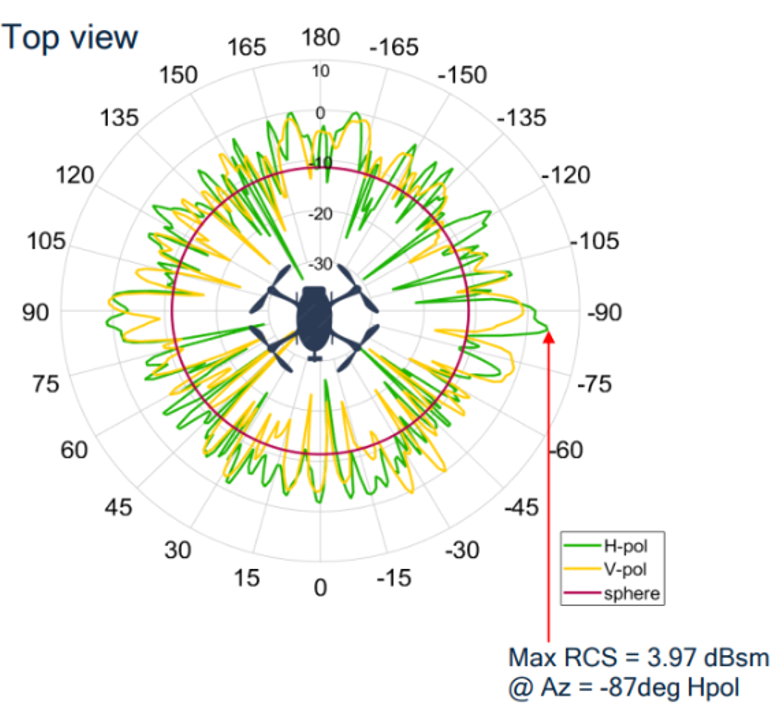}}
    \subfloat[]{\includegraphics[width=0.46\linewidth]{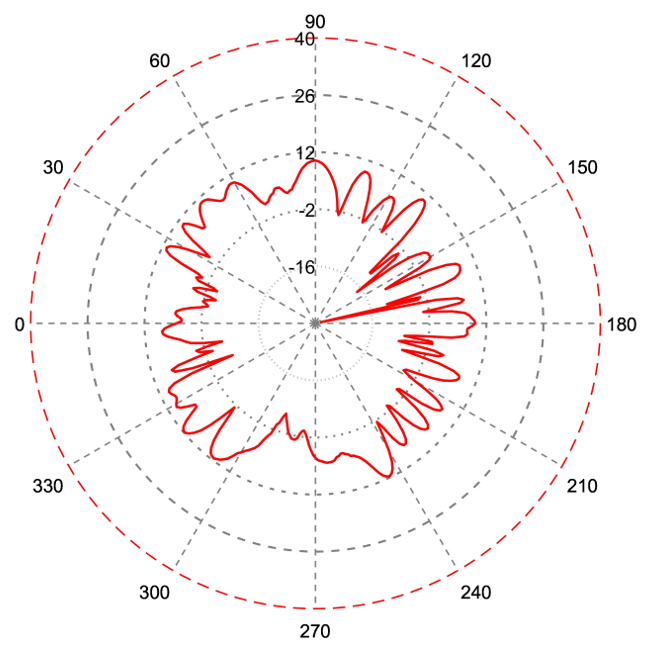}}
    \caption{Monostatic RCS distribution of different STs (a) small UAV \cite{ref_高通120提案} and (b) human under RCS model 1 \cite{ref_小米120b提案}.}
    \label{fig:smallUAV&human1}
\end{figure}

Fig. \ref{Different ST RCS} (a) presents the measured monostatic RCS result of a vehicle \cite{ref_北邮119提案}. Significant scattering peaks are observed on the front, left, rear, and right sides, as well as on the roof of the vehicle. Furthermore, contributions from 3GPP participating organizations exhibit large RCS values in specific scattering angles for UAV with large size, AGV, and human under RCS model 2, as shown in Fig. \ref{Different ST RCS} (b), (c) and (d) \cite{ref_北邮119提案}, \cite{ref_北邮121提案}; while UAV with small size (as depicted by Qualcomm Incorporated in Fig. \ref{fig:smallUAV&human1} (a) \cite{ref_高通120提案}) and human under RCS model 1 (as depicted by Xiaomi in Fig. \ref{fig:smallUAV&human1} (b) \cite{ref_小米120b提案}) exhibit angular independence in their scattering patterns. 

Moreover, the RCS of all STs demonstrates random fluctuations. Based on experimental target scattering characteristics, Beijing University of Posts and Telecommunications (BUPT), Xiaomi Corporation et al. propose RCS models for all SPSTs. The RCS coefficient $\sigma_{\text{RCS}}$ for an SPST comprises three multiplicative components\cite{zhang2025unifiedRCS}:
\begin{equation}\label{eq:rcs}
\sigma_{\text{RCS}} = \sigma_M \cdot \sigma_D \cdot \sigma_S.
\end{equation}
It is worth noting that if the model is transformed into the dB domain, the relationship can be expressed as:
\begin{equation}\label{eq:rcs_dB}
\sigma_{RCS\_dB} = \sigma_{M\_dB} + \sigma_{D\_dB} + \sigma_{S\_dB}.
\end{equation}
The first component $\sigma_M$ represents a large-scale parameter, which is a deterministic value for each SPST. Both the second component $\sigma_D$ and the third component $\sigma_S$ are small-scale parameters. Depending on whether the target scattering characteristic is omnidirectional or angular dependent, $\sigma_D$ can be modeled as unity or as an angular dependent function. Meanwhile, $\sigma_S$ follows a log-normal distribution. To ensure normalization, the mean $\mu_{\sigma_{S\_{dB}}}$ and standard deviation $\sigma_{\sigma_{S\_{dB}}}$ must satisfy the following constraint: $u_{\sigma_{s\_dB}}=\frac{-\ln(10)}{20}\sigma_{s\_dB}^{2}$.

For the UAV with small size and human under RCS model 1, their monostatic scattering characteristics are characterized as an SPST model. The angle-dependent component $\sigma_\mathrm{D}$ is fixed to 1, while $\sigma_\mathrm{M}$ represents the linear average RCS under monostatic sensing. The RCS parameter logarithmic values for the UAV with small size and human under RCS model 1 are provided in Table \ref{tab:RCS model 1}.


\begin{table*}[htbp]
\centering
\caption{Parameters on RCS for the STs with angular independent monostatic RCS values}
\begin{tabular}{m{4cm}<{\centering}|m{4cm}<{\centering}|m{4cm}<{\centering}|m{4cm}<{\centering}}
    \toprule
    \hline
    \textbf{Target} & $10\lg(\sigma_M)$ [dBsm] & $10\lg(\sigma_D)$ [dB] & $10lg(\sigma_s)$ [dB] \\
    \hline
    UAV with small size & -12.81 & 0 & 3.74 \\
    \hline
    Human under RCS model 1 & -1.37 & 0 & 3.94 \\ 
    \hline
    \bottomrule
\end{tabular}
\label{tab:RCS model 1}
\end{table*}

The scattering characteristics of UAV with large size and human under RCS model 2 under monostatic sensing conditions are similarly characterized by single SPST model. For Vehicle and AGV, both single SPST and multiple SPST models are applicable. As recommended by the 3GPP standard, the multiple SPST model for these targets is characterized by five scatter points located at the front, left, back, right and roof side of the vehicle or AGV. For each SPST of the UAV with large size, human under RCS model 2, vehicle, and AGV, the component $\sigma_D$ is angle-dependent and can be modeled using a pattern similar to an antenna radiation pattern. The values/pattern $10\lg(\sigma_M\sigma_D)$, denoted as $\sigma_{MD\_dB}(\theta,\phi)$, of the RCS for a SPST is deterministic based on the incident angle $(\theta,\phi)$.
\begin{equation}
\label{equ_mono RCS model 2}
\sigma_{MD\_dB}(\theta,\phi) = 
G_{\text{max}} - \min\left\{ -\left( \sigma_{\text{dB}}^V(\theta) + \sigma_{\text{dB}}^H(\phi) \right), \sigma_{\max}\right\}
\end{equation}

\noindent where:
\begin{itemize}
    \item $\sigma_{\text{dB}}^V(\theta)$ is defined by:
    \begin{equation}
    \label{eq:vertical}
\sigma_{\mathrm{dB}}^{V}(\theta) = 
\min\biggl\{ -12\left( \frac{\theta - \theta_{\mathrm{center}}}{\theta_{\mathrm{3\,dB}}} \right)^{\!2}, -\sigma_{\max} \biggr\}
    \end{equation}
    \item $\sigma_{\text{dB}}^H(\phi)$is defined by:
    \begin{equation}
    \label{eq:horizontal}
    \sigma_{\mathrm{dB}}^{H}(\phi) = \min\biggl\{ -12\left( \frac{\phi -\phi_{\mathrm{center}}}{\phi_{\mathrm{3\,dB}}} \right)^{\!2}, -\sigma_{\max} \biggr\}
    \end{equation}
\end{itemize}

Due to space limitations, Table \ref{tab:rcs_params} only lists the values of parameters for the large-sized UAV's RCS, such as \(G_{\max}\), \(\sigma_{\max}\), \(\theta_{\mathrm{3dB}}\), \(\phi_{\mathrm{3dB}}\) and so on.

\subsection{Bistatic RCS}

\begin{table*}[ht]
    \centering
    \caption{Parameters on RCS for UAV with large size}
    \label{tab:rcs_params}
    \begin{tabular}{ m{1cm}<{\raggedright}m{1.5cm}<{\centering}m{1.5cm}<{\centering}m{1.5cm}<{\centering}m{1.5cm}<{\centering}m{1cm}<{\centering}m{1cm}<{\centering}m{2.5cm}<{\centering}m{2.5cm}<{\centering}}
        \toprule
        \hline
        \multirow{2}{*}{} & \multicolumn{7}{c}{$ 10\lg(\sigma_{M}\sigma_{D}) $ [dBsm]} \\
        \cmidrule(lr){2-9} 
        & $ \phi_{\text{center}} $ in $ ^{\circ} $ & $ \phi_{3dB} $ in $ ^{\circ} $ & $ \theta_{\text{center}} $ in $ ^{\circ} $ & $ \theta_{3dB} $ in $ ^{\circ} $ & $ G_{\max} $ & $ \sigma_{\max} $ & Range of $ \theta $ in $ ^{\circ} $ & Range of $ \phi $ in $ ^{\circ} $ \\
        \midrule 
        Left & 90 & 7.13 & 90 & 8.68 & 7.43 & 14.30 & [45,135) & [45,135) \\
        \cmidrule(lr){1-9} 
        Back & 180 & 10.09 & 90 & 11.43 & 3.99 & 10.86 & [45,135) & [135,225) \\
        \cmidrule(lr){1-9} 
        Right & 270 & 7.13 & 90 & 8.68 & 7.43 & 14.30 & [45,135) & [225,315) \\
        \cmidrule(lr){1-9} 
        Front & 0 & 14.19 & 90 & 16.53 & 1.02 & 7.89 & [45,135) & [-45,45) \\
        \cmidrule(lr){1-9} 
        Bottom & - & - & 180 & 4.93 & 13.55 & 20.42 & [135,180] & [0,360) \\
        \cmidrule(lr){1-9} 
        Roof & - & - & 0 & 4.93 & 13.55 & 20.42 & [0,45) & [0,360) \\
        \hline
        \bottomrule
    \end{tabular}
\end{table*}

For the case of bistatic sensing, the RCS is extended based on the monostatic sensing scenario. For targets such as UAV with small size and human under RCS model 1, where the angle-dependent component $\sigma_{D}=1$, their RCS in logarithmic scale \(10\log_{10}(\sigma_M \sigma_D)\) is denoted as \(\sigma_{\mathrm{MD}\_{\mathrm{dB}}}(\theta_i, \phi_i, \theta_s, \phi_s)\), which is defined as follows:
\begin{equation} \label{equ_bi RCS model 1 }
    \begin{aligned}
        \sigma_{\mathrm{MD}\_\mathrm{dB}}(\theta_i, \phi_i, \theta_s, \phi_s) 
        &= \max\biggl\{ 
            10\log_{10}(\sigma_M) - 3\sin\left(\frac{\beta}{2}\right), \\
        &\quad \sigma_{\mathrm{FS}}(\theta_i, \phi_i, \theta_s, \phi_s) 
        \biggr\}
    \end{aligned}
\end{equation}

\noindent where:
\begin{itemize}
    \item \((\theta_i, \phi_i)\): Incident ray direction
    \item \((\theta_s, \phi_s)\): Scattered ray direction
    \item \(\sigma_{\mathrm{FS}}\): Forward scattering effect (set to \(-\infty\) in Release 19)
    \item \(\beta\): Angle between incident and scattered rays, $\beta \in [0^\circ, 180^\circ]$
\end{itemize}
The values of $\sigma_M$, $\sigma_D$, $\sigma_S$ still refer to Table \ref{tab:RCS model 1}.

For large UAVs with a single scattering point, vehicles with single / multiple SPSTs, and AGVs with single / multiple SPSTs, their logarithmic scale RCS \(10\log_{10}(\sigma_M \sigma_D)\) is still denoted as \(\sigma_{\mathrm{MD}\_{\mathrm{dB}}}(\theta_i, \phi_i, \theta_s, \phi_s)\), which is defined as follows:

\begin{equation} \label{equ_bi RCS model 2}
    \begin{aligned}
        &\sigma_{\mathrm{MD\_dB}} (\theta_i, \phi_i, \theta_s, \phi_s) =\\
        &\max\bigg\{ 
            G_{\max} - \min\left\{ -\left( \sigma^V_{\mathrm{dB}}(\theta) + \sigma^H_{\mathrm{dB}}(\phi) \right), \sigma_{\max} \right\} -\\
            & k_1 \sin\left( \frac{k_2 \beta}{2} \right) + 5 \log_{10} \left( \cos\left( \frac{\beta}{2} \right) \right), \\
            & G_{\max} - \sigma_{\max},\sigma_{\mathrm{FS}} (\theta_i, \phi_i, \theta_s, \phi_s)
        \bigg\}
    \end{aligned}
\end{equation}
with $\sigma_{\text{dB}}^V(\theta)$ and $\sigma_{\text{dB}}^H(\phi)$ are defined by:
\begin{align}
    \sigma_{\mathrm{dB}}^{V}(\theta) &= 
    \min\biggl\{ -12\left( \frac{\theta - \theta_{\mathrm{center}}}{\theta_{\mathrm{3\,dB}}} \right)^{\!2}, -\sigma_{\max} \biggr\}
    \label{eq:vertical} \\
    \sigma_{\mathrm{dB}}^{H}(\phi) &= 
    \min\biggl\{ -12\left( \frac{\phi - \phi_{\mathrm{center}}}{\phi_{\mathrm{3\,dB}}} \right)^{\!2}, -\sigma_{\max} \biggr\}
    \label{eq:horizontal}
\end{align}

\noindent where:
\begin{itemize}
    \item \((\theta, \phi)\): zenith angle and azimuth angle of the bisector the incidence and scattered rays, whose zenith angles and azimuths are $(\theta_i, \phi_i)$ and $(\theta_s, \phi_s)$
    \item \(\sigma_{\mathrm{FS}}\): Forward scattering effect (set to \(-\infty\) in Release 19)
    \item \(\beta\): Angle between incident and scattered rays, $\beta \in [0^\circ, 180^\circ]$
    \item $k_1$, $k_2$ for different targets: 
        \begin{itemize} 
            \item $k_1= 6.05$, $k_2=1.33$ for UAV with large size.
            \item $k_1=0.5714$, $k_2=0.1$ for human under RCS model 2. 
            \item $k_1= 6$, $k_2=1.65$ for vehicle with single/multiple SPSTs.
            \item $k_1= 12$, $k_2=1.45$ for AGV with single/multiple SPSTs. 
        \end{itemize}
\end{itemize}




\subsection{Polarization Scattering Characteristics
}
During propagation, the transmitted signal undergoes reflection, diffraction, or scattering upon encountering objects. This interaction alters the polarization direction of the signal, with the extent of change influenced by the target's geometry, size, and electromagnetic material properties. Consequently, capturing and modeling the polarimetric scattering characteristics of STs can enhance target detection and recognition. TR 38.901 employs the CPM to represent polarization variations caused by ground reflections. During the Release 19 period, the polarimetric scattering characteristics of the targets are also characterized by the use of $CPM$.
The $CPM_{(sp,i)}$ of an SPST for a pair $i$ of incident/scattered angles is generally modeled by amplitude factors $\alpha_{i,1}$, $\alpha_{i,2}$, $\beta_{i,1}$, $\beta_{i,2}$, and initial random phases $\left\{\Phi_{s p, i}^{\theta \theta}, \Phi_{s p, i}^{\theta \phi}, \Phi_{s p, i}^{\phi \theta}, \Phi_{s p, i}^{\phi \phi}\right\}$, i.e.,
\begin{equation}
\label{equ_tar-sma}
C P M_{s p, i}=\left[\begin{array}{ll}
\alpha_{i, 1} \exp \left(j \Phi_{s p, i}^{\theta \theta}\right) & \beta_{i, 1} \exp \left(j \Phi_{s p, i}^{\theta \phi}\right) \\
\beta_{i, 2} \exp \left(j \Phi_{s p, i}^{\phi \theta}\right) & \alpha_{i, 2} \exp \left(j \Phi_{s p, i}^{\phi \phi}\right)
\end{array}\right].
\end{equation}
For UAV, human, vehicle or AGV, $\alpha_{i,1}=\alpha_{i,2}=1$, $\beta_{i,1}=\beta_{i,2}=\sqrt{\kappa_{s p, i}^{-1}}$, i.e.,
\begin{equation}
\label{equ_tar-sma}
C P M_{s p, i}=\left[\begin{array}{cc}
\exp \left(j \Phi_{s p, i}^{\theta \theta}\right) & \sqrt{\kappa_{s p, i}^{-1}} \exp \left(j \Phi_{s p, i}^{\theta \phi}\right) \\
\sqrt{\kappa_{s p, i}^{-1}} \exp \left(j \Phi_{s p, i}^{\phi \theta}\right) & \exp \left(j \Phi_{s p, i}^{\phi \phi}\right)
\end{array}\right]
\end{equation}
where $\kappa_{s p, i}$ is the cross-polarization ratios (XPRs) of the pair $i$ of incident/scattered angles. The XPR describes the ratio between the power of the co-polarized signal and the power of the cross-polarized signal, and is typically expressed in decibels (dB):
\begin{equation}
    \label{equ_defined xpr}
    \mathrm{XPR}=10log_{10}(\frac{P_{\mathrm{co}}}{P_{\mathrm{cross}}})[\mathrm{dB}]
\end{equation}

\begin{itemize}
\item $P_{co}$: The received power in the co-polarized direction
\item $P_{cross} $: The received power in the cross-polarized direction
\end{itemize}
According to the contributions of companies, $\kappa_{s p, i}$ is randomly generated by log-normal distribution per target type defined in Table \ref{tab:xpr}, and $\left\{\Phi_{s p, i}^{\theta \theta}, \Phi_{s p, i}^{\theta \phi}, \Phi_{s p, i}^{\phi \theta}, \Phi_{s p, i}^{\phi \phi}\right\}$ is uniformly distributed within $(-\pi,\pi]$.

\begin{table}[htbp]
    \caption{Statistical parameters of XPR for different STs}
    \centering
    \begin{tabular}{m{2.5cm}<{\centering}|m{2.5cm}<{\centering}m{2.5cm}<{\centering}}
        \toprule
        \hline
        \textbf{Target} & $\mu_{\mathrm{XPR}}$ & $\sigma_{\mathrm{XPR}}$ \\
        \hline
        UAV     & 13.75 & 7.07  \\
        Human   & 19.81 &  4.25 \\
        Vehicle & 21.12 &  6.88 \\
        AGV     & 9.6   &  6.85 \\
        \hline
        \bottomrule
    \end{tabular}
    \label{tab:xpr}
\end{table}

\section{Target Channel Characteristics and Modeling } \label{section4}

\begin{figure*}[h]
    \centering
    \subfloat[]{\includegraphics[width=2.3in]{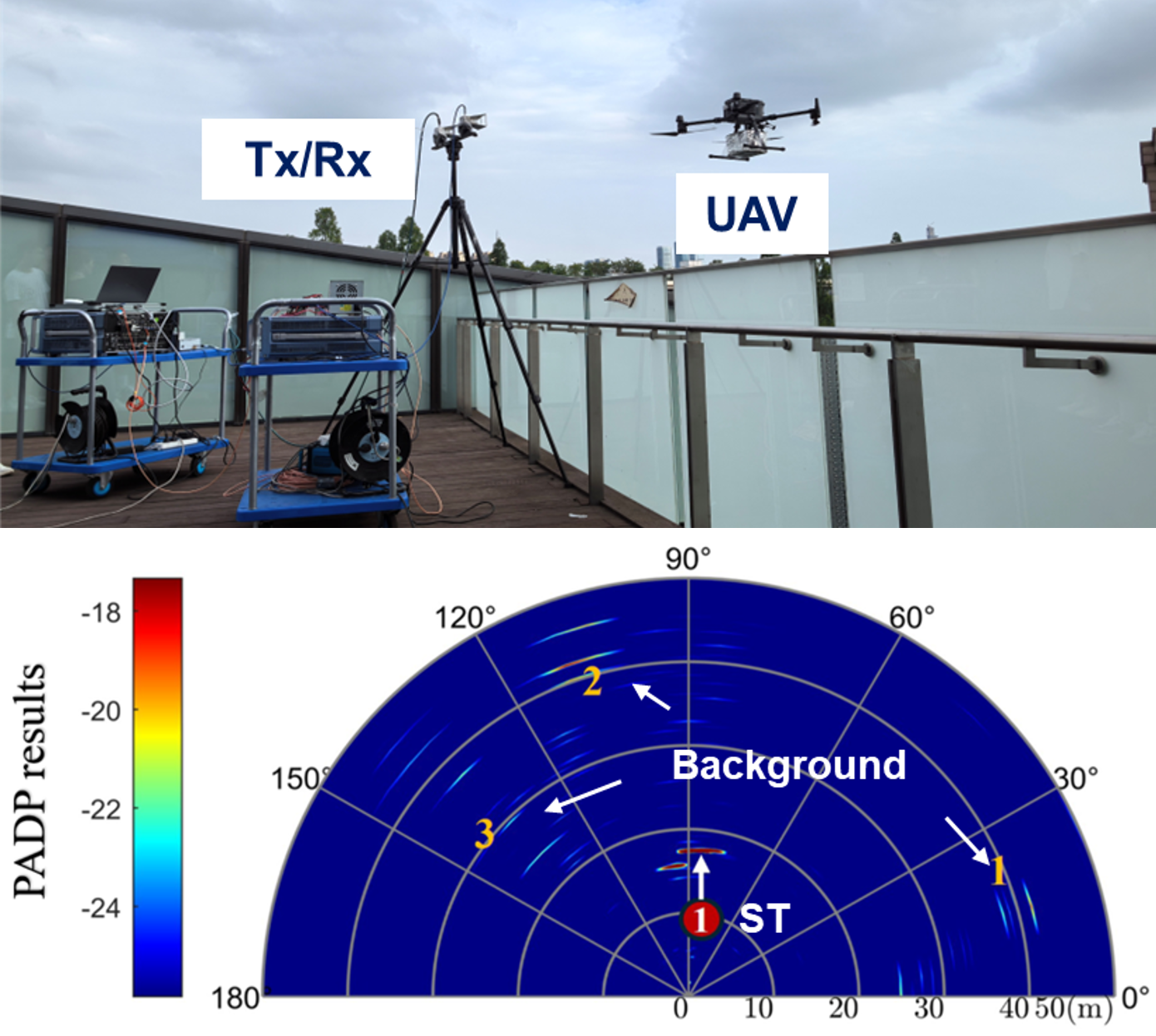}\label{fig:uav}}~
    \subfloat[]{\includegraphics[width=2.3in]{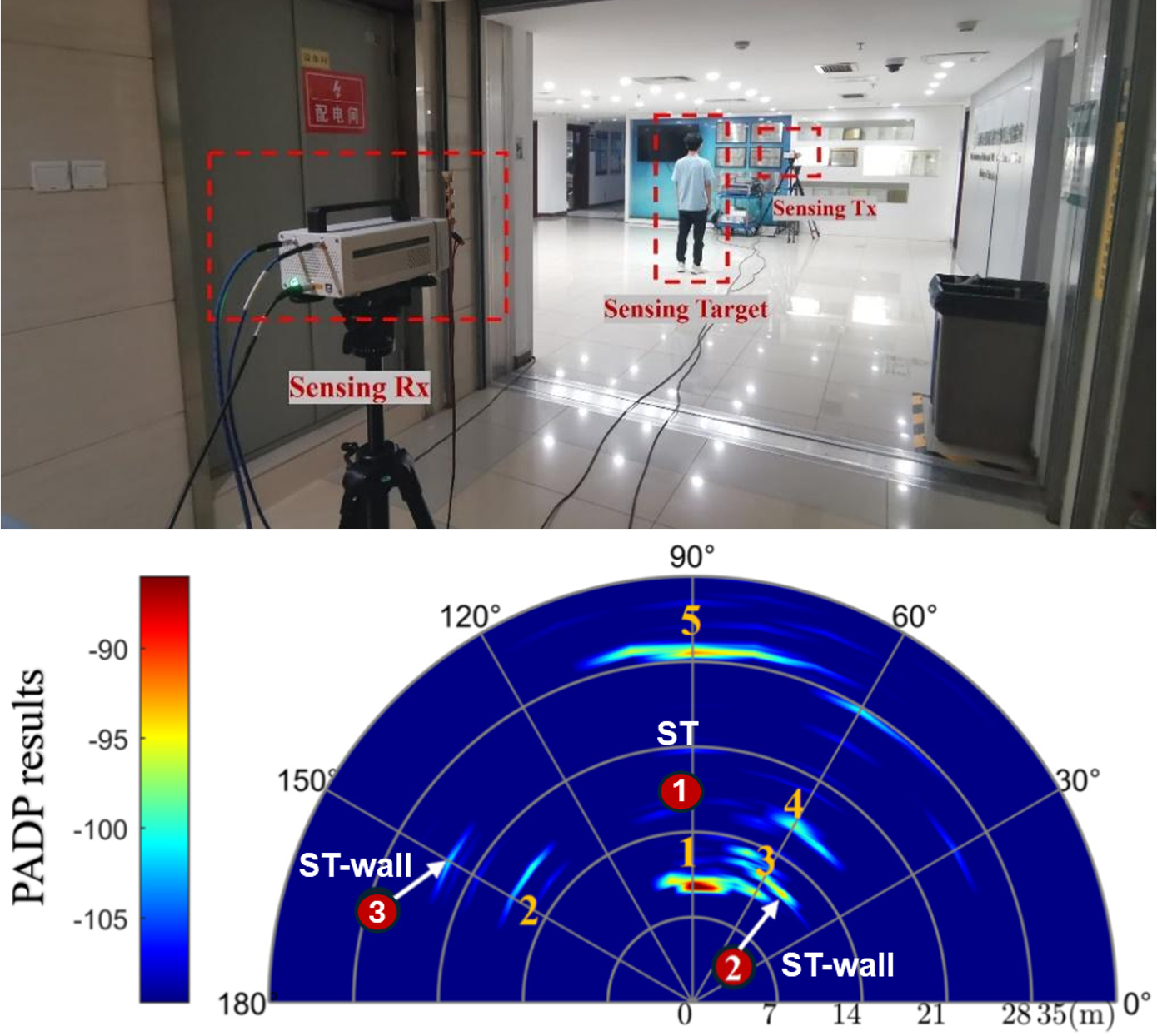}\label{fig:human}}~
    \subfloat[]{\includegraphics[width=2.3in]{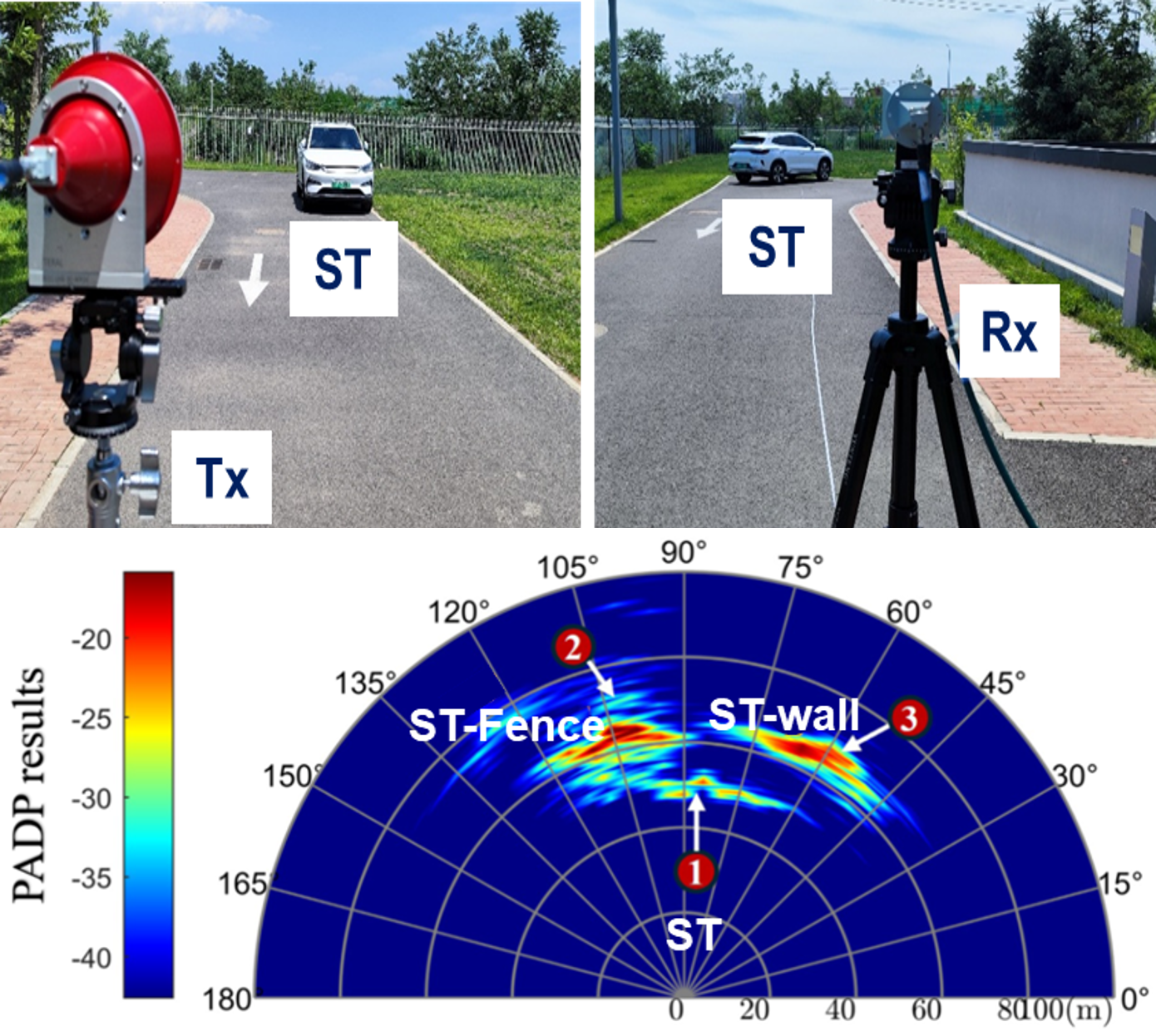}\label{fig:vehicle}}
    \caption{Comparative analysis of sensing target channel characteristics in three measurement scenarios, (a) UMa-UAV, (b) Indoor-Human, and (c) UMi-Vehicle \cite{zhang2025research}.}
    \label{fig_sce}
\end{figure*}

In 3GPP Release 19, substantial technical discussions were devoted to the modeling of target channels. This section presents a systematic survey from four key perspectives: direct and indirect paths, LoS probability, channel concatenation, and multiple targets.

\subsection{Direct and Indirect Paths}
For discussion purpose, the \#117 meeting \cite{ref_北邮117提案} in Japan approves the proposal to categorize the propagation paths in the target channel as: \textit{(1) the direct path}, i.e., LoS ray from Tx to target + LoS ray from target to Rx, \textit{(2) the indirect paths}, i.e., any propagation path other than the direct path, including LoS + NLoS, NLoS + LoS, and NLoS + NLoS rays. 

For the direct path of a target, the propagation parameters can be deterministically generated based on TR 38.901. Specifically,  Azimuth angle of Arrival (AoA)/ Zenith angle of Arrival (ZoA) at Rx and Azimuth angle of Departure (AoD) / Zenith angle of Departure (ZoD) at target can be generated based on the 3D location of target and Rx in the global coordinate system, AoD / ZoD at Tx, AoA / ZoA at target are generated based on the 3D location of Tx and target. The absolute propagation delay is expresses as $\tau=(d_1+d_2)/c$, with c denoting the speed of electromagnetic wave. The phase difference caused by the mobility of the target or Tx/Rx can be calculated as 
\begin{equation}
\label{equ_framework}
\Delta \varphi=\frac{2 \pi}{\lambda} \cdot\left[d_1 \cos \left(\Delta \theta_1\right)+d_2 \cos \left(\Delta \theta_2\right)\right],
\end{equation}
where $\Delta \theta_1$ represents the angle between the lines connecting the Tx and the target positions before and after the relative movement. $\Delta \theta_2$ represents the angle between the lines connecting the target and the Tx positions before and after the relative movement. 
The mobilit velocity of the target also more directly influences the Doppler frequency of the direct paths, as detailed in Section \ref{section6}-C.
Therefore, the direct path carries the most target information, enabling localization functionality. To simplify simulations or when considering open scenarios such as UAVs, some literature only discusses the direct path\cite{gomez2021air}. In \cite{ref_北邮117提案}, an 28 GHz monostatic channel measurement is conducted with a UAV as the target, where the Tx and Rx deployed on a rooftop. As shown in the Fig. \ref{fig:uav}, the PADP measurement results exhibit a strong dominant direct path, while the indirect paths are negligible, with the remaining paths representing background echoes. 

However, in complex scenarios such as indoor environments or vehicular networks, indirect paths are indeed present, as demonstrated by empirical measurements \cite{ref_北邮117提案,ref_爱立信117提案}. In \cite{ref_爱立信117提案}, an outdoor channel measurement is conducted with a pedestrian as the target, where the Tx and Rx are located in a square and on top of a building, respectively. The results show that the direct path exhibits significantly higher received power than the indirect (scattered) path, allowing it to be neglected based on the 25 dB power threshold criterion. In \cite{ref_北邮117提案}, an ISAC channel measurement campaign is carried out at an indoor hall and an outdoor road in BUPT. As shown in the Fig. \ref{fig:human} and Fig. \ref{fig:vehicle}, the power of indirect paths in the target channel reaches significant levels, which demonstrates the necessity of indirect path modeling. In summary, in complex indoor or outdoor scenarios, indirect paths can be observed from channel measurements, but their power proportion is closely related to the specific layout of the scenario.



Recent studies have placed significant emphasis on the modeling of indirect paths, primarily including methods based on stochastic clusters and based on EOs. On the one hand, when the stochastic cluster is used to generate the indirect paths in the target channel of a target, the simulation flow and parameter tables in section 7, TR 38.901 can be reused with some modifications.
On the other hand, considering the model's universality and low complexity, the ISAC channel model still uses the statistical modeling method GBSM as a starting point. However, since sensing function requires the realization of target positioning or tracking, deterministic parameters have been considered to introduce into ISAC channel modeling. To address the modeling requirements for target localization and tracking in sensing function, the concept of EO is introduced, defined as a non-target object with known location. The indirect paths generated by EOs are demonstrated to significantly enhance positioning accuracy. In the discussion, type-1 EO has comparable physical characteristics as a sensing target while type-2 EO has extremely large size. In a channel simulation, the multipath parameters of Tx-EO-target or target-EO-Rx can be generated according to the geometric relationship after modeling EO's position. Further details on modeling indirect paths using EOs are still being continuously discussed in the ISAC channel standardization process.

\subsection{LoS Probability}


Traditional communication channels consider whether the propagation conditions between Tx and Rx are LoS or NLoS condition, and generate different channel parameters accordingly. Since the target divides the sensing channel into Tx-target and target-Rx links, each link may have different propagation conditions. The \#116 bis meeting classifies the propagation conditions of the target channel into the following four categories: \textit{(1)} LoS condition in both Tx-target and target-Rx links, \textit{(2)} LoS condition in Tx-target and NLoS condition in target-Rx link, \textit{(3)} NLoS condition in Tx-target and LoS condition in target-Rx link, \textit{(4)} NLoS condition in both Tx-target and target-Rx links.

In TR 38.901, the LoS probability in scenarios of Urban Macro (UMa), Urban Micro (UMi) - Street canyon, Rural Macro (RMa), Indoor - Mixed office, Indoor - Open office, and Indoor factory (InF) with different clutter density and BS height is already defined to decide the LoS condition for Tx (BS) to Rx (UT) link. It depends on parameters such as the heights of the BS and UT, the 2D distance between the BS and UT. A straightforward way to determine the LoS conditions for ISAC channels is to reuse the existing distance-dependent LoS/NLoS condition decision probability model from TR 38.901. For bistatic sensing, the LoS probabilities of the Tx-target and target-Rx links are calculated separately, whereas for monostatic sensing, the LoS probability of Tx-target or target-Rx link is calculated. 

\begin{figure}[!htbp]
    \centering
    \subfloat[]{\includegraphics[width=0.5\columnwidth]{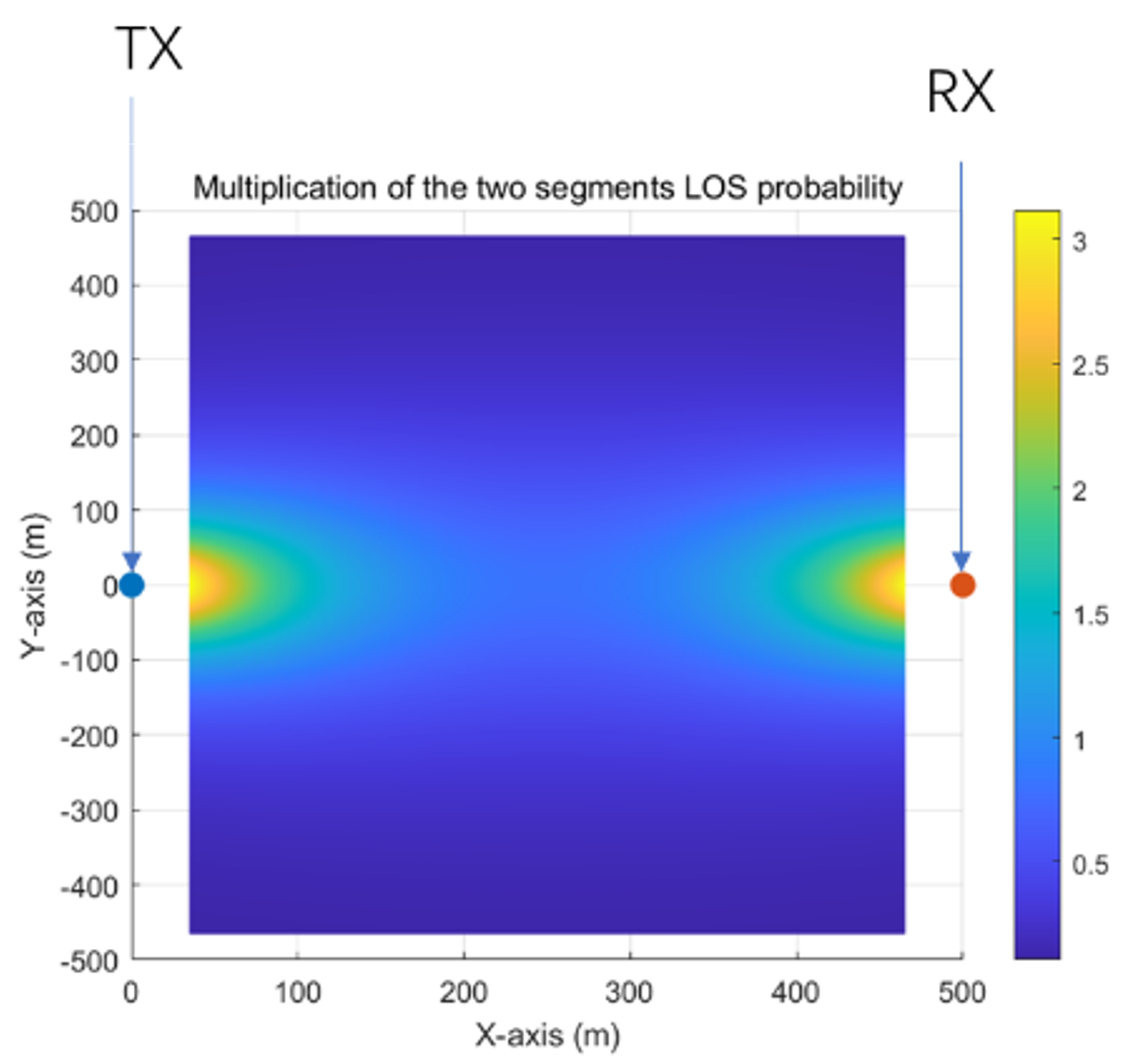}}
    \hfill
    \subfloat[]{\includegraphics[width=0.5\columnwidth]{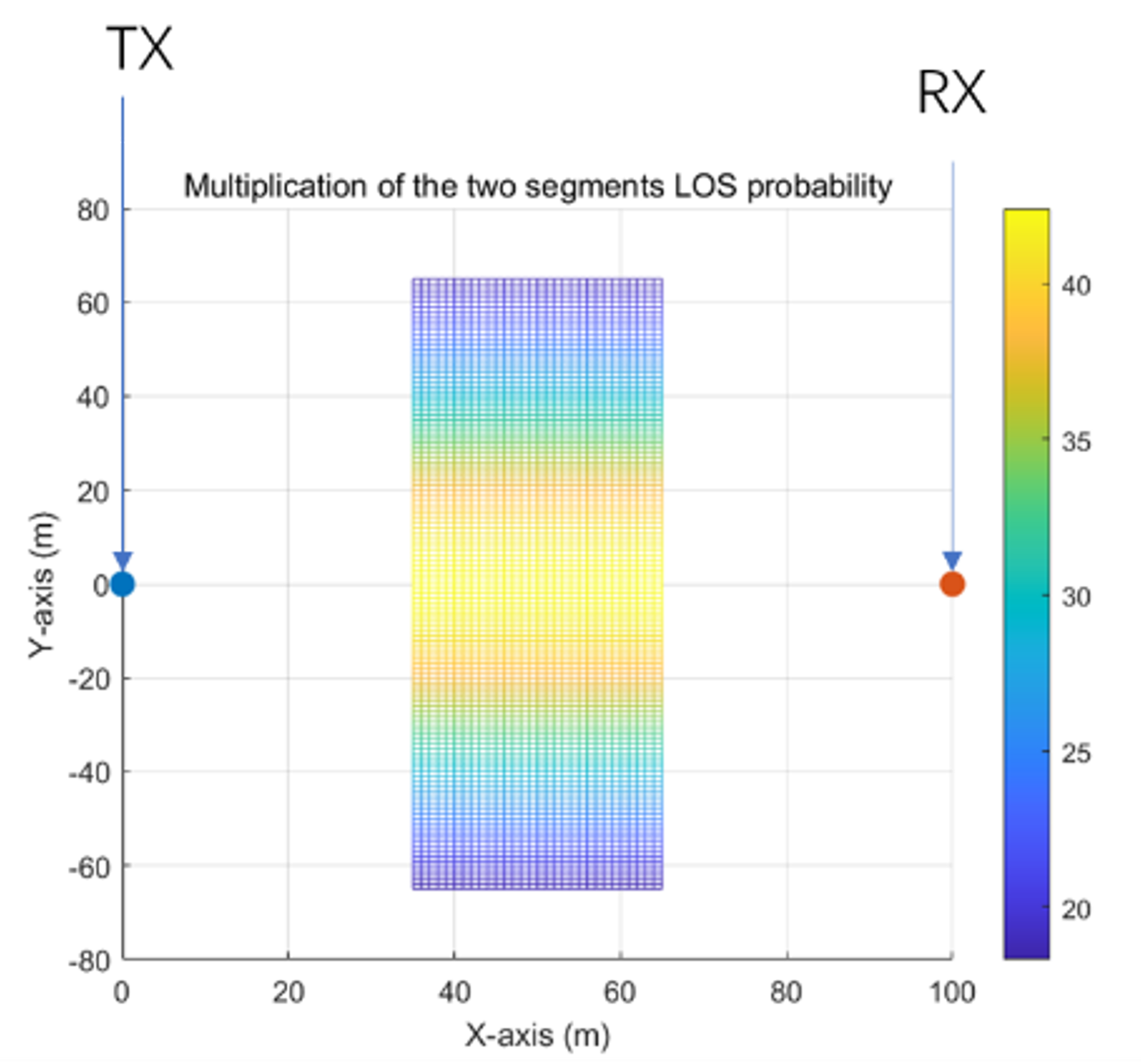}}\\
    \subfloat[]{\includegraphics[width=0.5\columnwidth]{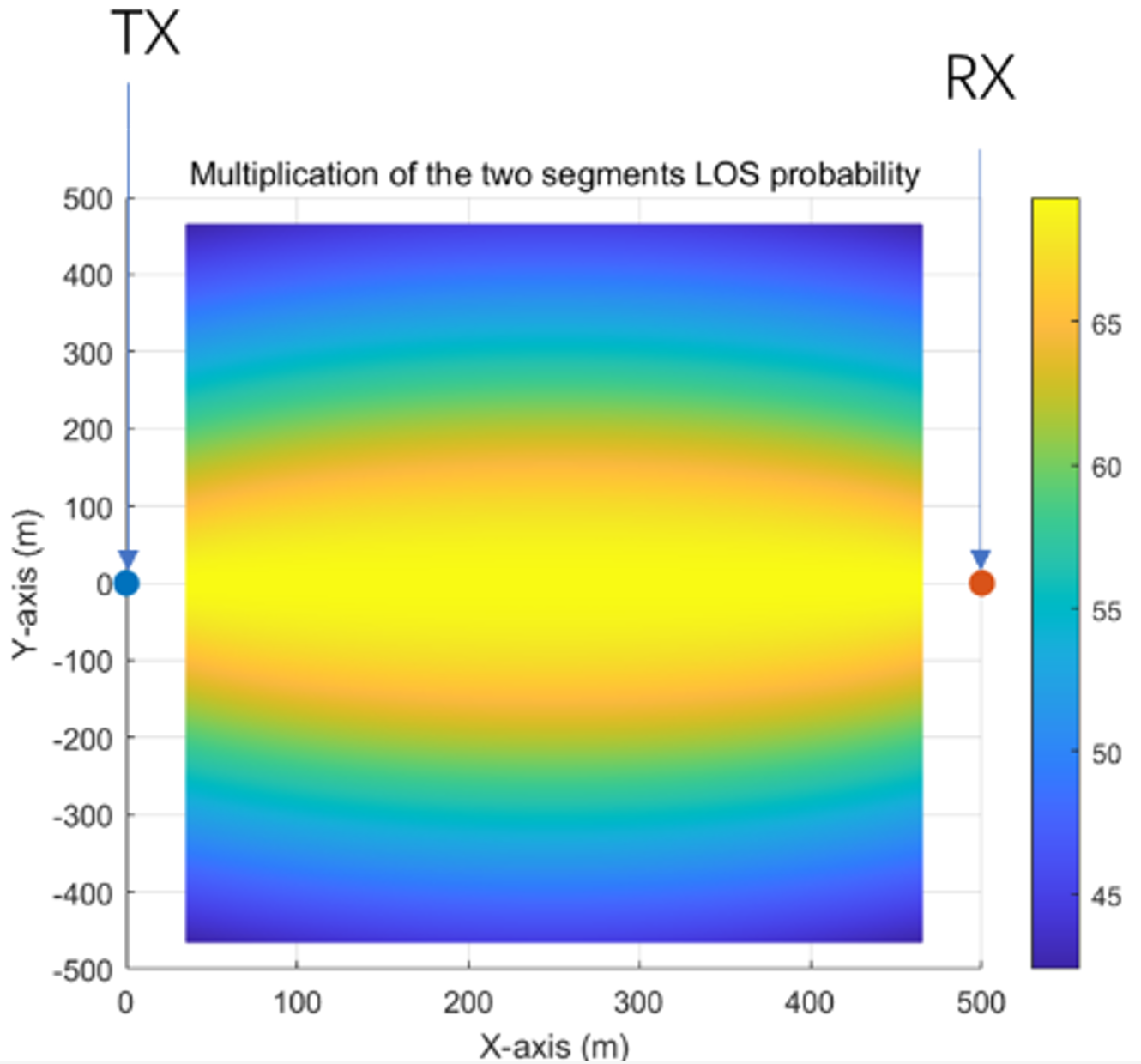}}
    \hfill
    \subfloat[]{\includegraphics[width=0.5\columnwidth]{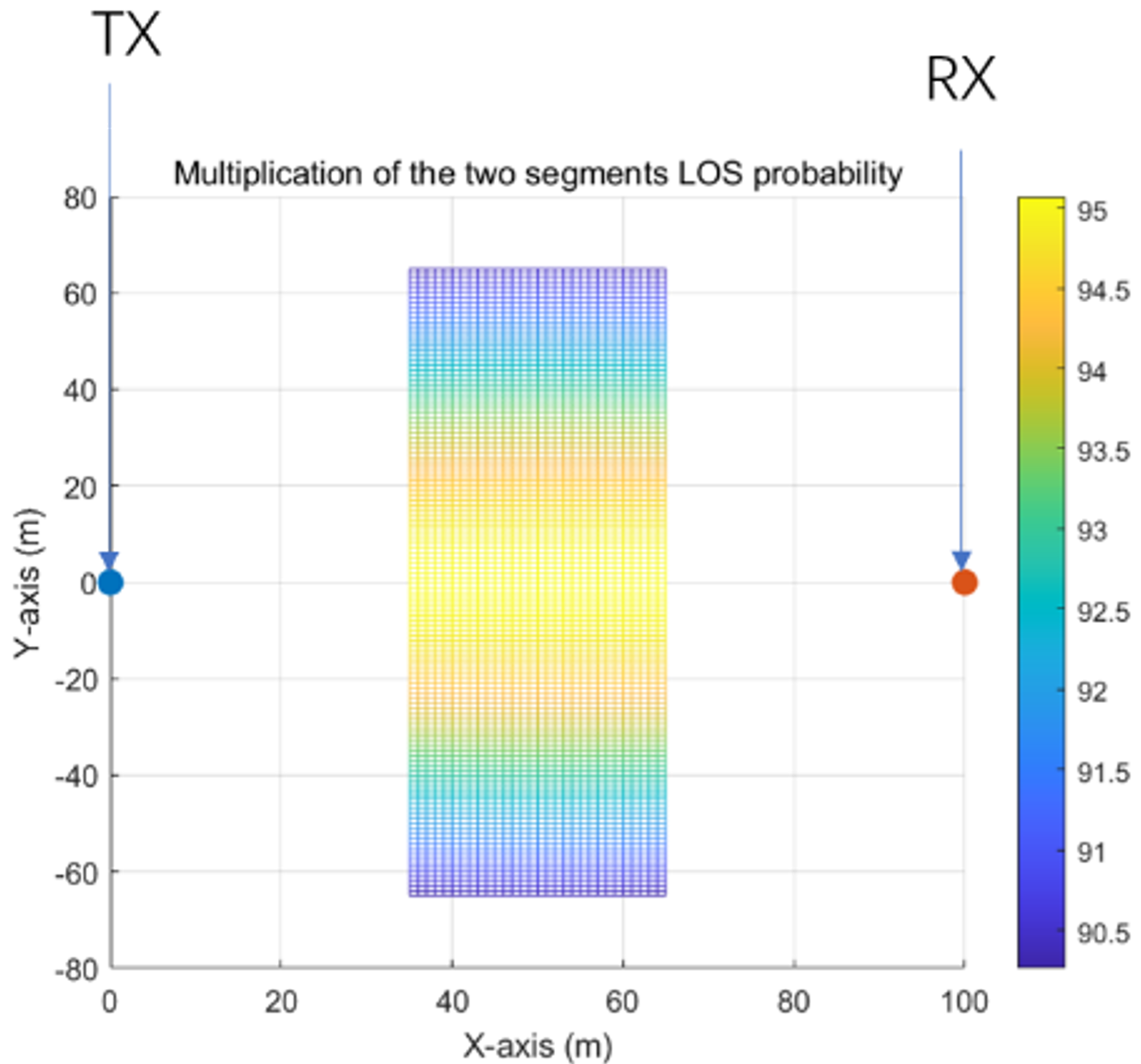}}
    \caption{The LoS probability calculated from the multiplication of two segments. (a) In UMa scenario, the x range is [35 m,465 m], y range is [-465 m, 465 m], (b) In UMa scenario, the x range is [35 m,65 m], y range is [-65 m, 65 m], (c) In UMa-UAV scenario, the x range is [35 m, 465 m], y range is [-465 m, 465 m], and (d) In UMa-UAV scenario, the x range is [35 m,65 m], y range is [-65 m, 65 m] \cite{ref_移动116提案}.}
    \label{figs_移动116提案双场景LoS概率图}
\end{figure}

The work in \cite{ref_移动116提案} simulates the multiplied LoS probability of UMa scenario and UMa-UAV scenario for bistatic sensing. As shown in Fig. \ref{figs_移动116提案双场景LoS概率图}, it can be found that the LoS probability of Tx-target-Rx link decreases sharply as the distance increases, which will significantly reduce the probability of successful target detection. The literature \cite{ref_移动116提案} suggests that decreasing the Tx-Rx distance is a candidate method to improve the LoS probability, while changing the LoS probability is another way with extra measurement validation. Literature \cite{ref_三星118提案} suggests that the EO blockage also affect the LoS/NLoS conditions of ISAC links. Moreover, considering the limitations of BS and UT heights in the communication standards, the LoS probability model needs to be adjusted based on measurements for various heights of sensing targets such as pedestrians, vehicles, and UAVs \cite{ref_小米116提案,ref_展讯117提案}.

\subsection{Channel Concatenation}

The traditional communication channel model focuses on the statistical characteristics between Tx and Rx, without considering the specific attributes of intermediate scatterers. However, Sensing requires explicit information about the target's position, velocity, RCS, etc. to enable functions such as localization and environmental reconstruction. To describe the sensing target in ISAC channel model, some perspectives advocate modeling the sensing channel, especially the indirect paths, as concatenation  \cite{ref_刘亚萌WCL,ref_移动116提案,ref_小米116提案}, while other works believe that traditional non-concatenated models are sufficient to support sensing evaluations\cite{ref_北交综述和簇替换,ref_Oppo116提案,ref_诺基亚116提案,ref_联想116提案}.

\begin{figure}[ht]
    \centering
    \includegraphics[width=3.0in]{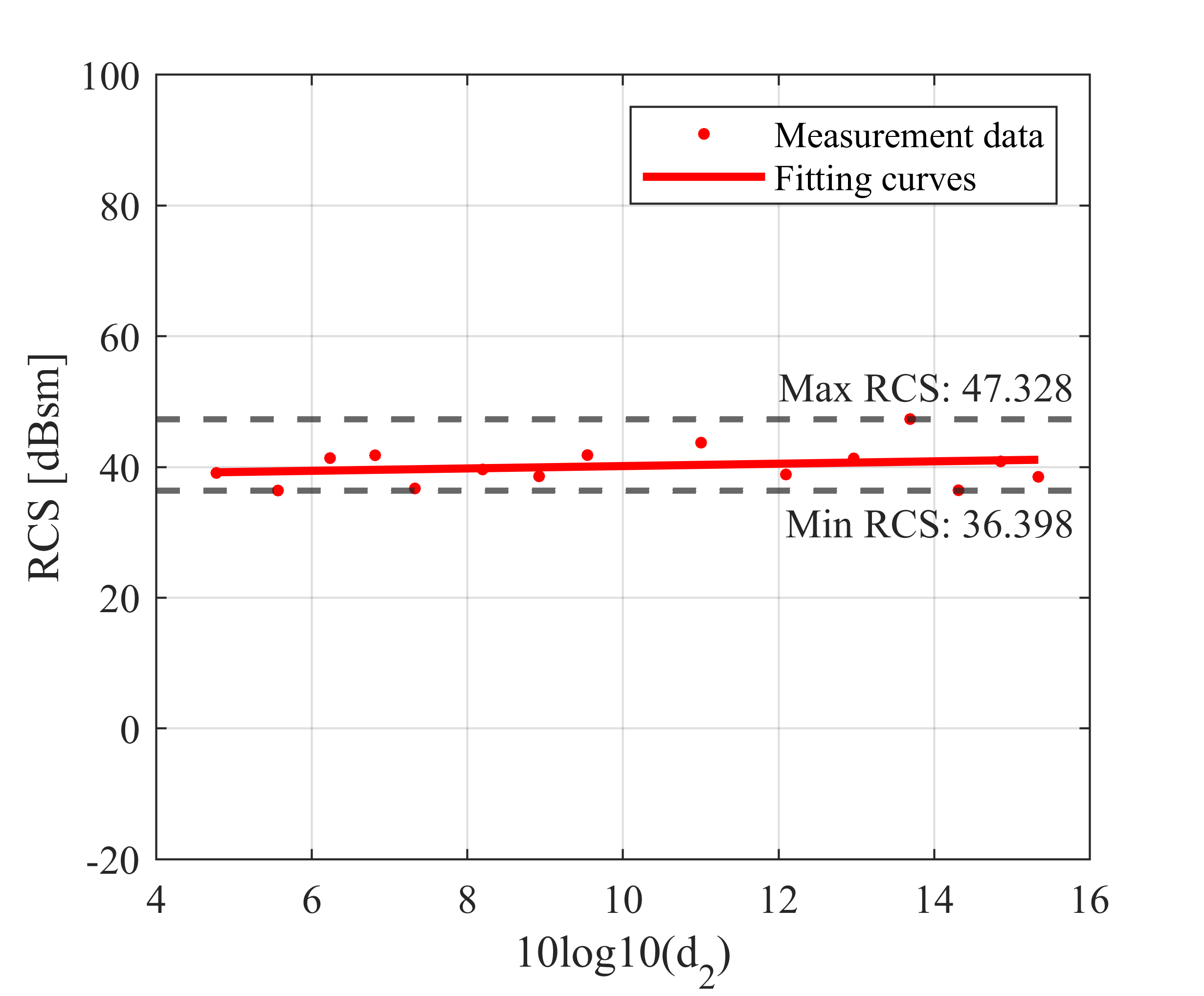}
    \caption{RCS calculated based on the concatenation model of path loss at different distances \cite{ref_小米120b提案}.}
    \label{fig_胡锡栋大尺度RCS}
\end{figure}

In the concatenated modeling solutions, the sensing target channel path loss derived from the Friis transmission formula is widely accepted and used for evaluating the target's link budget.
Specifically, Friis transmission formula is expressed as
\begin{equation}
\label{equ_friis}
P_R=\frac{P_T G_T {\sigma_{\text{RCS}}} G_R \lambda^2}{(4 \pi)^3 d_{1}^2 d_{2}^2},
\end{equation}
where $P_R$ and $P_T$ demote the receving and transmitting power, respectively. $G_R$ and $G_T$ are the power gain of the antennas or amplifiers at the Tx and Rx side, respectively. The wavelength of the electromagnetic wave is expressed as $\lambda$. $d_1$ and $d_2$ represent the 3D distances of Tx-target and target-Rx links. $\sigma_{\text{RCS}}$ denotes the RCS related coefficient of a sensing target at the large-scale level. Considering modeling the sensing target channel by adding the path loss of the Tx-target link and TAR-Rx link ($P L_{d B}\left(d_1\right)+P L_{d B}\left(d_2\right)$), the impact of the antenna aperture is calculated twice as much but missing the impact of target RCS. Consequently, the path loss of the sensing target channel can be modeled as:
\begin{align}
\label{equ_pl}
P L_{d B}\left(d_1, d_2\right)=& P L_{d B}\left(d_1\right)+P L_{d B}\left(d_2\right)\\ \notag
&+10 \log _{10}\left(\frac{\lambda^2}{4 \pi}\right)-10 \log _{10}\left(\sigma_{\text{RCS}}\right).
\end{align}
In the channel simulation, $P L_{d B}\left(d_1\right)$ and $P L_{d B}\left(d_2\right)$ can reuse the model and parameters given in TR 38.901/36.777/37.885 \cite{3gpp38901} considering different scenarios. 
The work in \cite{ref_小米116提案} has conducted channel measurements in an L-shaped corridor for Tx-target, target-Rx, and Tx-target-Rx links. As shown in Fig. \ref{fig_胡锡栋大尺度RCS}, the red dots represent measurement data taken under constant $d_1$ conditions, meaning the RCS is approximately constant. The difference between $P L_{d B}\left(d_1, d_2\right)$ and $P L_{d B}\left(d_2\right)$ is measured at 15 different $d_2$, with the vertical axis is in dB. The slope of the fitted curve is approximately 0, which validates the correctness of (\ref{equ_pl}).

As for the concatenation-based small-scale characteristics in target channel, CMCC Co., Ltd. and several other organizations in 3GPP \cite{ref_移动117提案,ref_北邮117提案} considering the propagation mechanism, propose modeling the clusters/multiples of the Tx-target-Rx link as the convolution of the clusters/multiples of the Tx-TAR and TAR-Rx links.

\begin{figure}[!htbp]
    \centering
    \subfloat[]{\includegraphics[width=2.5in]{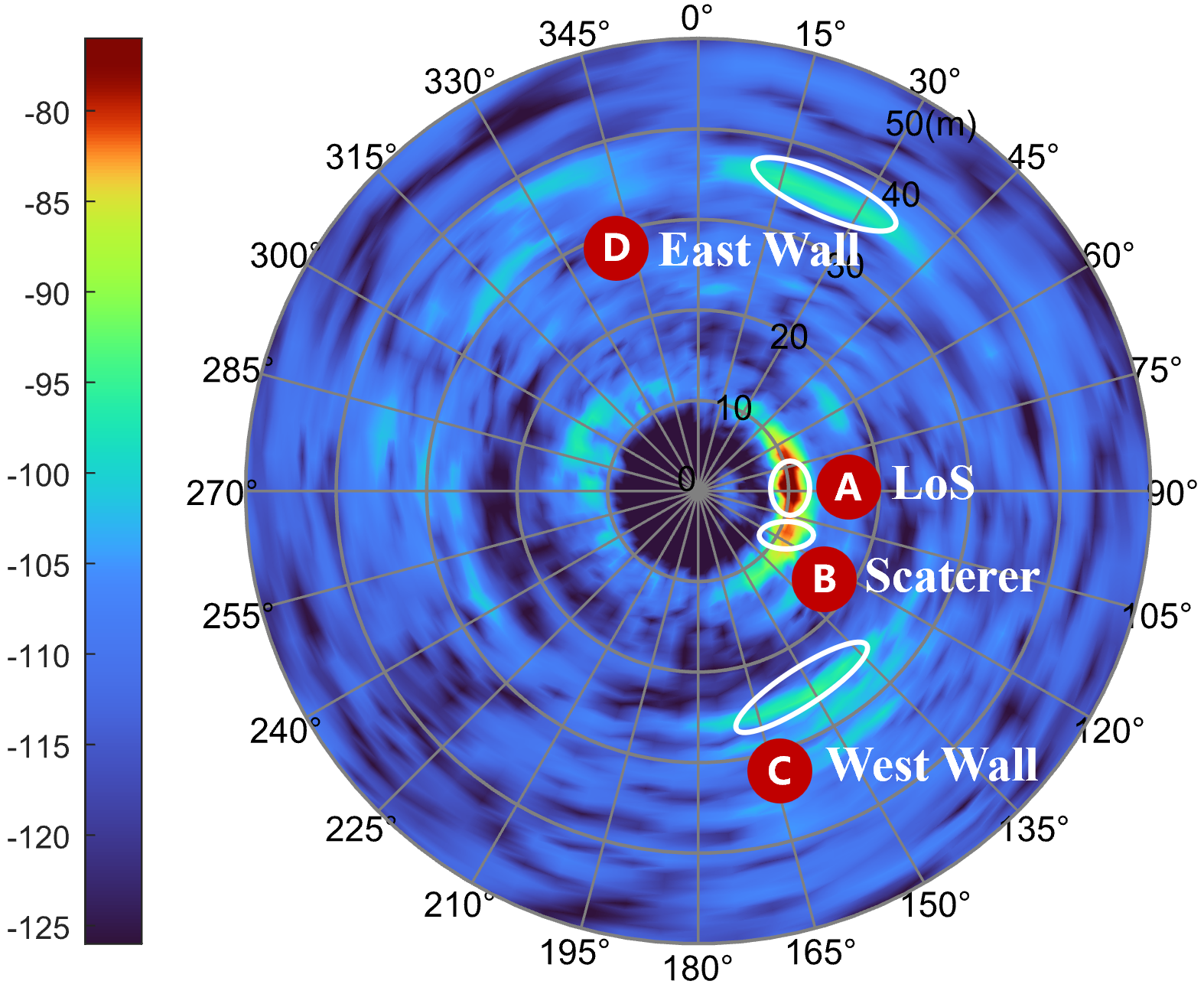}}
    \hfill
    \subfloat[]{\includegraphics[width=2.5in]{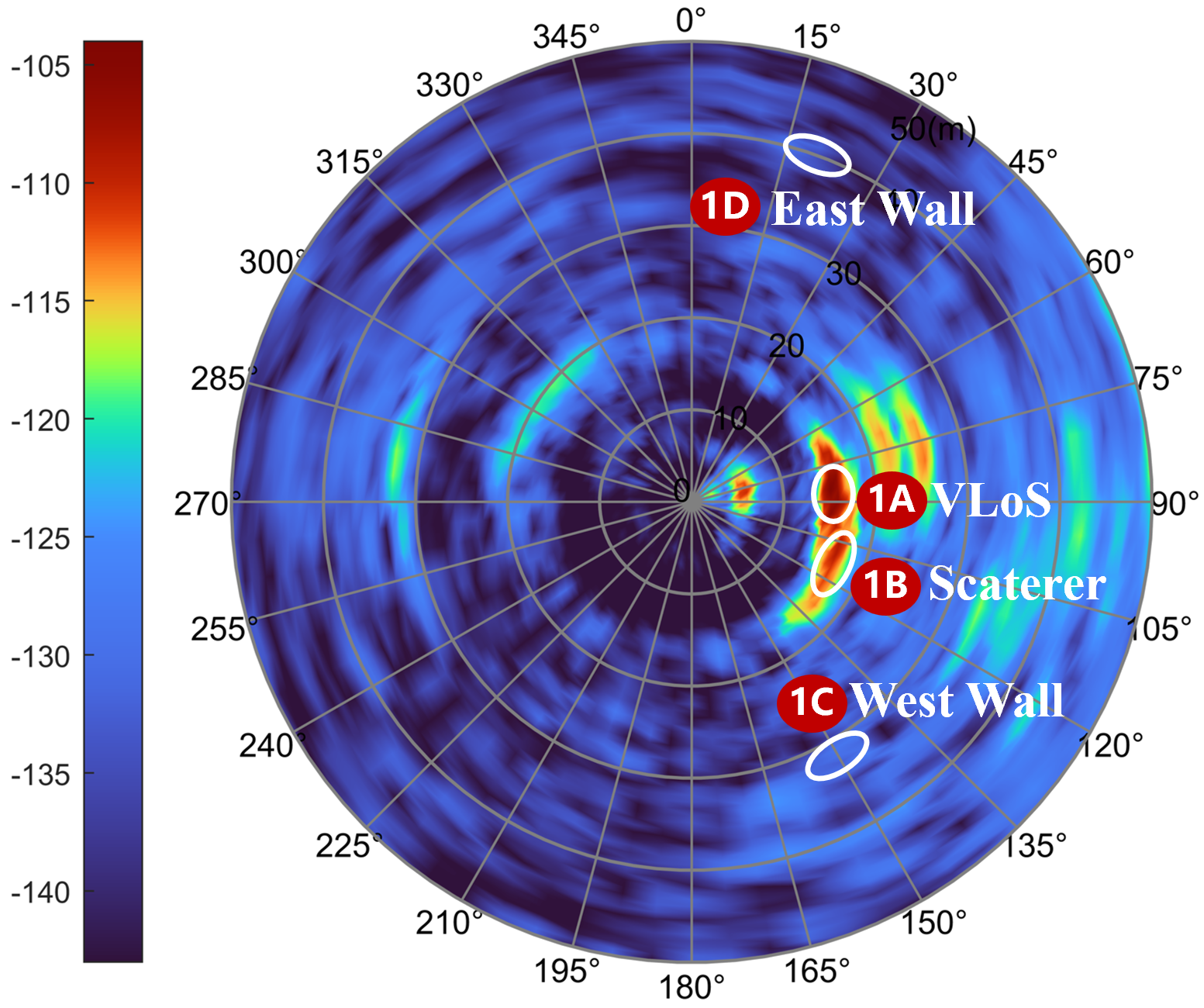}}
    \caption{PADPs of (a) target-Rx measurement, (b) Tx-target-Rx measurement \cite{ref_张骥威glob}.}
    \label{figs_小尺度级联验证PADP}
\end{figure}

In \cite{ref_北邮118提案}, channel measurements are conducted in a IIoT scenario for the Tx-target, target-Rx, and Tx-target-Rx links, with a Reconfigurable Intelligence Surface (RIS) as the target. As shown in the Fig.~\ref{figs_小尺度级联验证PADP}, there are four multipaths A-D in the target-Rx link. For the measured path 1 in the Tx-target link, Tx-target-Rx link appears as 4 concatenated paths, 1A-1D, at the corresponding delay accordingly. \cite{ref_北邮118提案} also presents the difference between their theoretical concatenated power and measured power values, validating equation~(\ref{equ_tar-sma}) in the power dimension.

To avoid an excessive number of cluster delays and high computational complexity, \cite{ref_Docomo120提案} suggests using constraint of incident angle and scattering angle to reduce the ray combinations. In \cite{ref_巩汇文WCL}, a power threshold can be set for cluster / ray reduction according to methods described. In addition, results from several proposals \cite{ref_LGE120提案} have demonstrated that when the paths in the Tx–TAR and TAR–Rx links are 1-by-1 matched, the overall channel performance can be maintained while significantly reducing computational complexity.

In the non-concatenated modeling solutions, the indirect paths from Tx to target to Rx (Tx-TAR-Rx) are considered part of Tx-Rx propagation and modeled by Tx-Rx paths satisfying Tx-TAR-Rx geometry \cite{ref_Oppo116提案}. The indirect paths are applied on a cluster or sub-cluster level, the path loss can also be modeled with (\ref{equ_pl}) and their delays, powers, angles, and initial phases are generated following cluster generation in section 7, TR 38.901. The path loss model 
This method has lower complexity and maximizes the reuse of existing standards in TR 38.901. However, these indirect paths are not directly or indirectly associated with the sensing target, which reduces their compatibility with actual environments. Additionally, further discussion is needed on how to select paths within the Tx-Rx models that satisfy the Tx-target-Rx geometry.

In summary, the advantages and disadvantages of the concatenation \& non-concatenation methods are as follows \cite{3gppRan117FL}:
\begin{itemize}
    \item {\textit{Concatenation} -Advantages: can accurately model all indirect paths of the target channel. -Disadvantages: high computational complexity and some of the modeled paths will likely not observable by experiment.}
    \item {\textit{Non-concatenation} -Advantages: simple modelling and low complexity. -Disadvantages: modeling results are not strongly associated with the target, affecting the accuracy of performance evaluation.}
\end{itemize}

\subsection{Multiple Targets}

3GPP RAN WG1 \#117 meeting adopts the proposal that multiple sensing targets can be modeled in the ISAC channel of a pair of sensing Tx and sensing Rx, and the same sensing target can be modeled in the ISAC channels of multiple pairs of sensing Tx and Rx. A key point for subsequent discussion involves whether to model a propagation path from Tx to Rx interacting with more than one sensing target. \cite{ref_爱立信117提案,ref_三星118提案} think the modeling for Tx-target1-target2-Rx path is the similar issue with the modeling for Tx-target-EO-Rx or Tx-target-EO-Rx link (especially the type-1 EO), which should be characterized or ignored at the same time. \cite{ref_联想116提案} suggests at least the blockage of the one target by another target shall be considered. However, considering the complexity of modeling and the necessity of evaluation, most literature and companies \cite{ref_118b次会议} do not recommend modeling the correlation between targets. In summary, the core controversy in current discussion revolves around whether and how to model propagation paths involving multiple STs, which necessitates a careful trade-off between modeling complexity and the potential gains in sensing performance.

\section{Background Channel Characteristics and Modeling } \label{section5}

Although the target channel involves ST information directly. However, the background channel multipaths actually constitute a larger fraction of the ISAC channel, which has been practically validated in \cite{zhang2025research}. The work in \cite{liu2025coupling} highlights the importance of the background channel and analyzes the coupling effects between the background component and the ST. \cite{wang2020small} investigates small target tracking in satellite videos by introducing background information into training. In theoretical analysis and design, simplified ideal assumptions about the background channel have also been initially applied. For example, in \cite{xiong2023fundamental,xiong2024torch}, the ST-unrelated component is considered as additive white Gaussian noise, with ISAC system performance derived accordingly. Similar channel assumptions are employed in \cite{yu2022location} and \cite{chen2021code}, where beamforming and system design are explored, respectively. Although the background channel does not carry direct ST information, its accurate modeling is crucial for evaluating sensing performance, especially in the complex environments \cite{zhang2021overview,liu2025coupling}.  

\subsection{Bistatic Background Channel}

ISAC background signals exhibit two propagation modes: bistatic and monostatic \cite{liu2022survey,zhang2023integrated}. The bistatic background channel, involves a spatial separation between the Tx and Rx, similar to communication configurations. The conventional GBSM generates propagation channels based on the relative positions of the transceivers as well as statistical parameter distributions \cite{3gpp38901}, making it readily applicable to bistatic background sensing \cite{luo2024channel,yang2024integrated}. 




During the discussions at the 3GPP RAN1 meeting, the corresponding scenarios and networks for TRP-TRP, TRP-UT, UT-TRP, and UT-UT dual-base sensing modes were configured according to the 3GPP specification \cite{3gpp38901}. Following the communication standardizations, large-scale parameters (such as path loss and shadow fading) and small-scale parameters (such as delay, power, and angle XPR) between the sensing Tx and Rx pairs were generated, along with the channel coefficients. Note that for the ISAC bistatic background channel, the absolute arrival time from Clause 7.6.9 of \cite{3gpp38901} must be applied to enable the sensing functionality. 

\subsection{Monostatic Background Channel}

The monostatic background channel, where the co-located Tx$\&$Rx acquire channel information by receiving its own transmitted signals reflected through environmental scatterers. The absence of independent anchors (e.g., Rx or ST) relative to the Tx renders existing GBSMs are not applicable. While radar research \cite{lampropoulos1999high,addabbo2021learning} has modeled monostatic clutter using statistical distributions, it fails to achieve compatibility with communication standards, making it challenging to perform unified performance evaluations for ISAC systems. 

\begin{figure}[t]
\centering
\subfloat[]{\includegraphics[width=3in]{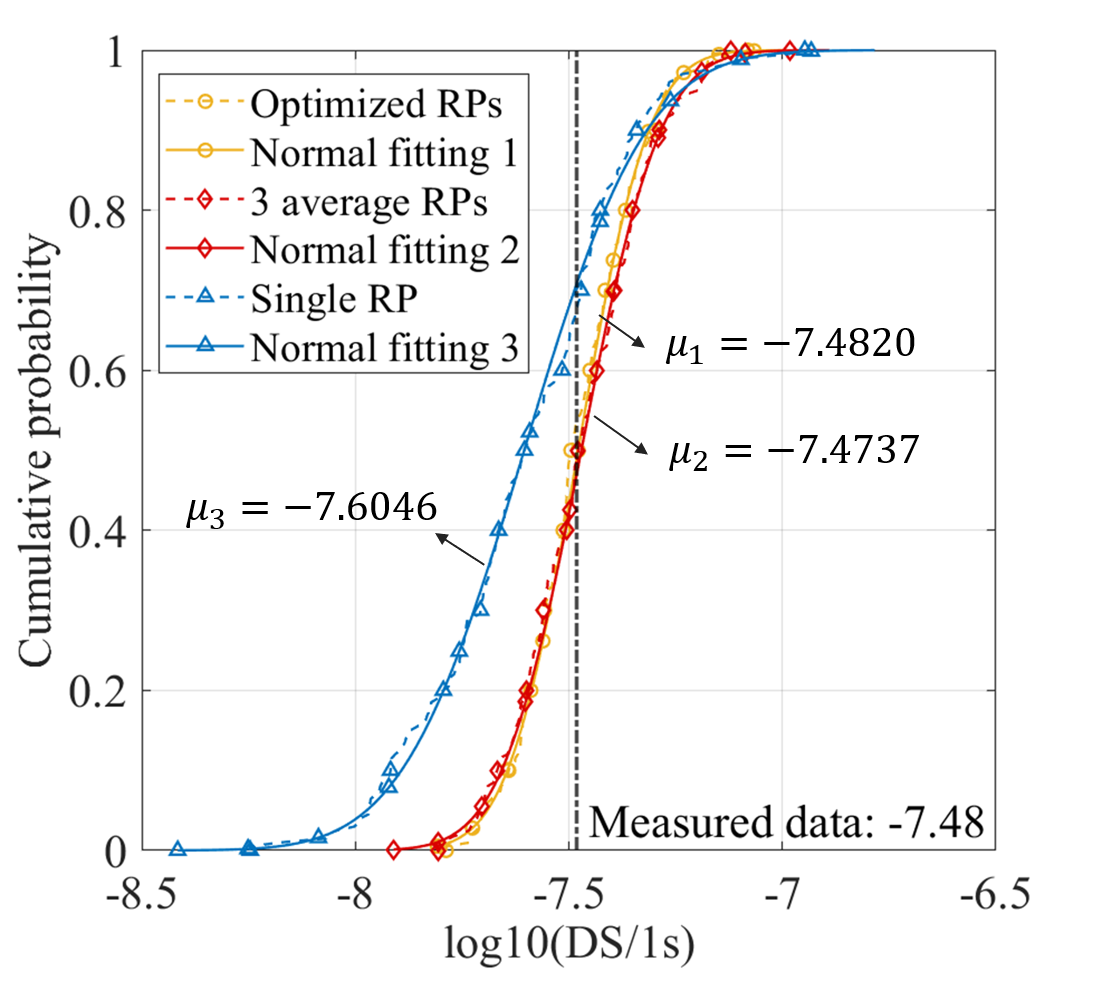}}\\
\subfloat[]{\includegraphics[width=3in]{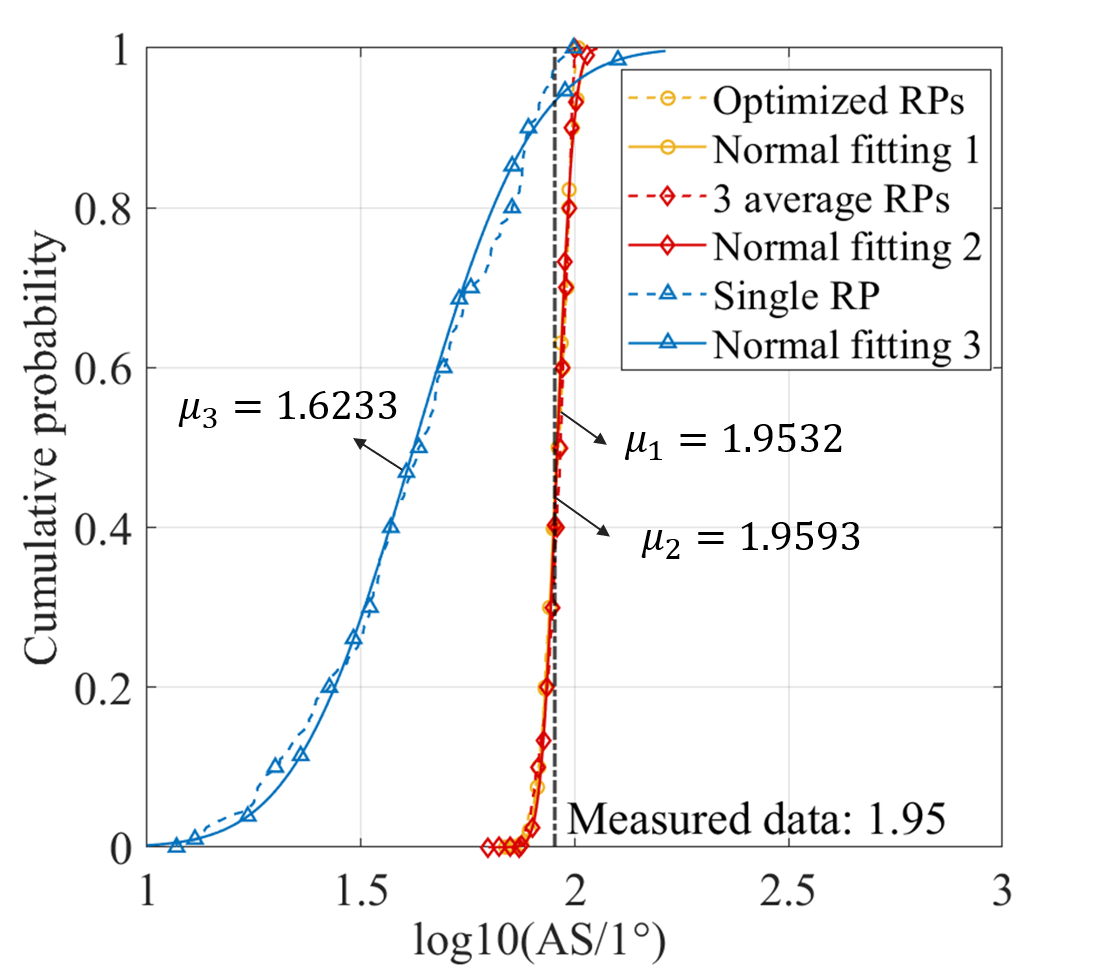}}
\caption{Simulated CDFs of modeled channel parameter under 200 simulations, where (a) $\rm{log}_{10}(DS/1s)$ and (b) $\rm{log}_{10}(AS/1^\circ)$ of AoDs with optimized multi-RPs, simplified 3 average multi-RPs, and single RP. The black lines with -7.48 and 1.95 mark the corresponding measured values 32.92 ns and 89.98$^\circ$, which serves as a baseline for evaluating the accuracy of the proposed model \cite{liu2025novel}.}
\label{fig_bacRP}
\end{figure}

To capture realistic characteristics of the monostatic background channel, state-of-the-art studies \cite{zhang2025research, chen2024empirical, ali2020leveraging} have conducted several measurements, demonstrating that propagation parameters are influenced by the scenario condition, such as the density of environmental scatterers around the Tx$\&$Rx. In \cite{barneto2022millimeter}, a geometric model is proposed based on an assumed small distance between the monostatic Tx and Rx. However, the high complexity of this approach limits its applicability for standardization. To enhance compatibility with existing standards, \cite{jiang2024novel} proposes an interesting virtual communication Rx approach, where the channel from the Tx$\&$Rx to single virtual Rx, can be referred to as a Referece Point (RP), is reused 3GPP TR 38.901 \cite{3gpp38901}.
In this work, monostatic Ray-tracing simulations are applied, and the strongest path is identified as the RP position. However, measurement results in \cite{liu2023shared} indicate that monostatic background channels exhibit a more discrete multipath distribution compared to communication channels, with significantly larger Delay Spread (DS) and Angular Spread (AS), parameters that traditional GBSMs with single Rx (i.e., RP) fail to capture. As a result, toward 3GPP ISAC standardization, accurately and practically developing an extended GBSM framework for ISAC monostatic background channels remains an open challenge, highlighting the urgent need for further methodological advancements and more validation.

In \cite{liu2025novel}, we conduct ISAC monostatic background channel measurements for an indoor scenario at 28 GHz. Realistic channel parameters are extracted, revealing pronounced single-hop propagation and discrete multipath distribution. Inspired by these properties, each monostatic background path is mapped to a virtual anchor, with nearby ones represented by a sub-channel from a communication Rx-like Reference Point (RP). A novel stochastic model is then proposed, characterizing the ISAC monostatic background channel as the superposition of sub-channels between the monostatic Tx$\&$Rx and multi-RPs. 
Based on the measurements and the proposed parameterization algorithm, the optimal number of multi-RPs is determined to be 3, with evenly distributed angles and equal distances considered as a simplification \cite{liu2025novel}. Results validate that the proposed model accurately matches the measured ISAC channel, with DS and AS errors reduced by over 20\% and 50\%, respectively, compared to reusing communication parameters.

To obtain the multi-RP parameters for typical ISAC scenarios, we conducted extensive monostatic background channel measurements and Ray-tracing simulations in collaboration with companies such as ZTE Corporation, Huawei Technologies Co., Ltd., and OPPO Corp., Ltd.\cite{ref_中兴120b提案}. These studies revealed that the multi-RPs follow a Gamma distribution with respect to the horizontal distance ($d_{mrp}$) from the monostatic Tx\&Rx and the vertical height ($h_{mrp}$) from the ground, expressed as
\begin{align}
\label{equ_rp}
d_{mrp} \sim \Gamma(\alpha_d,\beta_d)+c_d,\\
h_{mrp} \sim \Gamma(\alpha_h,\beta_h)+c_h,
\end{align}
where $\alpha_d$ and $\alpha_h$ denote the shape parameters, $\beta_d$ and $\beta_h$ denote the rate parameters, and $c_d$ and $c_h$ are the offsets of the corresponding Gamma distribution.
Together with companies, we also derived the parameter distributions for typical ISAC scenarios, including UMa, UMi, RMa, indoor office, and indoor factory, as well as different TRP and UT modes. Taking the UMi scenario as an example, which we exclusively provided and incorporated into the 3GPP standardization, the results are shown in Table \ref{tab:mrp}.

\begin{table*}[htbp]
    \caption{Multi-RP parameters of background channel for monostatic sensing \cite{ref_北邮121提案,3gppRan121sum}}
    \centering
    \begin{tabular}{m{4cm}<{\centering}|m{2cm}<{\centering}|m{3cm}<{\centering}m{3cm}<{\centering}m{4cm}<{\centering}}
        \toprule
        \hline
        \multicolumn{2}{c|}{\textbf{Parameters}} & \textbf{UMi-TRP} & \textbf{UMi-UT} & \textbf{UMi-UAV} \\
        \hline
        \multirow{3}{*}{Distance (m)}  & $\alpha_d$ & 6.1996  & 10.0220 & 0.0156$h$+5.5399\\
         & $\beta_d$ &  0.1558 & 1.2522 & 40.4517/($h$+254.6318)\\
         & $c_d$ &  15.2697 & 11.0040 & 0.0140$h$+15.1184\\
         \hline
        \multirow{3}{*}{Height (m)} & $\alpha_h$ &  12.0487 & 3.0487 & 0.0123$h$+11.9569\\
         & $\beta_h$ &  2.3261 & 1.9128 & 17.8047/($h$-0.2202)\\
         & $c_h$ & 0.0157  & 0.1785 & 0.0532$h$-0.0120\\
        \hline
        \multicolumn{5}{l}{Note: $h$ is the height of the aerial UT.}\\
        \hline
        \bottomrule
    \end{tabular}
    \label{tab:mrp}
\end{table*}






\section{Additional New Features in ISAC Channel Modeling} \label{section6}

\subsection{Spatial Consistency}
In the study of spatial consistency within geometry-based stochastic channel models (GSCMs) for communication channel, the relative positioning among simulated nodes plays a crucial role. To ensure realism, key channel parameters—such as path loss, large-scale parameters (LSPs), and small-scale parameters (SSPs)—must exhibit spatial correlation that aligns with the distances between these nodes. For instance, users located in close proximity are likely to experience comparable propagation characteristics due to their shared scattering environment and similar distances to the base station. This phenomenon is referred to as spatial consistency or spatial correlation. Maintaining spatial consistency is particularly vital in advanced communication techniques such as full-dimension MIMO (FD-MIMO), beamforming, and beam tracking, where accurate angular information and user positioning are essential.
In ISAC scenarios, moving targets and terminals are commonly encountered, whether in indoor human sensing or outdoor vehicular networking applications\cite{liu2021analyzing}. Therefore, accurate channel modeling of the continuous variation of channel parameters in dynamic environments is of critical importance. On one hand, the parameters perceived by sensing systems in practice exhibit continuous changes; on the other hand, such continuity is essential for accurately estimating the trajectories of moving targets. Moreover, tracking continuously evolving multipath components reflected by environmental scatterers can further enhance sensing accuracy and assist in improving positioning performance.

To ensure spatial consistency between LSPs and SSPs, existing research has established methods for LSPs. However, SSPs currently lack spatial correlation mechanisms. In conventional geometry-based stochastic channel modeling approaches, SSPs are typically generated by independently assigning random variables to each spatial position. Consequently, despite potential correlations in LSPs and path loss characteristics, the absence of spatial correlation in SSPs leads to fundamentally inconsistent channel behaviors.

To introduce spatial consistency in SSPs, in the WINNER II model two approaches are proposed \cite{winner2007winner}. The first one is based on the cluster death-birth process and has been resolved in the quasi deterministic radio channel generator (QuaDRiGa) model \cite{jaeckel2014quadriga}. The second approach is based on the appearance and disappearance of multipath components, according to a Markov process. Since for its application, the parameters have not yet been extracted from measurements, this approach was never adjusted. Differently, in the COST 2100 channel model, a global set of scatterers is shared by all users through so-called visibility regions \cite{liu2012cost}. Even though this type of channel model supports spatial consistency, it is currently not widely accepted due to its high complexity and its limited support on propagation scenarios.

 Additionally, the few existing scenarios are parametrized only for a small range of carrier frequencies. Furthermore, according to \cite{medbo2016radio}, this model is not suitable to be used with large antenna arrays since the spatial variation that comes with large antenna arrays is not considered by this model. Recently, the 3GPP in the study item \cite{3gpp38901} specified a new 3D radio channel model, feasible for frequencies of future mobile networks ranging from 0.5 to 100 GHz, that accounts also for spatial consistency. However, the spatial consistency model as described in \cite{3gpp38901}, is not very explicit and leaves room for various interpretations. In particular, the spatial correlation is only defined as a 2D random process based on the parameter-specific decorrelation distances. Further details on what a 3D random process represents in \cite{ademaj2019spatial1}.

Concurrently, in the work presented in \cite{ademaj2017modeling} propose the model for correlating SSPs indicates a high correlation over distance and a saturation in terms of correlation, where after approximately 30 m distance the model exhibits the same correlation regardless of the input parameters applied. In a further investigation in the work \cite{ademaj2018ray}, by performing Ray-tracing simulations, it is shown that abrupt changes of channel parameters occur, are realistic and reflect the changes in geometry of surrounding objects along the propagation path, and thus have to be included in the modeling of spatial consistency. \cite{ademaj2019spatial2} therefore carefully further develop our modeling approach and provide a detailed description together with a statistical validation and a full parametrization of our spatial consistency model for SSPs.

While the spatial consistency framework in conventional communication scenarios provides a solid foundation, ISAC scenarios introduce additional challenges such as dynamic targets, complex environmental scatterers, and multiple link types across heterogeneous network nodes. These aspects necessitate an extended and more refined spatial consistency model. The following subsections summarize the current specifications in standardization and highlight their applicability and limitations for ISAC environments.

The spatial consistency procedure in ISAC channel is used to generate the random variables for the sensing Tx-SPST links, the SPST-sensing Rx links and the background channels. The spatial consistency procedures in 3GPP ISAC channel are reused to handle the links between TRPs and STs/UTs. What's more, for sensing scenario UMi-AV, UMa-AV and RMa-AV, the 2D random process (in the horizontal plane) can be extended to 3D random process considering vertical correlation distance. 

It is worth emphasizing that the spatial consistency procedure is employed in ISAC scenarios to generate the random variables for the STx-SPST links, SPST-SRx links, and background channels, ensuring spatial correlation across different links.
For sensing scenarios such as UMi-AV, UMa-AV, and RMa-AV, the conventional 2D random process—defined in the horizontal plane—is extended to a 3D random process to account for variations in aerial UT height, as specified in \cite{3gppTR36777}. In this extension, the spatial correlation distance remains the same as the horizontal correlation distance defined in \cite{3gpp38901}. 

\begin{itemize}
    \item Link-correlated: parameters of any two links of UT-UT/SPST links are correlated, subjected to correlation distance.
    \item All-correlated: all UT-UT/SPST links are correlated.
\end{itemize}

Table \ref{tab:spatial} summarizes the correlation types for each parameter in this updated spatial consistency model.

\begin{table}[htbp]
    \centering
    \caption{Correlation type among UT-UT/SPST links}
    \begin{tabular}{m{5cm}<{\centering}m{3cm}<{\centering}}
        \toprule
        \hline
        \textbf{Parameters} & \textbf{Correlation type} \\
        \hline
        Delays  & Link-correlated \\
        Cluster powers  & Link-correlated \\
        AoA/ZoA/AoD/ZoD offset   & Link-correlated \\
        AoA/ZoA/AoD/ZoD sign  & Link-correlated \\
        Random coupling & Link-correlated \\
        XPR  & Link-correlated \\
        Initial random phase  & Link-correlated \\
        LoS/NLoS states  & Link-correlated \\
        Blockage (Model A) & All-correlated \\
        O2I penetration loss   & All-correlated \\
        Indoor distance    & All-correlated \\
         Indoor states  & All-correlated \\
        \hline
        \bottomrule
    \end{tabular}
    \label{tab:spatial}
\end{table}
The spatial consistency across the links between the STx/SRx and multiple SPSTs of a single ST is modeled as if the multiple SPSTs were individual independent STs. Spatial consistency is not modeled between the STx-SPST links, SPST-SRx links, and background channels under certain conditions. First, spatial consistency does not apply when the involved links belong to different link types, such as outdoor LoS, outdoor NLoS, or O2I. It is also not applicable when the UTs are located on different floors, again based on the definitions in the same table.In terms of sensing modes, for monostatic sensing, spatial consistency is not applied to the background channel. For bistatic sensing, spatial consistency is not modeled for any of the STx–SPST links, SPST–SRx links, or the background channel.
Furthermore, spatial consistency is not supported for links associated with non-co-located TRPs or when the links are generated using channel models with parameters from different communication scenarios. Similarly, in monostatic UT sensing, background channels across different UTs or different RPs of the same UT are considered spatially inconsistent. For sensing scenarios such as UMi, InH, and InF, spatial consistency is not modeled between TRP–target/UT links and target/UT–UT links. In addition, spatial consistency is not applied between the TRP-TRP links and any other links in the ISAC channel model. In the case of TRP monostatic sensing, background channels across different TRPs or different RPs of the same TRP are also excluded from spatial consistency modeling.

In the future research of ISAC systems, enhancing the design of perception algorithms and improving the quality of deep learning training datasets are two key directions. By utilizing spatial consistency modeling, more accurate prior information can be provided to perception algorithms, helping them better understand signal propagation patterns in complex environments, thereby improving robustness and positioning accuracy in dynamic scenarios. At the same time, the performance of deep learning models is heavily dependent on high-quality datasets. Spatial consistency theory enables the construction of more realistic training data, allowing models to achieve better generalization and optimized performance in real-world environments. In conclusion, optimizing perception algorithms and enhancing dataset quality will enable ISAC systems to achieve more efficient and accurate perception capabilities in practical applications.

\subsection{Environment Object}
ISAC-based Environmental Reconstruction enables base stations and user terminals to simultaneously execute high-rate communication tasks while performing high-resolution sensing of the surrounding environment. A prominent application of ISAC is Vehicle-to-Everything (V2X), which is critical for autonomous driving systems in smart transportation. Among various V2X functionalities, high-precision vehicle positioning stands out as a key capability, relying on ISAC-enabled environmental reconstruction for dynamic map updates and advanced fusion algorithms for robust localization in complex scenarios.

Conventional positioning algorithms such as Time of Arrival (TOA) and Time Difference of Arrival (TDOA) are effective under ideal Line-of-Sight (LoS) conditions \cite{kuutti2018survey}. However, in urban vehicular environments, the presence of obstacles like buildings and trees introduces rich multipath effects, severely degrading positioning accuracy\cite{bader2022nlos}. Based on ray tracing simulations in vehicular environments, Samsung also observed that the interactions between the ST and other objects (e.g., EOs or other STs) cause an additional path loss of 11-15 dB for the indirect path (Tx-Target-type 1/2 EO-Rx) compared to the direct path (Tx-Target-Rx), as summarized in Table \ref{tab:target-EO}. To address this challenge, \cite{al2002ml} proposes leveraging Non-Line-of-Sight (NLoS) paths that undergo single-bounce reflections, by jointly estimating the AoA and TOA from both ends of the link. Additionally, \cite{chen2024and} demonstrates through Ray-tracing simulations that NLoS-based positioning primarily relies on fixed, structured environmental scatterers such as walls. These are referred to as EOs, while multi-bounce paths are generally considered less reliable and should be discarded.

\begin{table*}[!h]
    \centering
    \caption{Preliminary simulation results of one bounce and two bounces for the target channel\cite{ref_三星118提案}}
    \begin{tabular}{m{4.5cm}<{\centering}|m{6.0cm}<{\centering}|m{6.0cm}<{\centering}}
        \toprule
        \hline
        \textbf{Paths} & \textbf{Path gain with type-1 EO [dB]} & \textbf{Path gain with type-2 EO [dB]} \\
        \hline
        Tx-Target-Rx & -104.1 & -103.7 \\
        \hline
        Tx-Target-EO-Rx or Tx-EO-Target-Rx & -114.9 & -118.8 \\
        \hline
        \bottomrule
    \end{tabular}
    \label{tab:target-EO}
\end{table*}

To ensure algorithm feasibility from practical physical scenarios, accurate channel modeling and real-world experimental validation are essential. Prior studies indicate that a considerable portion of received signal power comes from NLoS paths, significantly impacting system performance. For example, \cite{zhang2016measurement} reports that in an urban street environment at 2.6 GHz, the strongest first-order reflection from the ground is about 10 dB weaker than the LoS path, affecting the angular spread of incoming signals. Similarly, \cite{blaunstein2006signal} observes that in urban micro (UMi) scenarios with no clear LoS, 80\% to 90\% of the received power originates from reflected paths. At higher frequencies, such as 28 GHz. \cite{zhong2019outdoor} finds that in outdoor-to-indoor (O2I) settings, a reflected wall path can be just 8 dB weaker than the direct LoS path. Moreover, when the LoS is partially obstructed, the reflected path may even surpass the LoS in strength, highlighting the importance of NLoS components.

\begin{figure}[h]
    \centering
    \subfloat[]{\includegraphics[width=8.5cm]{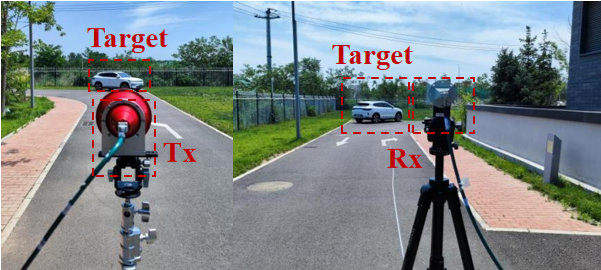}}\\
    \subfloat[]{\includegraphics[width=8.5cm]{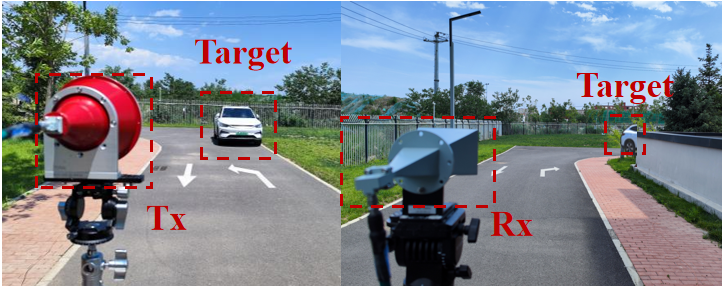}}
    \caption{The measurement campaign in urban street using car as a sensing target. (a) Case1 and (b) Case2 scenarios \cite{jiang2025novel}.}
    \label{fig:two_figures}
\end{figure}

In ISAC scenarios, the role of NLoS paths becomes even more critical with the presence of sensing targets. For example, \cite{liu2023shared} identifies environmental scatterers such as walls and columns when detecting a sensing target. Likewise, \cite{chen2024empirical} shows that in an indoor scenario at 105 GHz, human body detection is strongly influenced by indirect paths reflected from surrounding walls. Extending to outdoor scenarios, \cite{jiang2025novel} investigates NLoS effects in vehicular ISAC positioning through urban street measurements at 26GHz (Fig. \ref{fig:two_figures}), revealing that EOs contribute significantly to received power through strong reflections. These findings emphasize the necessity of explicitly modeling EOs in the target channel.

\subsection{Macro and Micro-Doppler}

\begin{table*}[t]
    \caption{Summary of measured macro-Doppler and micro-Doppler phenomena}
    \centering
        \begin{tabular}{m{1.5cm}<{\centering}|m{2cm}<{\centering}|m{2.5cm}<{\centering}|m{2.5cm}<{\centering}|m{2cm}<{\centering}|m{5cm}<{\centering}}
            \toprule
            \hline
                 \textbf{Reference}                                 &
                 \textbf{Type}                                      &
                 \textbf{Sensing target}                            &
                 \textbf{Scenario}                                  &
                 \textbf{Frequency}                                 &
                 \textbf{Conclusion} 
             \\ \hline
                 \cite{szwoch2018suppression}                       &
                 macro-Doppler                                            &
                 Vehicle                                            &
                 Outdoor Highway                                    &
                 24.125 GHz                                         &
             Improving the accuracy of vehicle speed measurements 
             \\ \hline
                \cite{wang2018road}                                 & 
                micro-Doppler                                       & 
                Vehicle                                             &
                Outdoor Highway                                     &
                24, 77 GHz                                          &
                Improved recognition accuracy of moving targets 
            \\ \hline
                \cite{wei2019classification}                        & 
                micro-Doppler                                       &
                Human (using cell phone)                            &
                Outdoor/Indoor                                      &
                77 GHz                                              & 
                Improving the accuracy of human hand behavior recognition 
            \\ \hline
                \cite{lin2017performance}                           & 
                micro-Doppler                                       &
                Human (walking and running)                         &
                Outdoor/Indoor                                      &
                5.8 GHz                                             & 
                Enhanced performance in walking and running activity classifications
            \\ \hline
                \cite{rabbani2020wireless}                          & 
                micro-Doppler                                       &
                Human (respiratory and heart rates)                 &
                Outdoor/Indoor                                      &
                60 GHz                                              & 
                Enhanced sensitivity for respiration rate and heart rate detection
            \\ \hline
                \cite{hwang2022motion}                              & 
                macro-Doppler                                             &
                UAV (own movement)                                  &
                Outdoor Aerial                                      &
                60 GHz                                              & 
                Improved accuracy in estimating the speed of the UAV itself
            \\ \hline
                \cite{wang2021lightweight}                          & 
                micro-Doppler                                       &
                UAV (Recognition, especially in hovering)           &
                Outdoor Aerial                                      &
                24-26 GHz                                           & 
                Enhancing the identification of small UAVs
            \\ \hline
                \cite{renga2014ship}                                & 
                macro-Doppler                                       &
                Ship                                                &
                Outdoor Marine                                      &
                5.2 GHz                                             & 
                Improving the accuracy of ship speed estimation
            \\ \hline
                \cite{wu2022measurement}                            & 
                micro-Doppler                                       &
                Ship (Such as the propeller blades)                 &
                Outdoor Marine                                      &
                1 MHz                                               & 
                Enhanced detection and identification of underwater targets
            \\ \hline
            \bottomrule
        \end{tabular}
    \label{tab:doppler}
\end{table*}

In ISAC systems, sensing performance faces heightened requirements. To meet these demands, systems must accurately capture not only the basic target information, such as size, position, and velocity, but also delve into the detailed analysis of the target's motion properties. Since the description of these motion properties relies on the Doppler effect, it becomes essential to expand the Doppler-related components in the existing channel models. This expansion should encompass both macro-Doppler and micro-Doppler elements to provide a more comprehensive representation of the target's motion properties.
The macro-Doppler effect means when the target moves with a constant velocity, the carrier frequency of the returned signal will be shifted\cite{chen2006micro}. In contrast, the micro-Doppler effect means if the target or any structure on the target has mechanical vibration or rotation in addition to its bulk translation, it might induce a frequency modulation on the returned signal that generates sidebands about the target's Doppler frequency shift\cite{chen2006micro}.

\begin{table*}
    \large
    \begin{align} \label{equ_doppler}
        f_{D, n^{\prime}, m^{\prime}, n, m}^{k, p}(t)=&\frac{\hat{r}^T_{r x, k, p, n^{\prime}, m^{\prime}}(\tilde{t}) \bar{v}_{r x}(\tilde{t})+\hat{r}_{k, p, n^{\prime}, m^{\prime}}^T(\tilde{t}) \bar{v}_{k, p}(\tilde{t})+2 \alpha_{r x, n^{\prime}, m^{\prime}}^{k,p} D_{r x, n^{\prime}, m^{\prime}}^{k, p}}{\lambda_0}\\ \notag 
        &+\frac{\hat{r}_{t x, k, p, n, m}^T(\tilde{t}) \bar{v}_{t x}(\tilde{t})+\hat{r}_{k, p, n, m}^T(\tilde{t}) \bar{v}_{k, p}(\tilde{t})+2 \alpha_{t x, n, m}^{k, p} D_{t x, n, m}^{k, p}}{\lambda_0}.
    \end{align}
\end{table*}

\begin{figure}[h]
    \centering
    \subfloat[]{\includegraphics[width=3.3in]{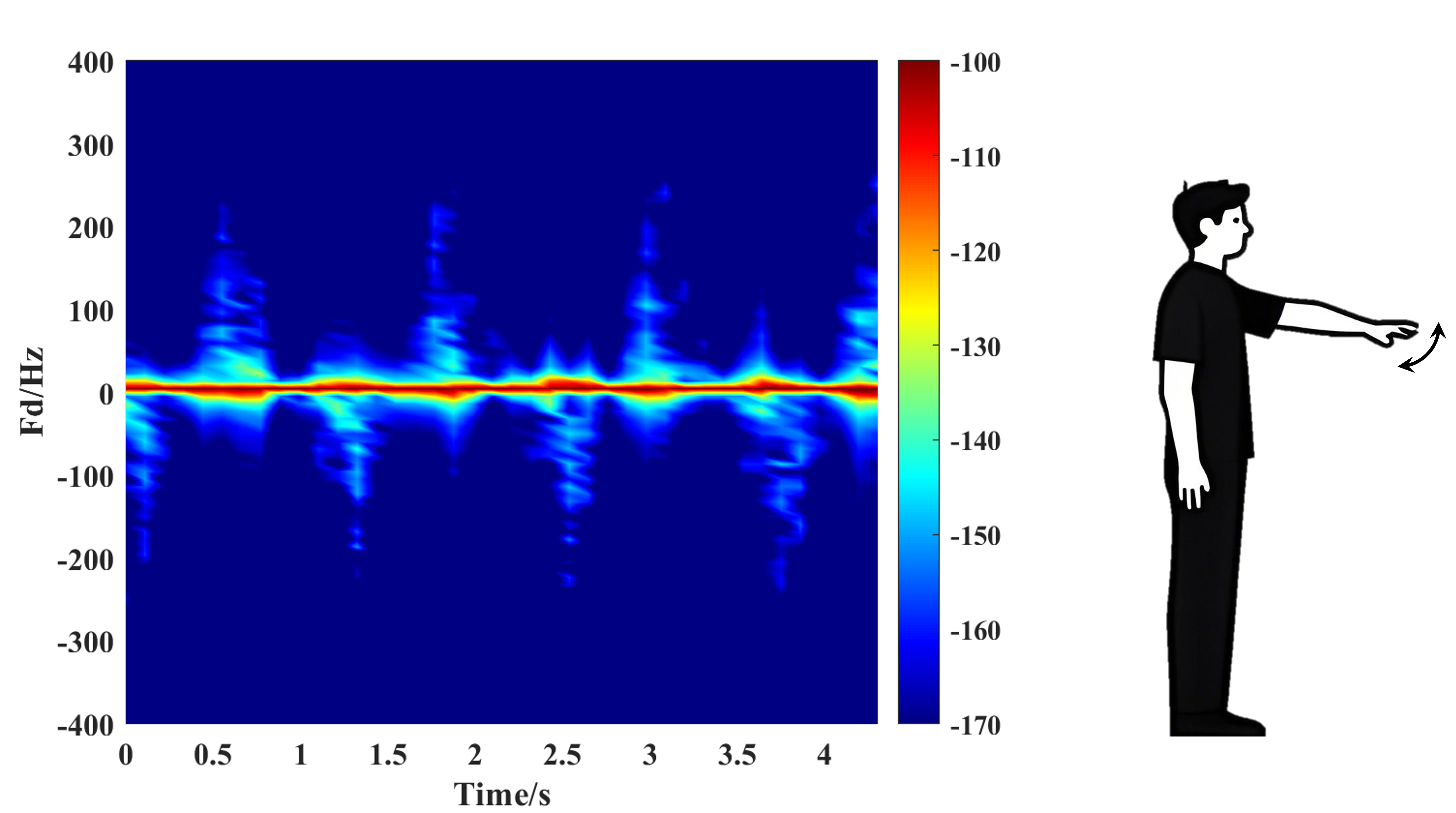}}\\
    \subfloat[]{\includegraphics[width=3.3in]{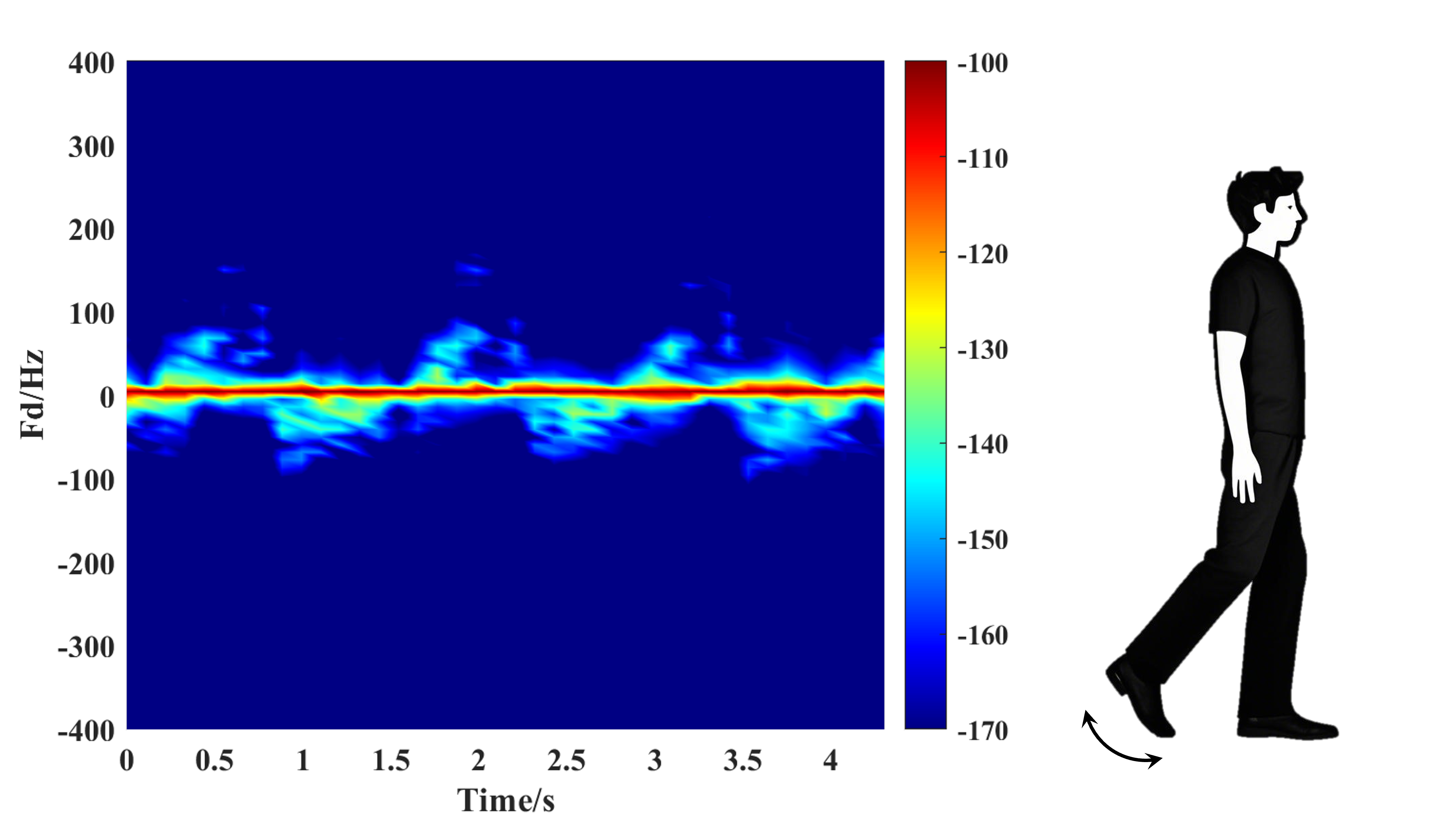}}
    \caption{Illustration of the micro-Doppler shift for (a) human arms and (b) legs \cite{ref_北邮119提案,ref_维沃119提案}.}
    \label{fig:Doppler}
\end{figure}

These two effects have distinct focuses in applications: the macro-Doppler effect is used to measure the target's own velocity, while the micro-Doppler effect captures the relative speeds of different parts within the target, thereby aiding in the detection and identification of micro-motions. Current research has successfully observed macro-Doppler and micro-Doppler frequency shifts phenomena for vehicles, humans, UAVs, and ships through various measurement simulations, as summarized in TABLE~\ref{tab:doppler}. In practice, capturing the micro-Doppler effect is more challenging than capturing the macro-Doppler effect. However, it provides more detailed information about the target's motion. For example, through theoretical derivation, vivo has demonstrated that the swinging of human body parts (e.g., arms and legs) in the horizontal direction can be approximated as a simple harmonic oscillation, and the resulting micro-Doppler shift can be modeled using a sinusoidal function \cite{ref_维沃119提案}. BUPT conducted human micro-Doppler measurements in an anechoic chamber. The experimental results obtained at 15 GHz (Fig. \ref{fig:Doppler}) show clear periodic micro-Doppler shifts associated with the arms and legs, which can be fitted using the proposed sinusoidal functions \cite{ref_北邮119提案}. This demonstrates the necessity of incorporating micro-Doppler modeling in ISAC systems.


Consequently, in 3GPP ISAC channel modeling standardization, both macro-Doppler and micro-Doppler effects for sensing targets can be modeled using a unified formulation \cite{3gpp38901}. During signal propagation, the movement of transceiver antennas, sensing targets, and non-sensing scatterers in the environment can all cause Doppler shifts. Thus, the Doppler frequency component $f_{D,n',m',n,m}^{k,p}(t)$ of a propagation path can be expressed as Eq. (\ref{equ_doppler}). The first term in the equation represents the Doppler effect of the SPST-SRx propagation path, and the second term that of the STx - SPST propagation path. Each term consists of three parts. The first two parts respectively represent the Doppler effect caused by the LoS propagation paths of SPST - SRx and STx - SPST. The last part of each term represents the Doppler shift caused by the movement of non - target scatterers in the environment. The specific meanings of the variables in the equation are as follows:
\begin{itemize}
    \item $\hat{r}^{T}_{rx,k,p,n^{\prime},m^{\prime}}$ is the spherical unit vector at receiver for the link from Rx to the scattering point.
    \item $\hat{r}^{T}_{tx,k,p,n,m}$ is the spherical unit vector at transmitter for the link from Tx to the scattering point.
    \item $\hat{r}^{T}_{k,p,n^{\prime},m^{\prime}}$ is the spherical unit vector at the scattering point for the link from the scattering point to Rx.
    \item $\hat{r}^{T}_{k,p,n,m}$ is the spherical unit vector at the scattering point for the link from the scattering point to Tx.
    \item $D_{rx, n^{\prime}, m^{\prime}}^{k, p}$ and $D_{tx, n, m}^{k, p}$ are random variables from $-v_{scatt}$ to $v_{scatt}$, and $v_{scatt}$ is the maximum speed of the clutter.
    \item $\alpha_{rx, n^{\prime}, m^{\prime}}^{k,p}$ and $\alpha_{tx, n, m}^{k,p}$ are random variables of Bernoulli distribution with mean $p'$ and $p$. Parameter $p'$ and $p$ determines the proportion of mobile scatterers and can thus be selected to appropriately model statistically larger number of mobile scatterers (higher $p'$ and $p$) or statistically smaller number of mobile scatterers (e.g. in case of a completely static environment: $p'=0$ and $p=0$ results in all scatteres having zero speed).
\end{itemize}

Current 3GPP specifications lack explicit definitions for sensing targets requiring Doppler/micro-Doppler characterization, and motion-state classifications with functional modeling remain incomplete. Consequently, future enhancements to $\bar{v}_{k, p}(\tilde{t})$ in Eq. (\ref{equ_doppler}) are essential for accurately modeling these properties across diverse sensing targets.

\subsection{Shared Cluster}

In communication research, the sharing feature has been observed in several multi-link channel measurements, and the effects on correlation and capacity of the shared clusters (contributed by the shared scatterers) are studied \cite{poutanen2010significance,poutanen2011multi}. The sharing of communication and sensing channels is necessary to leverage channel information obtained by one system to enhance the performance of the other. Hence, unlike existing models that independently generate communication and sensing channels, it is imperative to carefully consider the practical sharing feature of scatterers and clusters in ISAC channel modeling.

\begin{figure}[h]
    \centering
    \subfloat[]{\includegraphics[width=2.7in]{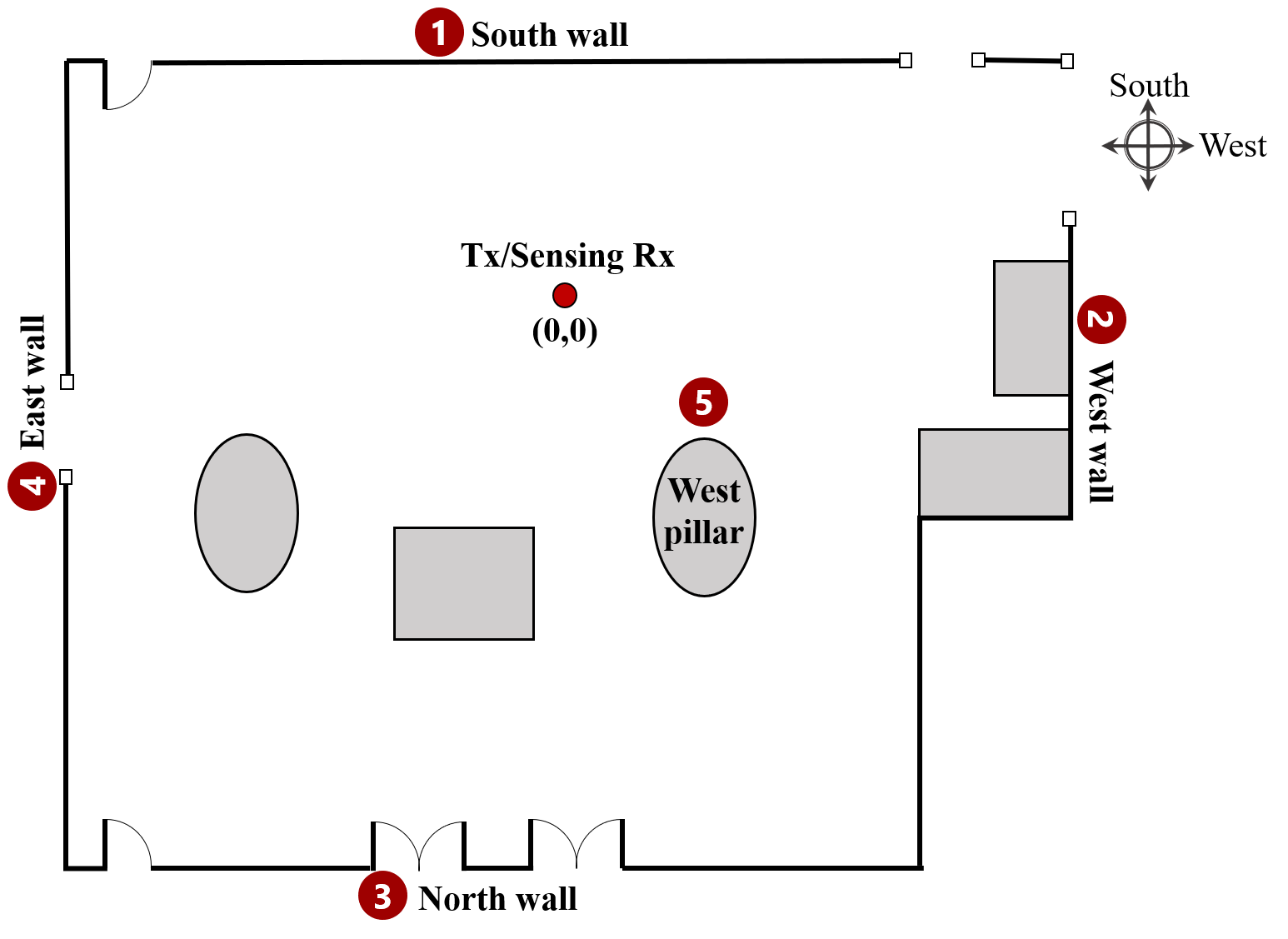}}\\
    \subfloat[]{\includegraphics[width=2.5in]{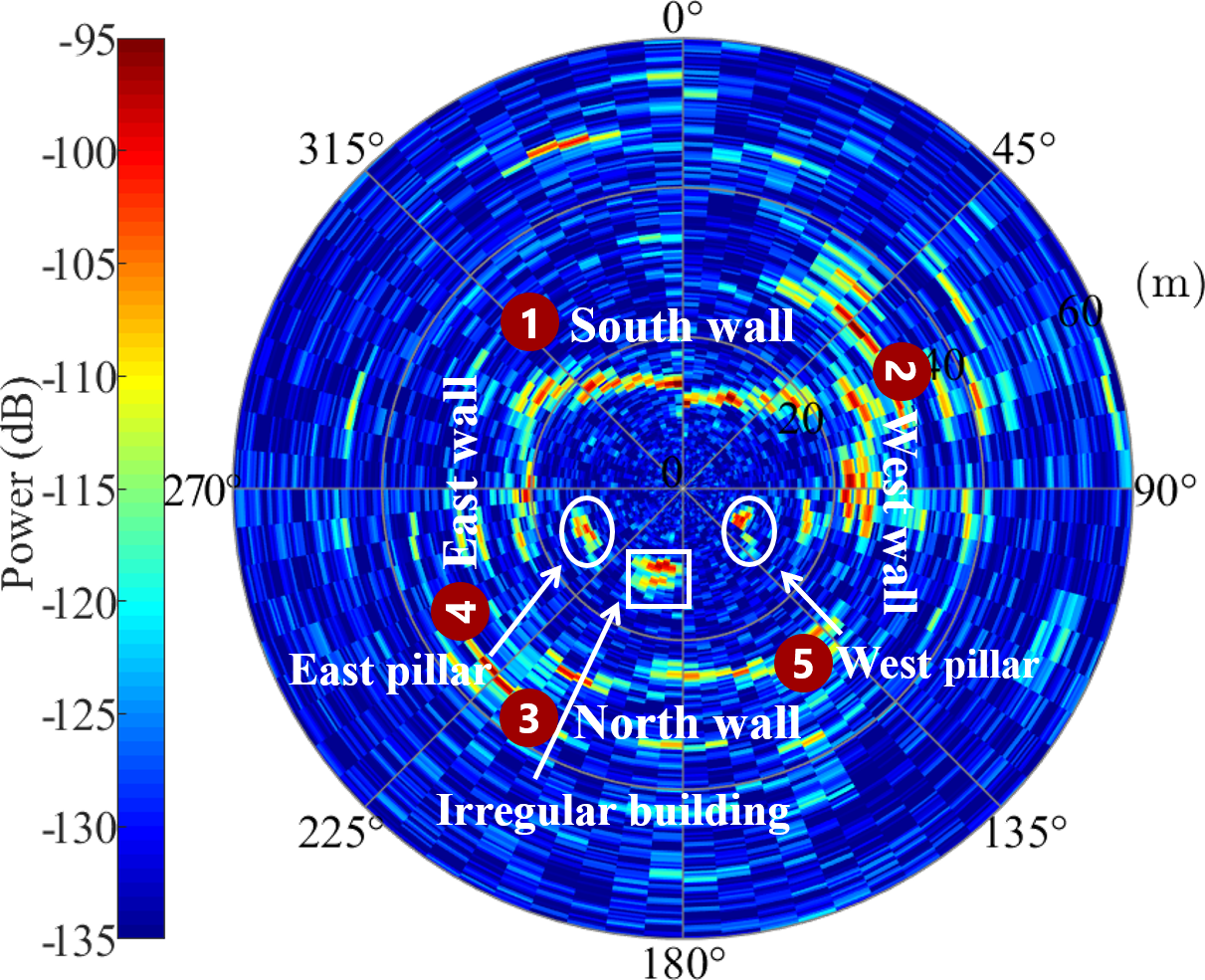}}\\
    \subfloat[]{\includegraphics[width=2.5in]{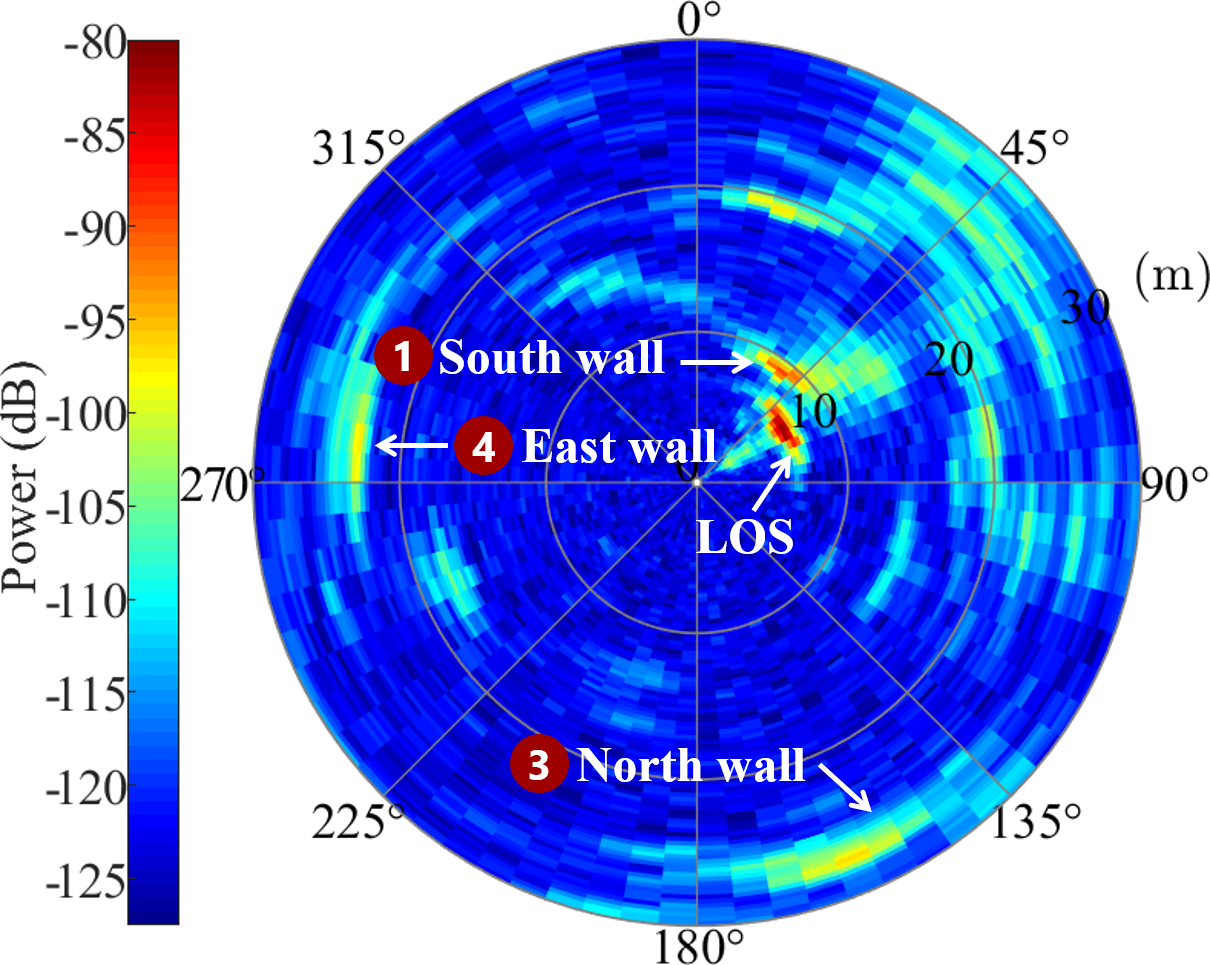}}
    \caption{Measurement campaign and results of shared scatterers, including (a) sensing and communication measurement environment in the indoor hall, and (b) sensing PADP and (c) communication PADP \cite{liu2023shared}.}
    \label{fig_7_4_mea}
\end{figure}


To investigate the above ISAC channel characteristics, we conduct realistic ISAC channel measurement campaigns in a typical indoor hall at BUPT and firstly observe the shared scatterers in ISAC channel \cite{liu2023shared}. Fig. \ref{fig_7_4_mea}(a) illustrates the layout of the ISAC channel measurement scenario, with primary scatterers identified by red sequence numbers. Based on the CIRs obtained through measurements, the power-angular-delay profiles (PADPs) of sensing and communication channels are calculated, as shown in Fig. \ref{fig_7_4_mea}(b) (c). As can be seen from Fig. \ref{fig_7_4_mea}, the sensing multipaths corresponding to the scatterers exhibit the realistic environment, while the communication multipaths indicate the dominant scatterers that affected in the communication signal propagation. Compared Fig. \ref{fig_7_4_mea}(b) with Fig. \ref{fig_7_4_mea}(c), it can be observed that some scatterers are shared between the sensing channel and communication channel, i.e. the south wall, the north wall, and the east wall. Moreover, the shared clusters generated by these shared scatterers reveal the realistic sharing feature of communication and sensing channels, which need to be adequately considered in channel modeling for ISAC systems. Based on the observed sharing feature, we defined the clusters contributed by the shared scatterers as the shared clusters. Then, we propose a shared cluster-based ISAC channel, where shared and non-shared clusters between communication and sensing channels are superimposed \cite{liu2023shared}, as shown in Fig. \ref{fig_7_4_mea}.

Nowadays, an increasing number of studies have followed and expanded the research on the sharing feature of ISAC channels. In \cite{Wang2024Millimeter}, channel measurement campaign in typical indoor factory (InF) environments at 28 GHz and 38 GHz are conducted, and the shared cluster characteristics are analyzed. In \cite{li2024characteristics},  the sharing feature in the UAV ISAC scenarios, including the angle, time–varying characteristics, and SD, are analyzed. The work in \cite{li2024characteristics} considers the shared clusters between communication and sensing channels, and simulates the channel statistical properties. \cite{Xiong2023Channel} studies the  time-varying sharing feature of a heterogeneous vehicular ISAC system, and also analyze the channel propagation properties.





\begin{figure*}[t]
\centering
\includegraphics[width=6.5in]{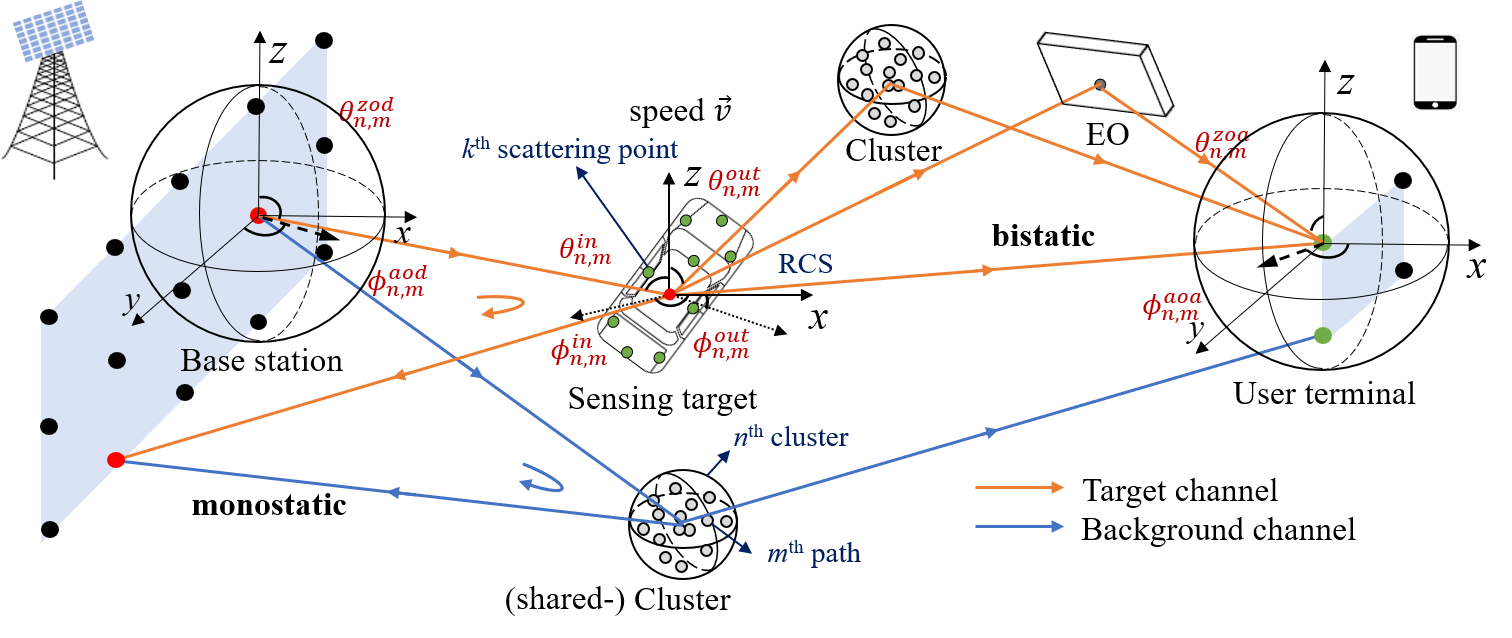}
\caption{Illustration of the unified ISAC channel model based on E-GBSM.}
\label{fig_EGBSM}
\end{figure*}

\begin{table*}[!b]
    \large
    \begin{align} \label{eq:EGBSM}
        \boldsymbol{h}_{p/q}^{\mathrm{Tx/Rx-Tar}}\left(\boldsymbol{\Omega}^{\mathrm{in/out}},\boldsymbol{\tau}\right)=&\sum_n^{N_1+N_s}\boldsymbol{S}_{p/q,n}^{\mathrm{Tx/Rx}}\cdot\sum_m^{M_1}\boldsymbol{A}_{p/q,n,m}^{\mathrm{Tx/Rx}}\cdot\boldsymbol{I}_{n,m}\cdot\sqrt{\boldsymbol{P}_{n,m}}\notag\\&\cdot\boldsymbol{F}_{p/q}\left(\boldsymbol{\Omega}_{n,m}^{\mathrm{Tx/Rx}}\right)\boldsymbol{e}^{j\frac{2\pi}{\lambda}\hat{d}_p\cdot\hat{c}_p}\boldsymbol{\delta}\left(\boldsymbol{\Omega}^{\mathrm{in/out}}-\boldsymbol{\Omega}_{n,m}^{\mathrm{in/out}}\right)\boldsymbol{\delta}\left(\boldsymbol{\tau}-\boldsymbol{\tau}_{n,m}\right)
    \end{align}
\end{table*}
\begin{table*}[!b]
    \large
    \begin{align} \label{eq:isacchannel}
        h_{u,s}^{sen}(t,\tau)=&\sqrt{1-K_{EO}}\cdot\Big[\sum_{n_1,m_1}^{N_1+N_S,M_1}\sum_{n_2,m_2}^{N_2,M_2}\sqrt{\frac{P_{n_1,m_1}P_{n_2,m_2}}{n_1n_2}}F_u^{\mathrm{Rx}}(\phi_{\mathrm{Rx}}^{n_2,m_2})\sqrt{\sigma(\phi_{\mathrm{out}}^{n_{2},m_{2}},\phi_{\mathrm{in}}^{n_{1},m_{1}})}F_{s}^{\mathrm{Tx}}(\phi_{\mathrm{Tx}}^{n_{1},m_{1}})\notag\\&\notag\cdot\exp\left(j\frac{2\pi}{\lambda}\left(\mathbf{r}_{\mathrm{Rx},n_{2},m_{2}}^{\mathrm{T}}\cdot\mathbf{d}_{\mathrm{Rx},u}+\mathbf{r}_{\mathrm{Tx},n_{1},m_{1}}^{\mathrm{T}}\cdot\mathbf{d}_{\mathrm{Tx},s}\right)\right)\exp\left(j2\pi v_{n_{1},m_{1},n_{2},m_{2}}t\right)\\&\delta(\tau-\tau_{n_{1},m_{1}}-\tau_{n_{2},m_{2}})+O_{isac}\cdot h_{u,s}^{back}(t,\tau)\Big]+\sqrt{K_{EO}}\cdot\sum_{k=1}H_{u,s,k}^{EO}(t)\cdot\delta\big(\tau-\tau_{EO,k}\big)
    \end{align}
\end{table*}

\section{Unified ISAC Channel Modeling under the Extended Geometry-Based Stochastic Model}\label{section7} 

\subsection{E-GBSM for 6G Channel}

With the global advancement of cutting-edge research on the 6G mobile communication systems, both industry and academia have proposed numerous novel technologies, frequency bands, and applications. These include ISAC, Extra-Large Scale Multiple-Input Multiple-Output (XL-MIMO), multi-band communications, RIS, and integrated space-air-ground-sea networks. These innovations aim to deliver ultra-high-speed, ultra-low-latency, and ultra-reliable mobile communication services with extended coverage.
The 6G channel model serves as the foundational research for system design, technology evaluation, and network deployment planning and optimization. Channel model standardization is among the first critical tasks in 6G standardization. Emerging 6G technologies such as ISAC, XL-MIMO, and RIS introduce new channel characteristics that require specialized modeling approaches:
\begin{enumerate}
    \item ISAC, whose key new features have been thoroughly discussed in the preceding Sections \ref{section3}-\ref{section6}.
    \item XL-MIMO, with its enlarged antenna array aperture, leads to spatial non-stationarity and near-field effects.
    \item RIS requires the characterization of the cascaded Tx-RIS and RIS-Rx channels.
    \item New frequency bands (e.g., 6–24 GHz, millimeter-wave, and sub-THz) demand the accurate representation of frequency-dependent cluster-sparse propagation properties.
\end{enumerate}

\begin{table*}[!t]
    \centering  
    \caption{Comparative analysis of 6G channel simulators}
    \begin{tabular}{m{2.5cm}<{\centering}|m{2.5cm}<{\centering}|m{9.5cm}<{\centering}|m{2cm}<{\centering}}
        \toprule 
        \hline
        \textbf{Channel simulator} & \textbf{Developer} & \textbf{Channel simulation capability}  & \textbf{Release time} \\ \hline 
        
        BUPTCMCC CMG-IMT2030 &  ARTT Lab of BUPT  &This channel model, developed based on measurements spanning from 0.5 to 330 GHz \cite{tang2021channel}, supports advanced 6G features and technologies including ISAC, near-field communications, SnS, RIS, NTN, among others. The simulator demonstrates full compatibility with standardized channel models such as ITU-R M.2412, 3GPP TR 38.900/901, TR 36.777, TR 36.873, and TR 37.885, with calibration performed using TR 38.901 as the baseline. The researchers have been actively engaged in the 3GPP Release 19 ISAC and 7-24 GHz channel modeling standardization activities. Notably, the simulator incorporates the latest Release 19 channel simulation capabilities, representing state-of-the-art implementation of current 3GPP channel modeling standards. & June 2023 \\ \hline  
        
        QuaDRiGa v2.8.1 & Fraunhofer Heinrich Hertz Institute (HHI) \cite{CMG_Ref2} & The simulator supports standardized channel models including 3GPP TR 36.873, TR 37.885, TR 38.901 TDL/CDL, mmMAGIC framework, satellite channels, and spatial consistency modeling for accurate channel evolution representation.  & December 2023 \\ \hline
        
        NYUSIM v4.0 & NYU WIRELESS \cite{NYUSIM} & Based on measured statistical channel models, supporting UMi, UMa, RMa, InH, InF scenarios simulation from 0.5 to 150 GHz. & June 2023  \\ \hline
        
        Thor Simulator & Horizon 2020 Joint EU-Japan project \cite{thor} & A simulator for 300 GHz backhaul links. & June 2022  \\ \hline
        
        KUCG v1.0.1 & Kyoto University \cite{kucg} & Used to generate channel impulse responses for the 60 GHz (millimeter wave), 95 GHz and 105 GHz (sub-THz) frequency bands. These responses are based on actual channel measurements conducted by the Orihara Laboratory of Kyoto University. & May 2024 \\ \hline 
        
        SimRIS v2.0.1 & Koç University \cite{CMG_SimRIS} & Can be used for channel modeling in RIS-assisted MIMO systems, with adjustable working frequency, terminal position, number of RIS components, and environment. & August 2023  \\ \hline 
        
        NirvaWave & Princeton University \cite{nirvawave}&  A near-field channel simulator based on scalar diffraction theory and Fourier principles, accurately modeling propagation evolution of arbitrary EM signals through complex user-defined wireless mediums.  & September 2024 \\ \hline
        \bottomrule
    \end{tabular}
    \label{Sum of simulators}
\end{table*}

The 3D GBSM has been widely adopted in 5G channel modeling due to its strong scalability. Given its proven effectiveness, this approach is expected to remain the foundation for standard channel models in 6G, as endorsed by the ITU and 3GPP.
Building upon the 3D GBSM framework, we proposes an Extended GBSM (E-GBSM) for 6G standardization, as expressed in equation (\ref{eq:EGBSM}). In (\ref{eq:EGBSM}), $h_{q}^{Tar-Rx}(\Omega^{out},\tau)$ and $h_{p}^{Tx-Tar}(\Omega^{in},\tau)$ represent channels of the link from the ST / RIS to the Rx and the link from the Tx to the ST / RIS, respectively. These sub-channels are expressed as
\begin{align}
        \label{eq:Subchannel}
        h_{p,q}^{\mathrm{Tx-Tar-Rx}}(\tau)=&\iint h_q^{\mathrm{Tar-Rx}}\left(\Omega^{\mathrm{out}},\tau\right)^*h_p^{\mathrm{Tx-Tar}}\left(\Omega^{\mathrm{in}},\tau\right)\notag\\&\cdot F^{\mathrm{Tar}}(\Omega^{\mathrm{out}},\Omega^{\mathrm{in}})\mathrm{d}\theta^{\mathrm{out}}\mathrm{d}\theta^{\mathrm{in}}.
\end{align}

The remaining symbols in this equation are defined as follows: $(\cdot)^T$ stands for matrix transposition.  $\lambda$ is the wavelength of the carrier frequency. $\tau$ is the delay of multipaths. $N_1$ represents the number of clusters of the new frequency band channels, used to characterize the sparsity of the new frequency band channels. $S_{p,n}^{Tx/Rx}$ describe the spatial non-stationary characteristics of the nth cluster in the $q$th receiving antenna and the $p$th transmitting antenna in the XL-MIMO channel. $A_{p,n,m}^{Tx/Rx}$ is the near-field steering vector of the $m$th ray in the $n$th cluster when the $p$th array element is received or the $q$th array element is emitted.  $I_{n,m}$ is the intra-cluster power allocation coefficient, which is used to describe the multi-path sparsity within each cluster of the new frequency band channels. When the signal carrier frequency is in the low-frequency range, $I_{n,m}=1$. $*$ represents the concatenated convolution relationship between the sensing target channel / RIS channel and the two adjacent sub-channels.
$F^{Tar}(\Omega^{out},\Omega^{in})$ is the equivalent radiation pattern of the RCS or RIS representing the target. $\Omega^{in}$ and $\Omega^{out}$ represent the incident and reflection angles of the target / RIS. $N_s$ is the number of shared clusters between communication channel and sensing channel.  $p$ and $q$ are the $p$-th transmit antenna and $q$-th receive antenna,respectively.

\begin{figure*}[!h]
    \centering
    \includegraphics[width=0.88\linewidth]{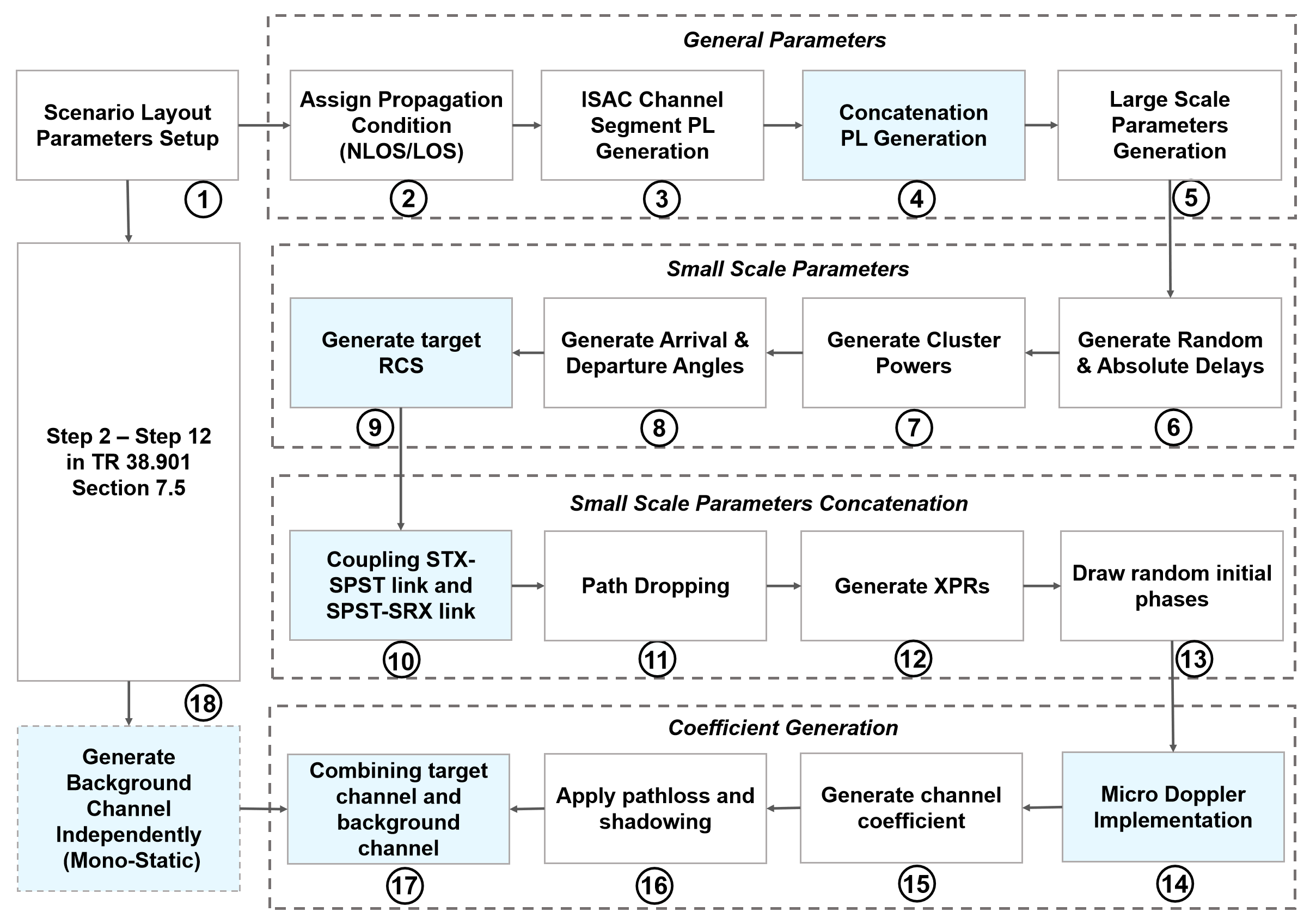} 
    \caption{Channel coefficient generation procedure \cite{3gpp38901}.}
    \label{sim flow}
\end{figure*}

\subsection{Unified ISAC Channel Modeling Methodology}

The proposed model retains the fundamental principles of GBSM while incorporating necessary enhancements to address 6G-specific propagation characteristics. Based on the E-GBSM model, a unified ISAC channel model considering all empirical features described in Section \ref{section3}-\ref{section6} will be introduced in this section. Fig. \ref{fig_EGBSM} illustrates a typical frequency-selective ISAC channel that includes both bistatic (consistent with communication configuration) and monostatic sensing configurations. In the propagation paths between the BS and UT, orange lines in Fig. \ref{fig_EGBSM} represent the target channel, while blue lines indicate the background channel. The ST, in this case a vehicle, exhibits RCS characteristics, motion-induced Doppler shifts, and spatial consistency. The environment also contains deterministic EOs. In scenarios where communication and sensing coexist, certain channel clusters are shared between the two functional channels.

The unified ISAC channel model is expressed as equation (\ref{eq:isacchannel}).
Here, $N_1$ and $N_2$ represent the number of clusters in the Tx-target and target-Rx sub-channels, respectively. $M_1$ and $M_2$ denote the number of rays within each cluster in the Tx-target and target-Rx sub-channels, respectively. $P_{n1,m1}$ and $P_{n2,m2}$ represent the power of the $n_1$-th cluster and $m_1$-th ray in the Tx-target link, and the $n_2$-th cluster and $m_2$-th ray in the target-Rx link, respectively. $F_{u}^{Rx}$ and $F_{s}^{Tx}$ are the antenna patterns of receive antenna u and transmit antenna s, respectively. $\phi_{out}$ and $\phi_{in}$ are the RCS outgoing and incident angles, respectively. $\phi_{Rx}$ and $\phi_{Tx}$ denote the angle of arrival at the receiver and angle of departure at the transmitter, respectively. $r_{Rx}^T$ and $r_{Tx}^T$ are the unit direction vectors at the receiver and transmitter, respectively. $d_{Rx,u}$ and $d_{Tx,s}$ are the position vectors of receive antenna $u$ and transmit antenna $s$, respectively. $v$ represents the Doppler shift, $\tau$ denotes the time delay. $O_{isac}$ is the background channel normalization factor. $h_{u,s}^{back}$ is the background channel. $K_{EO}$ is the EO power factor. $h_{u,s,k}^{EO}$ is the EO channel.

\begin{figure*}[!t]
  \centering
    \begin{minipage}[c]{\linewidth}
    \centering
      \subfloat[]{
        \includegraphics[width=0.45\linewidth]{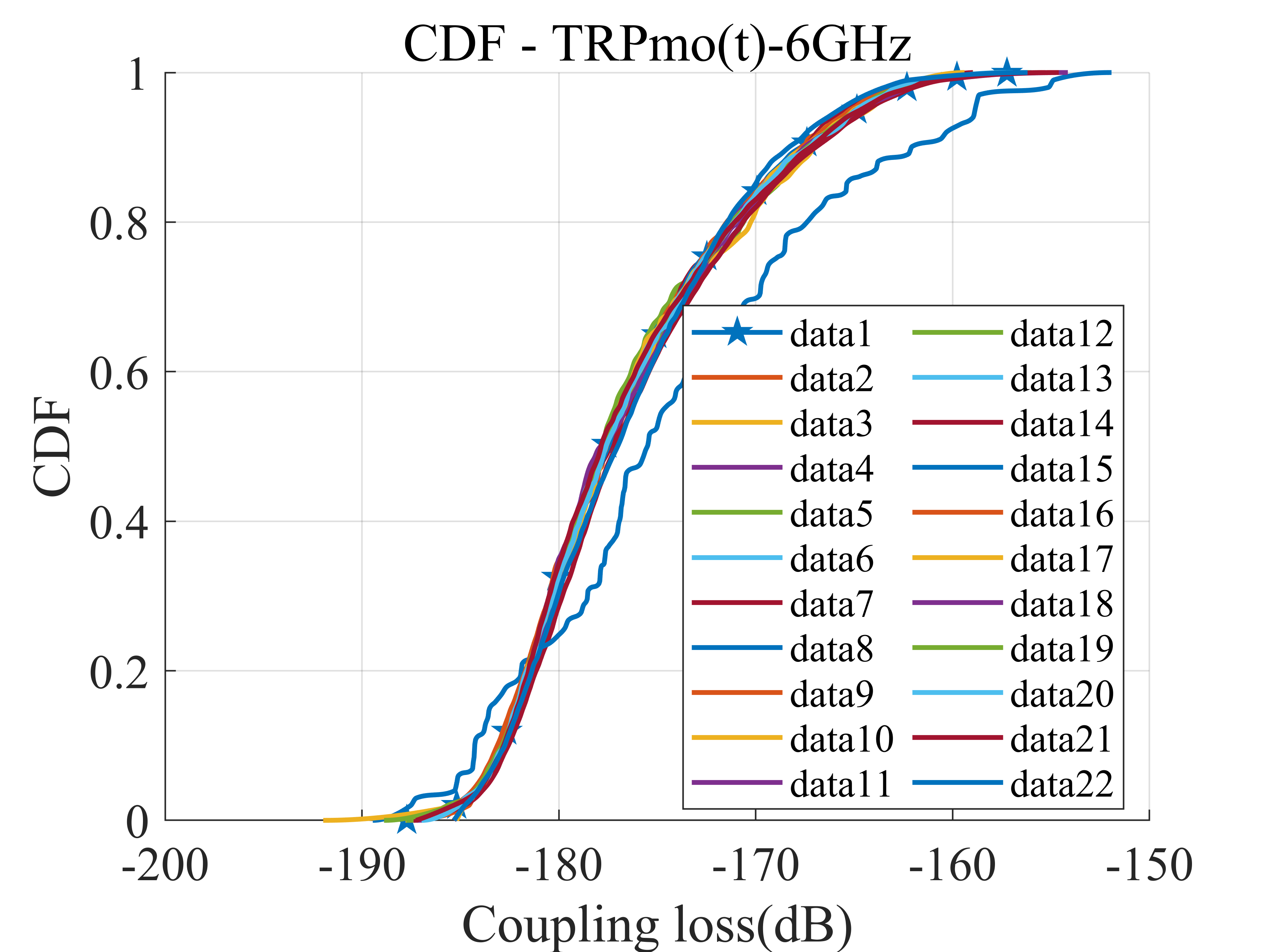}
      }
      \subfloat[]{
        \includegraphics[width=0.45\linewidth]{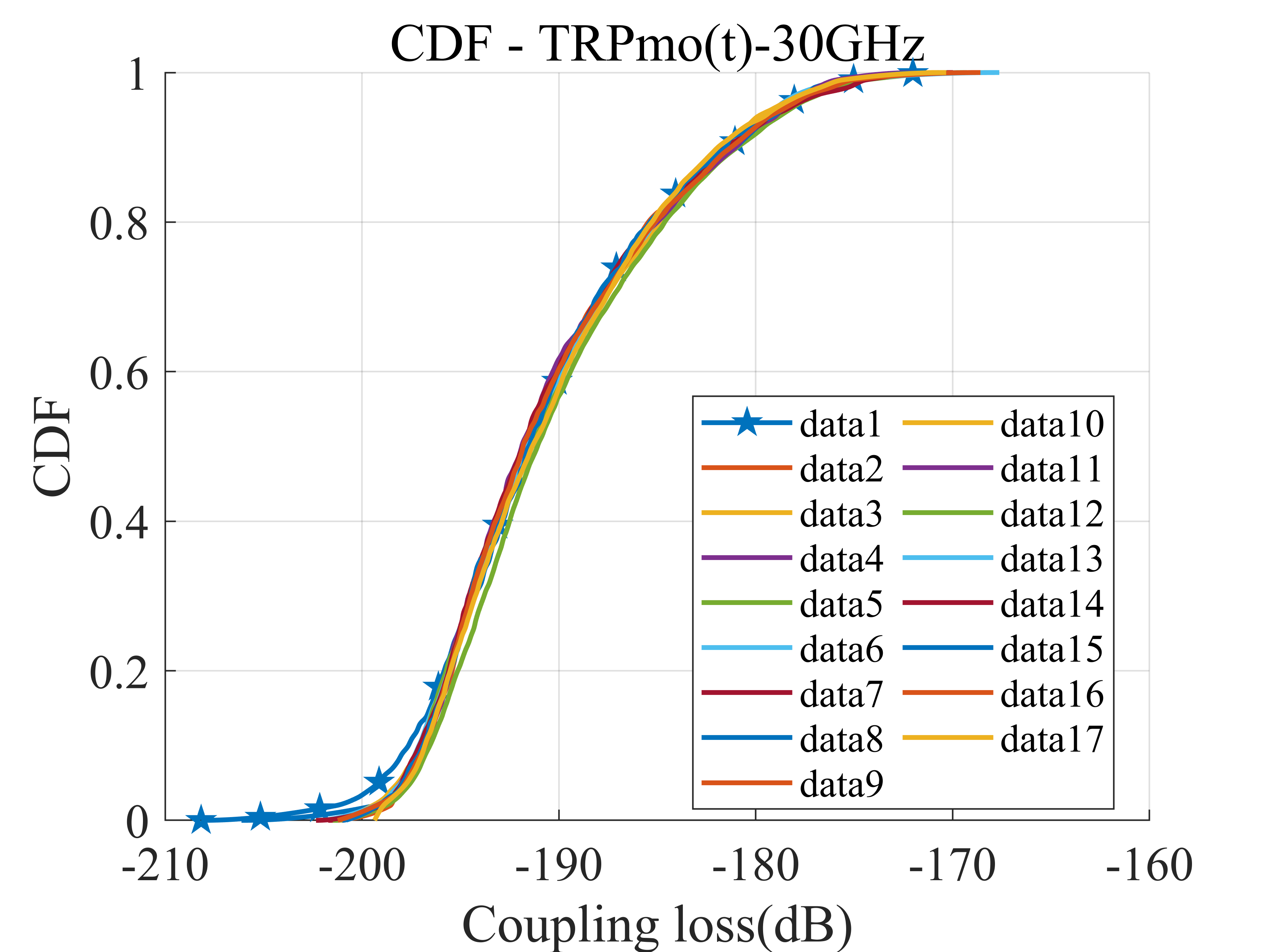}
      }

    \end{minipage}

    \begin{minipage}[c]{\linewidth}
    \centering
    \subfloat[]{
    \includegraphics[width=0.43\linewidth]{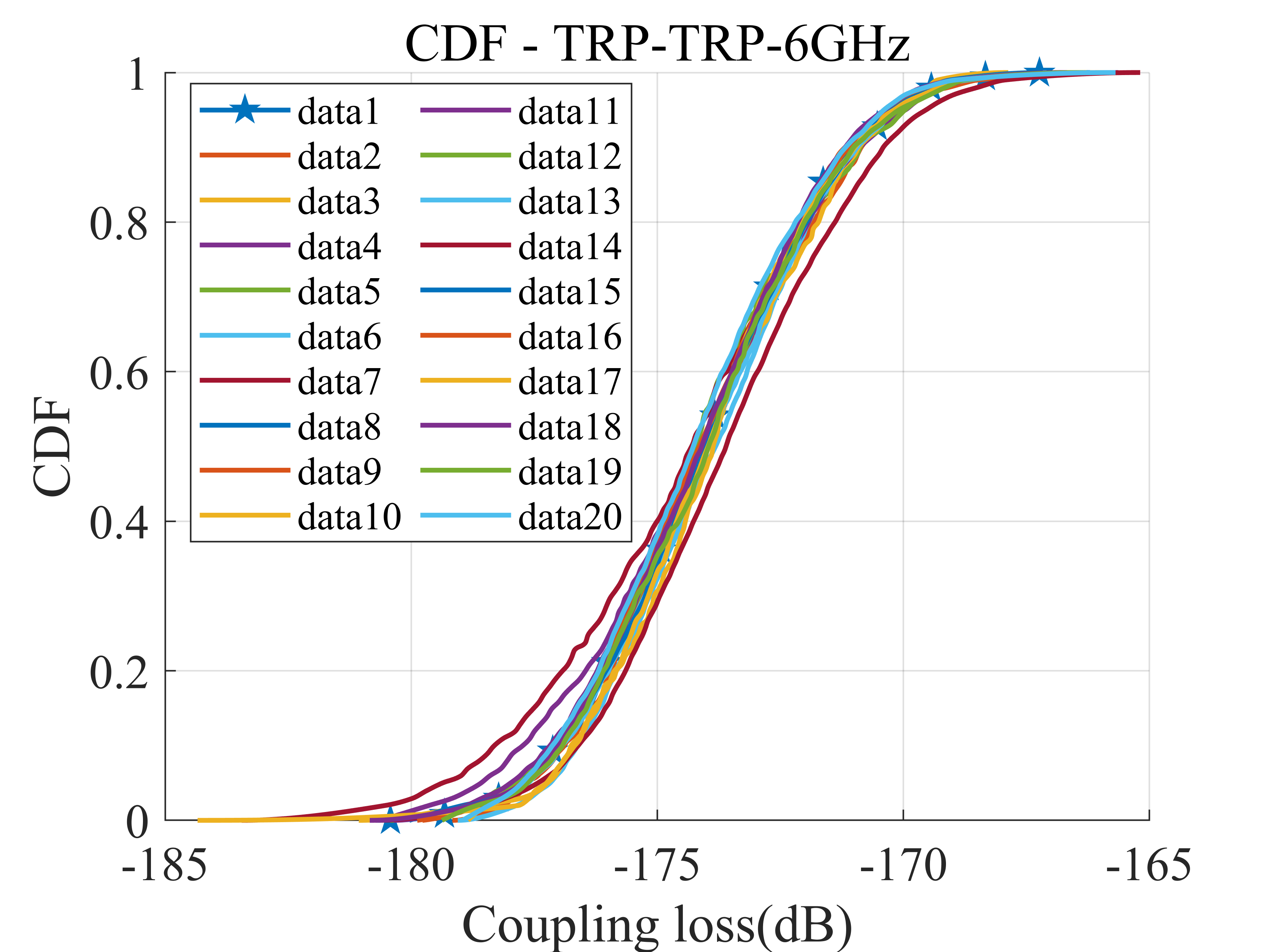}
      }
       \subfloat[]{
        \includegraphics[width=0.43\linewidth]{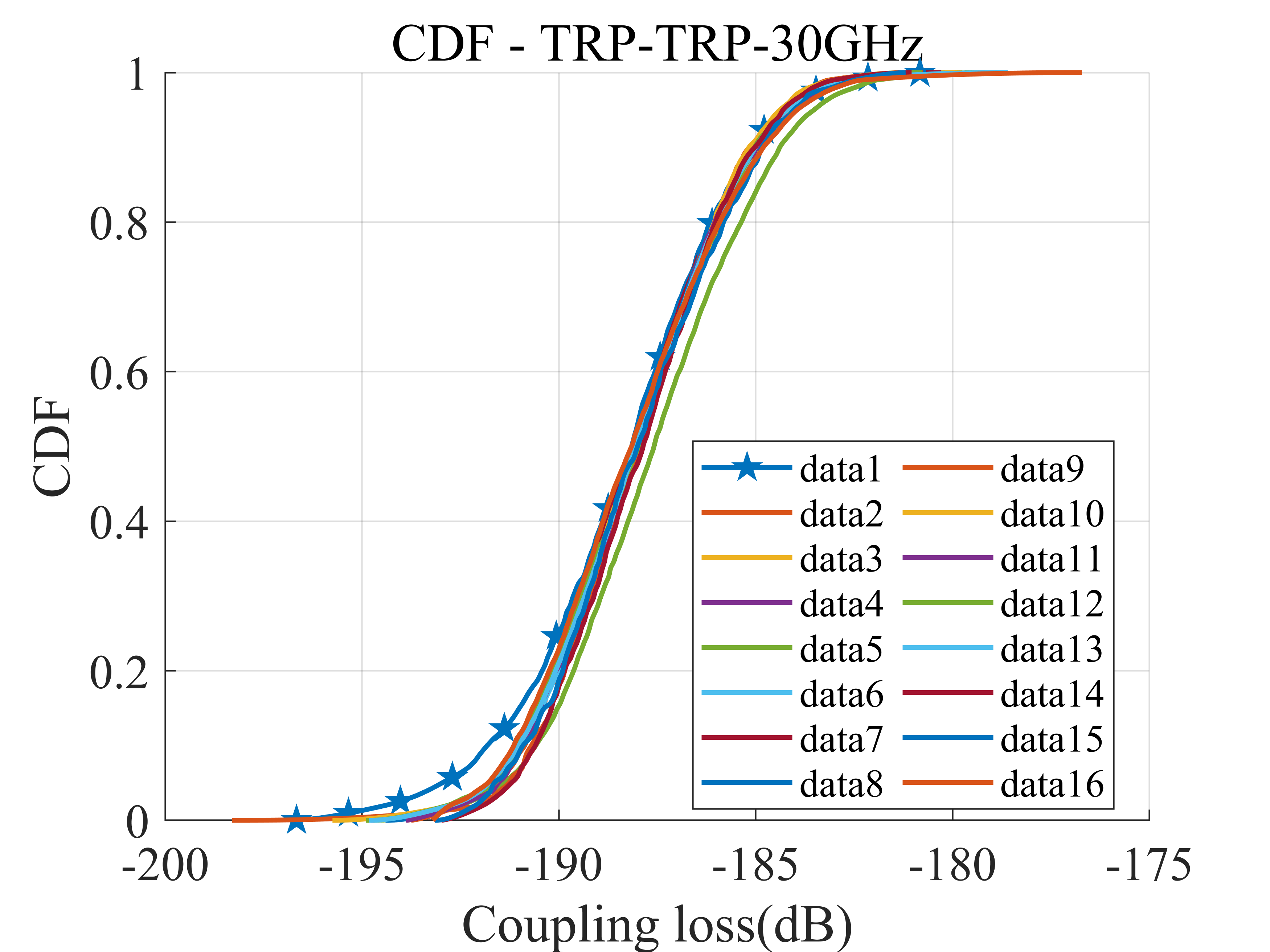}
      }
     \end{minipage}
    \caption{Comparison results with the calibration of 3GPP Release 19 ISAC channel model. TRPmo(t) represents the target channel in a monostatic case.}
    \label{fig:901}
\end{figure*}

\section{ISAC Channel Standardized Simulator and Calibration} \label{section8}

To support the research and evaluation of potential 6G technologies such as ISAC, XL-MIMO, THz (Terahertz), and RIS, various simulation platforms have been developed by research institutions and universities worldwide. Researchers from New York University compiled a comprehensive survey of simulators in 2023 \cite{NYUSIM}, encompassing channel simulators, link-level simulators, system-level simulators, and network simulators. In this work, we focus on a systematic survey of global 6G channel models and simulators. As summarized in Table \ref{Sum of simulators}, the majority of channel simulators target THz band characterization, whereas a minority address specific domains. For instance, NirvaWave specializes in near-field channel modeling, while SimRIS focuses on RIS channel emulation. Notably, BUPTCMCCCMG-IMT2030 stands out as one of the few simulators equipped with 3GPP Release 19 ISAC channel simulation capabilities.

\subsection{Simulation Workflow of the ISAC Channel Simulator}

The ISAC channel simulation framework of the proposed BUPTCMCCCMG-IMT2030 is designed to generate realistic time-varying channel coefficients by incorporating both sensing and communication components. The comprehensive simulation flowchart is depicted in Fig.~\ref{sim flow}, and the detailed procedure is described as follows.

\subsubsection{Large-Scale Parameter Generation}
The simulation process initiates by assigning a propagation condition, i.e., LoS or NLoS, for each channel segment. Subsequently, the path loss for the ISAC channel segment is generated, followed by the concatenation of path loss components across the entire link. Large-scale parameters, including shadow fading and distance-dependent attenuation, are then generated to characterize the propagation  effects of large-scale fading.

\subsubsection{Small-Scale Parameter Generation}
This module focuses on the small-scale fading characteristics. The key parameters are generated in the following sequence:
\begin{itemize}
    \item Generate random delays and calculate absolute delays for all paths.
    \item Generate the relative powers for each cluster.
    \item Generate the AoA and AoD for multipath components.
    \item Generate the RCS of the sensing target.
\end{itemize}

\subsubsection{Parameter Concatenation and Combining}
The parameters of STx-SPST and SPST-SRx links are coupled to form a complete ISAC concatenation channel. This stage involves:
\begin{itemize}
    \item Coupling the STx-SPST and SPST-SRx links.
    \item Implementing weak power paths-dropping.
    \item Generating XPRs.
    \item Drawing random initial phases for each path.
\end{itemize}
Concurrently, an independent monostatic background channel is generated to model environmental clutter and scattering.

\subsubsection{Channel Coefficient Synthesis}
The final CIR is synthesized in the combinatory modules. The operations within each combinator are executed as follows:
\begin{enumerate}

\item Implement Micro-Doppler: The micro-Doppler effect, induced by the motion of the target, is incorporated into the channel coefficients.

\item Generate Coefficients: The complex, time-varying channel coefficients $h(t, \tau)$ are generated.
    
\item Apply Large-Scale Effects: The combined channel is scaled by the total path loss and shadowing.

\item Combine Channels: The target channel and the background channel are coherently combined.
\end{enumerate}
This combinatory process is iterated across multiple instances to ensure statistical reliability and to capture the evolution of the channel over time and target movement.

\subsection{ISAC Channel Calibration}\label{section9}

To validate the reliability of the proposed model, this study presents comprehensive calibration results from the 3GPP Release 19 standardization phase, benchmarking our model against reference implementations from leading industry players in accordance with 3GPP TR 38.901 \cite{3gpp38901} specifications. The validation encompasses standardized test conditions including typical sensing modes, diverse target types (e.g., UAVs, vehicles, pedestrians), and multiple scenarios (UMa, InF, Urban grid). As an example, detailed simulation parameters for UAV targets are provided in Table \ref{tab:calibration_LSP} in Appendix A. Simulation results demonstrate close alignment with reference data reported by global telecom leaders in 3GPP documents, particularly in terms of large-scale calibration (coupling loss) and full calibration (DS, and AS of AoA, AoD, ZoA, and ZoD). The calibration follows a rigorous two-phase methodology: calibration of fundamental parameters through a progressive sequence from large-scale to full-scale parameter validation, ensuring comprehensive model fidelity across all critical dimensions.

The simulation results presented in Fig. \ref{fig:901} demonstrate the coupling loss characteristics across UAV-UMa scenario. A detailed comparative analysis with 3GPP reference data reveals remarkable consistency in our modeling approach. Notably, Fig. \ref{fig:901}(a) and (b) exhibit excellent agreement between our simulations and standardized results for TRP monostatic sensing mode at 6 GHz and 30 GHz respectively, while similar levels of accuracy are observed TRP-TRP bistatic sensing mode at 6 GHz and 30 GHz respectively (Fig. \ref{fig:901}(c) and (d)). This comprehensive validation across diverse propagation conditions, including both sensing modes  and and typical frequency, conclusively verifies the effectiveness of our large-scale fading modeling methodology. The close alignment with 3GPP benchmarks throughout all test scenarios underscores the model's reliability for practical 6G system design and performance evaluation. For the purposes of large scale calibration for UAV sensing targets, the following calibration parameters are provided in Table \ref{tab:calibration_LSP}.
The BUPTCMCCCMG-IMT2030 channel model has undergone rigorous calibration and validation against 3GPP Release 19. As demonstrated in Fig. \ref{fig:901}, the calibration results (data 1) are in excellent agreement with the measurement data from other industrial partners.


\section{Conclusion} \label{section9}

3GPP Release 19 RAN1 has completed the preliminary work on the ISAC standardized channel model, and the official standardization, 3GPP TR 38.901 V19.0.0, was officially released in June 2025. This paper presents a comprehensive survey and tutorial on ISAC standardized channel modeling, covering new features validated by empirical measurements, a unified modeling methodology based on E-GBSM, and the supporting channel simulator BUPTCMCCCMG-IMT2030, developed under the 3GPP GBSM framework. Specifically, we firstly provides a comprehensive overview of the key requirements and challenges in ISAC channel research, and the standardization workflow throughout of the ISAC study item in 3GPP Release 19 process are presented. Then, an in-depth discussion of critical channel aspects including RCS of physical objects, target channels, background channels, as well as additional features such as spatial consistency, EO, macro and micro-Doppler effect, and shared clusters, are provided based on extensive measurement-based analysis. To enable unified ISAC channel modeling, the E-GBSM is proposed, capturing all key ISAC channel characteristics and yielding statistically consistent time-varying coefficients. Based on the proposed E-GBSM, a standardized simulator BUPTCMCCCMG-IMT2030 is developed, and a two-phase calibration procedure aligned with 3GPP standardization is conducted to validate the framework across standardized scenarios and sensing modes. The results demonstrate high consistency with reference implementations at 6 GHz and 30 GHz under both TRP-monostatic and bistatic sensing configurations, thereby confirming the engineering applicability of the standardized channel model and simulator.
This paper provides a systematic survey of 3GPP Release 19 ISAC channel standardization and offers insights into best practices for new feature characterization, unified modeling methodology, and standardized simulator implementation, effectively supporting ISAC performance evaluation and future 6G standardization efforts.

\bibliography{reference}

@article{kumari2017ieee,
  title={{IEEE} 802.11 ad-based radar: An approach to joint vehicular communication-radar system},
  author={Kumari, Preeti and Choi, Junil and Gonz{\'a}lez-Prelcic, Nuria and Heath, Robert W},
  journal={IEEE Trans. Veh. Technol.},
  volume={67},
  number={4},
  pages={3012--3027},
  year={Nov. 2017},
  publisher={IEEE}
}

@article{nie2022predictive,
  title={A predictive {6G} network with environment sensing enhancement: From radio wave propagation perspective},
  author={Nie, Gaofeng and Zhang, Jianhua and Zhang, Yuxiang and Yu, Li and Zhang, Zhen and Sun, Yutong and Tian, Lei and Wang, Qixing and Xia, Liang},
  journal={China Commun.},
  volume={19},
  number={6},
  pages={105--122},
  year={Jun. 2022},
  publisher={IEEE}
}

@article{zhang2018multibeam,
  title={Multibeam for joint communication and radar sensing using steerable analog antenna arrays},
  author={Zhang, J Andrew and Huang, Xiaojing and Guo, Y Jay and Yuan, Jinhong and Heath, Robert W},
  journal={IEEE Trans. Veh. Technol.},
  volume={68},
  number={1},
  pages={671--685},
  year={Nov. 2018},
  publisher={IEEE}
}

@article{liu2022survey,
  title={A survey on fundamental limits of integrated sensing and communication},
  author={Liu, An and Huang, Zhe and Li, Min and Wan, Yubo and Li, Wenrui and Han, Tony Xiao and Liu, Chenchen and Du, Rui and Tan, Danny Kai Pin and Lu, Jianmin and others},
  journal={IEEE Commun. Surveys Tuts.},
  volume={24},
  number={2},
  pages={994--1034},
  year={Feb. 2022},
  publisher={IEEE}
}

@article{cui2021integrating,
  title={Integrating sensing and communications for ubiquitous {IoT}: Applications, trends, and challenges},
  author={Cui, Yuanhao and Liu, Fan and Jing, Xiaojun and Mu, Junsheng},
  journal={IEEE Netw.},
  volume={35},
  number={5},
  pages={158--167},
  year={Nov. 2021},
  publisher={IEEE}
}

@article{zhang2021overview,
  title={An overview of signal processing techniques for joint communication and radar sensing},
  author={Zhang, J Andrew and Liu, Fan and Masouros, Christos and Heath, Robert W and Feng, Zhiyong and Zheng, Le and Petropulu, Athina},
  journal={IEEE J. Sel. Topics Signal Process.},
  volume={15},
  number={6},
  pages={1295--1315},
  year={Sep. 2021},
  publisher={IEEE}
}

@article{pucci2022system,
  title={System-level analysis of joint sensing and communication based on 5{G} new radio},
  author={Pucci, Lorenzo and Paolini, Enrico and Giorgetti, Andrea},
  journal={IEEE J. Sel. Areas Commun.},
  volume={40},
  number={7},
  pages={2043--2055},
  year={Mar. 2022},
  publisher={IEEE}
}

@article{zhang2020channel,
  title={Channel measurements and models for {6G}: current status and future outlook},
  author={Zhang, Jianhua and Tang, Pan and Yu, Li and Jiang, Tao and Tian, Lei},
  journal={Front. Inf. Technol. Electron.
Eng.},
  volume={21},
  number={1},
  pages={39--61},
  year={Mar. 2020},
  publisher={Springer}
}

@article{yuan2022spatial,
  title={Spatial Non-stationary Near-field Channel Modeling and Validation for Massive {MIMO} Systems},
  author={Yuan, Zhiqiang and Zhang, Jianhua and Ji, Yilin and Pedersen, Gert Fr{\o}lund and Fan, Wei},
  journal={IEEE Trans. Antennas Propag.},
  volume={71},
  number={1},
  pages={921--933},
  year={Nov. 2022},
  publisher={IEEE}
}

@article{nguyen2022access,
  title={Access Management in Joint Sensing and Communication Systems: Efficiency {Versus} {Fairness}},
  author={Nguyen, Trung Thanh and Elbassioni, Khaled and Luong, Nguyen Cong and Niyato, Dusit and Kim, Dong In},
  journal={IEEE Trans. Veh. Technol.},
  volume={71},
  number={5},
  pages={5128--5142},
  year={Feb. 2022},
  publisher={IEEE}
}

@article{rahman2019framework,
  title={Framework for a perceptive mobile network using joint communication and radar sensing},
  author={Rahman, Md Lushanur and Zhang, J Andrew and Huang, Xiaojing and Guo, Y Jay and Heath, Robert W},
  journal={IEEE Trans. Aerosp. Electron. Syst.},
  volume={56},
  number={3},
  pages={1926--1941},
  year={Sep. 2019},
  publisher={IEEE}
}

@article{zhang2022integrated,
  title={Integrated sensing and communication waveform design with sparse vector coding: Low sidelobes and ultra reliability},
  author={Zhang, Ruoyu and Shim, Byonghyo and Yuan, Weijie and Di Renzo, Marco and Dang, Xiaoyu and Wu, Wen},
  journal={IEEE Trans. Veh. Technol.},
  volume={71},
  number={4},
  pages={4489--4494},
  year={Jan. 2022},
  publisher={IEEE}
}

@article{yuan2020spatio,
  title={Spatio-temporal power optimization for {MIMO} joint communication and radio sensing systems with training overhead},
  author={Yuan, Xin and Feng, Zhiyong and Zhang, J Andrew and Ni, Wei and Liu, Ren Ping and Wei, Zhiqing and Xu, Changqiao},
  journal={IEEE Trans. Veh. Technol.},
  volume={70},
  number={1},
  pages={514--528},
  year={Dec. 2020},
  publisher={IEEE}
}

@inproceedings{poutanen2010significance,
  title={Significance of common scatterers in multi-link indoor radio wave propagation},
  author={Poutanen, Juho and Haneda, Katsuyuki and Salmi, Jussi and Kolmonen, Veli-Matti and Tufvesson, Fredrik and Hult, Tommy and Vainikainen, Pertti},
  booktitle={Proc. 4th Eur. Conf. Antennas Propag. (EuCAP)},
  pages={1--5},
  year={Jul. 2010},
}

@article{poutanen2011multi,
  title={Multi-link {MIMO} channel modeling using geometry-based approach},
  author={Poutanen, Juho and Tufvesson, Fredrik and Haneda, Katsuyuki and Kolmonen, Veli-Matti and Vainikainen, Pertti},
  journal={IEEE Trans. Antennas Propag.},
  volume={60},
  number={2},
  pages={587--596},
  year={Mar. 2011},
  publisher={IEEE}
}

@article{chen2021code,
  title={Code-division {OFDM} joint communication and sensing system for 6{G} machine-type communication},
  author={Chen, Xu and Feng, Zhiyong and Wei, Zhiqing and Zhang, Ping and Yuan, Xin},
  journal={IEEE Internet Things J.},
  volume={8},
  number={15},
  pages={12093--12105},
  year={Feb. 2021},
  publisher={IEEE}
}

@article{zhang20173,
  title={3-{D} {MIMO}: {How} much does it meet our expectations observed from channel measurements?},
  author={Zhang, Jianhua and Zhang, Yuxiang and Yu, Yawei and Xu, Ruijie and Zheng, Qingfang and Zhang, Ping},
  journal={IEEE J. Sel. Areas Commun.},
  volume={35},
  number={8},
  pages={1887--1903},
  year={Jun. 2017},
  publisher={IEEE}
}

@article{chen2006micro,
  title={Micro-Doppler effect in radar: phenomenon, model, and simulation study},
  author={Chen, Victor C and Li, Fayin and Ho, S-S and Wechsler, Harry},
  journal={IEEE Trans. Aerosp. Electron. Syst.},
  volume={42},
  number={1},
  pages={2--21},
  year={2006},
  publisher={IEEE}
}

@inproceedings{szwoch2018suppression,
  title={Suppression of distortions in signals received from Doppler sensor for vehicle speed measurement},
  author={Szwoch, Grzegorz},
  booktitle={Proc. Signal Process.: Algorithms, Architectures, Arrangem., Appl. (SPA)},
  pages={16--21},
  year={2018},
  organization={IEEE}
}

@inproceedings{wang2018road,
  title={Road targets recognition based on deep learning and micro-Doppler features},
  author={Wang, Jie and Guo, Jianying and Shao, Xiang and Wang, Kai and Fang, Xulong},
  booktitle={Proc. Int. Conf. Sensor Netw. Signal Process. (SNSP)},
  pages={271--276},
  year={2018},
  organization={IEEE}
}

@inproceedings{wei2019classification,
  title={Classification of pedestrian motion based on micro-Doppler feature with LFMCW RADAR},
  author={Wei, Yinsheng and Zhang, Yun and Xu, Zhaoyang and Li, Xin},
  booktitle={Proc. IEEE Int. Conf. Signal, Inf. Data Process. (ICSIDP)},
  pages={1--5},
  year={2019},
  organization={IEEE}
}

@inproceedings{lin2017performance,
  title={Performance analysis of classification algorithms for activity recognition using micro-Doppler feature},
  author={Lin, Yier and Le Kernec, Julien},
  booktitle={Proc. 13th Int. Conf. Comput. Intell. Secur. (CIS)},
  pages={480--483},
  year={2017},
  organization={IEEE}
}

@inproceedings{rabbani2020wireless,
  title={Wireless health monitoring with 60 GHz-band beam scanning Micro-Doppler radar},
  author={Rabbani, Muhammad and Feresidis, Alexandros},
  booktitle={Proc. IEEE MTT-S Int. Microw. Biomed. Conf. (IMBioC)},
  pages={1--3},
  year={2020},
  organization={IEEE}
}

@inproceedings{hwang2022motion,
  title={Motion compensation for body-frame Doppler estimation of radar sensors on multi-rotor UAV platforms},
  author={Hwang, Seongbu and Yu, Taewoo and Nam, Sangwook},
  booktitle={Proc. 19th Eur. Radar Conf. (EuRAD)},
  pages={133--136},
  year={2022},
  organization={IEEE}
}

@inproceedings{wang2021lightweight,
  title={A lightweight UAV recognition algorithm based on micro-Doppler features},
  author={Wang, Yilin and Zhao, Caidan and Luo, Gege},
  booktitle={Proc. IEEE/CIC Int. Conf. Commun. in China (ICCC)},
  pages={6--10},
  year={2021},
  organization={IEEE}
}

@inproceedings{renga2014ship,
  title={Ship velocity estimation by Doppler Centroid analysis of focused SAR data},
  author={Renga, Alfredo and Moccia, Antonio},
  booktitle={Proc. IEEE Geosci. Remote Sens. Symp. (IGARSS)},
  pages={1809--1812},
  year={2014},
  organization={IEEE}
}

@inproceedings{wu2022measurement,
  title={Measurement and Extraction of Micro-Doppler Feature of Underwater Rotating Target Echo},
  author={Wu, Yongqing and Luo, Mingcheng and Li, Shengquan},
  booktitle={Proc. OCEANS 2022-Chennai},
  pages={1--5},
  year={2022},
  organization={IEEE}
}

@inproceedings{gomez2021air,
  title={Air-to-ground directional channel sounder with drone and 64-antenna dual-polarized cylindrical array},
  author={Gomez-Ponce, Jorge and Choi, Thomas and Abbasi, Naveed A and Adame, Aldo and Alvarado, Alexander and Bullard, Colton and Shen, Ruiyi and Daneshgaran, Fred and Dhillon, Harpreet S and Molisch, Andreas F},
  booktitle={Proc. IEEE Int. Conf. Commun. Workshops (ICC Workshops)},
  pages={1--6},
  year={2021},
  organization={IEEE}
}

@article{ref_北交综述和簇替换,
  title={A general channel model for integrated sensing and communication scenarios},
  author={Zhang, Zhengyu and He, Ruisi and Ai, Bo and Yang, Mi and Li, Chao and Mi, Hang and Zhang, Zhangdui},
  journal={IEEE Commun. Mag.},
  volume={61},
  number={5},
  pages={68--74},
  year={2022},
  publisher={IEEE}
}

@misc{ref_小米116提案,
     note= {R1-2400573 (Xiaomi, BUPT), Study on ISAC channel model, 3GPP RAN1 \#116 meeting, Feb 26th – Mar 1st, 2024}
}

@misc{ref_展讯117提案,
     note= {R1-2404039 (Spreadtrum Communications), Discussion on ISAC channel modeling, 3GPP RAN1 \#117 meeting, May 20th – May 24th, 2024}
}

@misc{ref_三星118提案,
    note= {R1-2406665 (Samsung), Discussion on ISAC channel modelling, 3GPP RAN1 \#118 meeting, Aug 19th – Aug 23rd, 2024}
}

@misc{ref_移动116提案,
    note= {R1-2400342 (CMCC, BUPT), Study on ISAC channel model, 3GPP RAN1 \#116 meeting, Feb 26th – Mar 1st, 2024}
}

@misc{ref_移动117提案,
    note= {R1-2404469 (CMCC, BUPT, SEU, PML), Discussion on channel modelling methodology for ISAC, 3GPP RAN1 \#117 meeting, May 20th – May 24th, 2024}
}

@misc{ref_118b次会议,
    note= {3GPP TSG RAN WG1 \#118b, Chair Notes, Oct. 14–Oct. 18, 2024. Available: https://www.3gpp.org/}
}

@misc{ref_Oppo116提案,
    note= {R1-2400617 (OPPO), Study on ISAC channel modelling, 3GPP RAN1 \#116 meeting, Feb 26th – Mar 1st, 2024}
}

@misc{ref_诺基亚116提案,
    note= {R1-2400649 (Nokia), Discussion on ISAC channel modelling, 3GPP RAN1 \#116 meeting, Feb 26th – Mar 1st, 2024}
}

@misc{ref_联想116提案,
    note= {R1-2400649 (Lenovo), Discussion on Channel Modelling for ISAC, 3GPP RAN1 \#116 meeting, Feb 26th – Mar 1st, 2024}
}

@misc{ref_北邮117提案,
     note= {R1-2404417 (BUPT, CMCC), Discussion on ISAC channel modeling, 3GPP RAN1 \#117 meeting, May 20th – May 24th, 2024}
}

@misc{ref_北邮118提案,
    note= {R1-2406107 (BUPT, CMCC, vivo), ISAC Channel Measurements and Modeling, 3GPP RAN1 \#118 meeting, Aug 19th – Aug 23rd, 2024}
}

@misc{ref_爱立信117提案,
    note= {R1-2405010 (Ericsson), Discussion on ISAC Channel Modelling, 3GPP RAN1 \#117 meeting, May 20th – May 24th, 2024}
}

@misc{ref_高通120提案,
    note= {R1-2501167 (Qualcomm Incorporated), Discussion on ISAC Channel modeling, 3GPP RAN1 \#120 meeting, Feb 17th – Feb 21st, 2025}
}

@misc{ref_小米120b提案,
    note= {R1-2502452 (Xiaomi, BJTU, BUPT), Discussion on ISAC Channel Model, 3GPP RAN1 \#120bis meeting, Apr 7th – Apr 11st, 2025}
}

@misc{ref_北邮120b提案,
    note= {R1-2502419 (BUPT, CMCC, vivo), ISAC Channel Modeling and Measurement Validation, 3GPP RAN1 \#120bis meeting, Apr 7th – Apr 11st, 2025}
}

@misc{ref_NIST119提案,
    note= {R1-2410136 (NIST), Discussion on ISAC Channel Modeling, 3GPP RAN1 \#119 meeting, Nov 18th – Nov 22nd, 2024}
}

@misc{ref_中兴118提案,
    note= {R1-2406960 (ZTE, Sanechips), Discussion on channel modelling for ISAC, 3GPP RAN1 \#118 meeting, Aug 19th – Aug 23rd, 2024}
}

@misc{ref_小米120提案,
    note= {R1-2500743 (Xiaomi, BJTU, BUPT), Discussion on ISAC channel model, 3GPP RAN1 \#120 meeting, Feb 17th – Feb 21st, 2025}
}

@misc{ref_华为119提案,
    note= {R1-2409394 (Huawei), Channel modelling for ISAC, 3GPP RAN1 \#119 meeting, Nov 18th – Nov 22nd, 2024}
}

@misc{ref_NIST120提案,
    note= {R1-2500681 (NIST), Discussion on ISAC Channel Modeling, 3GPP RAN1 \#120 meeting, Feb 17th – Feb 21st, 2025}
}

@misc{ref_维沃119提案,
    note= {R1-2409693 (vivo, BUPT), Views on Rel-19 ISAC channel modelling, 3GPP RAN1 \#119 meeting, Nov 18th – Nov 22nd, 2024}
}

@misc{ref_苹果120b提案,
    note= {R1-2502624 (Apple), Discussion on ISAC channel modelling, 3GPP RAN1 \#120bis meeting, Apr 7th – Apr 11st, 2025}
}

@misc{ref_Oppo120提案,
    note= {R1-2500463 (OPPO), Study on ISAC channel modelling, 3GPP RAN1 \#120 meeting, Feb 17th – Feb 21st, 2025}
}

@misc{ref_大唐119提案,
    note= {R1-2409953 (CATT, CICTCI), Discussion on ISAC channel modelling, 3GPP RAN1 \#119 meeting, Nov 18th – Nov 22nd, 2024}
}

@misc{ref_LGE118b提案,
    note= {R1-2408304 (LGE), Discussion on ISAC channel modelling, 3GPP RAN1 \#118bis meeting, Oct 14th – Oct 18th, 2024}
}

@misc{ref_LGE120提案,
    note= {R1-2501046 (LGE), Discussion on ISAC channel modelling, 3GPP RAN1 \#120 meeting, Feb. 17th – Feb. 21st, 2025}
}

@misc{ref_三星120b提案,
    note= {R1-2502379 (Samsung), Discussion on ISAC channel modelling, 3GPP RAN1 \#120bis meeting, Apr 7th – Apr 11st, 2025}
}

@misc{ref_华为120b提案,
    note= {R1-2502208 (Huawei), Channel modelling for ISAC, 3GPP RAN1 \#120bis meeting, Apr 7th – Apr 11st, 2025}
}

@misc{ref_维沃118提案,
    note= {R1-2406197 (vivo, BUPT), Study on ISAC channel modelling, 3GPP RAN1 \#118 meeting, Aug 19th – Aug 23rd, 2024}
}

@misc{ref_Docomo120提案,
    note= {R1-2501212 (NTT DOCOMO), Discussion on ISAC Channel Modelling, 3GPP RAN1 \#120 meeting, Feb. 17th – Feb. 21st, 2025}
}

@misc{ref_北邮119提案,
    note= {R1-2410659 (BUPT, CMCC, vivo), ISAC Channel Modeling and Measurement Validation, 3GPP RAN1 \#119 meeting, Nov 18th – Nov 22nd, 2024}
}

@misc{ref_华为118提案,
    note= {R1-2406975 (Huawei), Channel modelling for ISAC, 3GPP RAN1 \#118 meeting, Aug 19th – Aug 23rd, 2024}
}

@misc{ref_北邮121提案,
    note= {R1-2503859 (BUPT, CMCC, vivo, X-Net), ISAC Channel Modeling and Measurement Validation, 3GPP RAN1 \#121 meeting, May 19th – May 23rd, 2025}
}

@misc{ref_中兴120b提案,
    note= {R1-2502063 (ZTE Corporation, Sanechips, OPPO, BUPT, BJTU, CAICT, Xiaomi), Joint views on mono-static background channel modeling, 3GPP RAN1 \#120bis meeting, April 7th – 11st, 2025}
}

@misc{ref_docomo120b提案,
    note= {R1-2502776 (NTT DOCOMO, INC), Discussion on ISAC Channel Modelling, 3GPP RAN1 \#120bis meeting, April 7th – 11st, 2025}
}

@ARTICLE{ref_巩汇文WCL,
  author={Gong, Huiwen and Zhang, Jianhua and Zhang, Yuxiang and Zhou, Zhengfu and Liu, Guangyi},
  journal={IEEE Wireless Commun. Lett.}, 
  title={How to Extend {3-D GBSM} to {RIS} Cascade Channel With Non-Ideal Phase Modulation?}, 
  year={2024},
  volume={13},
  number={2},
  pages={555-559},
  doi={10.1109/LWC.2023.3336361}}

@INPROCEEDINGS{ref_张骥威glob,
  author={Zhang, Jiwei and Zhang, Yuxiang and Jiang, Tao and Gong, Huiwen and Xing, Hongbo and Tian, Lei},
  booktitle={Proc. IEEE Globecom Workshops (GC Wkshps)}, 
  title={Cascaded channel modeling and experimental validation for RIS assisted communication system}, 
  year={2024},
  volume={},
  number={},
  pages={1-6}
}

@ARTICLE{ref_刘亚萌WCL,
  author={Liu, Yameng and Zhang, Jianhua and Zhang, Yuxiang and Gong, Huiwen and Jiang, Tao and Liu, Guangyi},
  journal={IEEE Wireless Commun. Lett.}, 
  title={How to Extend {3-D GBSM} to Integrated Sensing and Communication Channel With Sharing Feature?}, 
  year={2024},
  volume={13},
  number={8},
  pages={2045-2049}
}

@article{zhang2023integrated,
  title={Integrated sensing and communication channel: Measurements, characteristics, and modeling},
  author={Zhang, Jianhua and Wang, Jialin and Zhang, Yuxiang and Liu, Yameng and Chai, Zeyong and Liu, Guangyi and Jiang, Tao},
  journal={IEEE Commun. Mag.},
  volume={62},
  number={6},
  pages={98--104},
  year={2023},
  publisher={IEEE}
}

@article{zhang2024latest,
  title={Latest progress for {3GPP} {ISAC} channel modeling standardization},
  author={Zhang, Yuxiang and Zhang, Jianhua and Pei, Yuanpeng and Liu, Yameng and Jiang, Tao},
  journal = {Sci. China Inf. Sci.},
  volume = {67},
  number = {11},
  pages = {217301},
  year={2024},
}

@article{luo2024channel,
  title={Channel Modeling Framework for Both Communications and Bistatic Sensing Under {3GPP} Standard},
  author={Luo, Chenhao and Tang, Aimin and Gao, Fei and Liu, Jianguo and Wang, Xudong},
  journal={IEEE J. Sel. Areas Commun.},
  year={2024},
  publisher={IEEE}
}

@article{yang2024integrated,
  title={Integrated sensing and communication channel modeling and measurements: Requirements and methodologies toward {6G} standardization},
  author={Yang, Wenfei and Chen, Yi and Cardona, Narcis and Zhang, Yunhao and Yu, Ziming and Zhang, Min and Li, Jian and Chen, Yan and Zhu, Peiying},
  journal={IEEE Veh. Technol. Mag.},
  year={2024},
  publisher={IEEE}
}

@article{lampropoulos1999high,
  title={High resolution radar clutter statistics},
  author={Lampropoulos, GA and Drosopoulos, A and Rey, N and others},
  journal={IEEE Trans. Aerosp. Electron. Syst.},
  volume={35},
  number={1},
  pages={43--60},
  year={1999},
  publisher={IEEE}
}

@article{addabbo2021learning,
  title={Learning strategies for radar clutter classification},
  author={Addabbo, Pia and Han, Sudan and Orlando, Danilo and Ricci, Giuseppe},
  journal={IEEE Trans. Signal Process.},
  volume={69},
  pages={1070--1082},
  year={2021},
  publisher={IEEE}
}

@misc{3gppRan102,
  note = {3GPP TSG RAN Meeting \#102, RP-234016, ``New Study Item Description (SID): Study on Channel Modelling for Integrated Sensing and Communication (ISAC) for NR,'' Dec. 2023. [Online]. Available: https://www.3gpp.org/}
}

@misc{3gpp22837,
  note = {3GPP, TR 22.837 V19.2.1, ``Feasibility Study on Integrated Sensing and Communication (ISAC),'' Feb. 2024. [Online]. Available: https://www.3gpp.org/}
}

@misc{3gppRan121sum,
  note= {3GPP TSG RAN WG1 \#121, Summary \#6 on ISAC channel modelling, May 19–Mar 23, 2025. Available: https://www.3gpp.org/}
}

@misc{3gpp38901,
  note= {3GPP, ``Study on channel model for frequencies from 0.5 to 100 {GHz}'', Sophia Antipolis Cedex, France, 3GPP TR 38.901, ver. 19.0.0, Jun. 2025. [Online]. Available: https://www.3gpp.org}
}

@misc{3gppTR38808,
  note = {3GPP, ``Study on supporting NR from 52.6 GHz to 71 GHz'', Sophia Antipolis Cedex, France, 3GPP TR 38.808, ver. 17.0.0, Mar. 2021. [Online]. Available: https://www.3gpp.org}
}

@misc{3gppTR37885,
  note = {3GPP, ``Study on evaluation methodology of new Vehicle‑to‑Everything (V2X) use cases for LTE and NR'', Sophia Antipolis Cedex, France, 3GPP TR 37.885, ver. 15.3.0, Jun. 2019. [Online]. Available: https://www.3gpp.org}
}

@misc{3gppTR38859,
  note = {3GPP, ``Study on expanded and improved NR positioning'', Sophia Antipolis Cedex, France, 3GPP TR 38.859, ver. 18.1.0, Jan. 2022. [Online]. Available: https://www.3gpp.org}
}

@misc{3gppTR36777,
  note = {3GPP, ``Enhanced LTE support for aerial vehicles'', Sophia Antipolis Cedex, France, 3GPP TR 36.777, ver. 15.0.0, Jan. 2018. [Online]. Available: https://www.3gpp.org}
}

@misc{3gpp117oppo,
  note= {R1-2404876 (OPPO), Study on ISAC channel modelling, 3GPP RAN1 \#117 meeting, May 20th – May 24th, 2024}
}

@misc{3gpp117nokia,
  note= {R1-2403995 (Nokia, Nokia Shanghai Bell), Discussion on ISAC channel modeling, 3GPP RAN1 \#117 meeting, May 20th – May 24th, 2024}
}

@misc{3gpp117bupt,
  note= {R1-2404417 (BUPT, CMCC), Discussion on ISAC channel modeling, 3GPP RAN1 \#117 meeting, May 20th – May 24th, 2024}
}

@misc{3gpp117vivo,
  note= {R1-2404190 (vivo), Views on Rel-19 ISAC channel modelling, 3GPP RAN1 \#117 meeting, May 20th – May 24th, 2024}
}

@article{winner2007winner,
  title={{WINNER II} channel models},
  author={Winner, II},
  journal={IST-4-027756 WINNER II, D. 1. 1. 2 V1. 2},
  year={2007}
}

@article{jaeckel2014quadriga,
  title={QuaDRiGa: A 3-D multi-cell channel model with time evolution for enabling virtual field trials},
  author={Jaeckel, Stephan and Raschkowski, Leszek and B{\"o}rner, Kai and Thiele, Lars},
  journal={IEEE Trans. Antennas Propag.},
  volume={62},
  number={6},
  pages={3242--3256},
  year={2014},
  publisher={IEEE}
}

@article{liu2012cost,
  title={The {COST 2100 MIMO} channel model},
  author={Liu, Lingfeng and Oestges, Claude and Poutanen, Juho and Haneda, Katsuyuki and Vainikainen, Pertti and Quitin, Fran{\c{c}}ois and Tufvesson, Fredrik and De Doncker, Philippe},
  journal={IEEE Wireless Commun.},
  volume={19},
  number={6},
  pages={92--99},
  year={2012},
  publisher={IEEE}
}

@inproceedings{ademaj2017modeling,
  title={Modeling of spatially correlated geometry-based stochastic channels},
  author={Ademaj, Fjolla and Mueller, Martin K and Schwarz, Stefan and Rupp, Markus},
  booktitle={Proc. IEEE 86th Veh. Technol. Conf. (VTC-Fall)},
  pages={1--6},
  year={2017},
  organization={IEEE}
}

@inproceedings{ademaj2018ray,
  title={Ray-tracing based validation of spatial consistency for geometry-based stochastic channels},
  author={Ademaj, Fjolla and Schwarz, Stefan and Guan, Ke and Rupp, Markus},
  booktitle={Proc. IEEE 88th Veh. Technol. Conf. (VTC-Fall)},
  pages={1--5},
  year={2018},
  organization={IEEE}
}

@inproceedings{ademaj2019spatial1,
  title={Spatial consistency of multipath components in a typical urban scenario},
  author={Ademaj, Fjolla and Schwarz, Stefan},
  booktitle={Proc. 13th Eur. Conf. Antennas Propag. (EuCAP)},
  pages={1--5},
  year={2019},
  organization={IEEE}
}

@article{ademaj2019spatial2,
  title={A spatial consistency model for geometry-based stochastic channels},
  author={Ademaj, Fjolla and Schwarz, Stefan and Berisha, Taulant and Rupp, Markus},
  journal={IEEE Access},
  volume={7},
  pages={183414--183427},
  year={2019},
  publisher={IEEE}
}

@inproceedings{liu2021analyzing,
  title={Analyzing {V2I} channel and spatial consistency through simulation},
  author={Liu, Mingmin and Wang, Yuyuan and Li, Huafu and Jing, Yanmei and Zhang, Gongsheng and He, Wenxue},
  booktitle={IEEE 7th Int. Conf. Comput. Commun. (ICCC)},
  pages={453--458},
  year={2021},
  organization={IEEE}
}

@article{medbo2016radio,
  title={Radio propagation modeling for 5G mobile and wireless communications},
  author={Medbo, Jonas and Kyosti, Pekka and Kusume, Katsutoshi and Raschkowski, Leszek and Haneda, Katsuyuki and Jamsa, Tommi and Nurmela, Vuokko and Roivainen, Antti and Meinila, Juha},
  journal={IEEE Commun. Mag.},
  volume={54},
  number={6},
  pages={144--151},
  year={2016},
  publisher={IEEE}
}

@INPROCEEDINGS{Wang2024Millimeter,
  author={Wang, Yang and Wang, Tianxiang and Zheng, Xiangquan and Liao, Xi and Zhang, Jie},
  booktitle={Proc. IEEE 99th Veh. Technol. Conf. (VTC2024-Spring)}, 
  title={Millimeter Wave Radio Propagation Measurements and Channel Characterization in Indoor Factory Environments for {ISAC}}, 
  year={2024},
  volume={},
  number={},
  pages={1-5}
  }

@article{li2024characteristics,
  title={Characteristics Analysis and Modeling of Integrated Sensing and Communication Channel for Unmanned Aerial Vehicle Communications},
  author={Li, Xinru and Liu, Yu and Zhang, Xinrong and Zhang, Yi and Huang, Jie and Bian, Ji},
  journal={Drones},
  volume={8},
  number={10},
  pages={538},
  year={2024},
  publisher={MDPI}
}

@INPROCEEDINGS{Xiong2023Channel,
  author={Xiong, Baiping and Zhang, Zaichen and Ge, Yingmeng and Wang, Haibo and Jiang, Hao and Wu, Liang and Zhang, Ziyang},
  booktitle={Proc. IEEE 98th Veh. Technol. Conf. (VTC2023-Fall)}, 
  title={Channel Modeling for Heterogeneous Vehicular ISAC System with Shared Clusters}, 
  year={2023},
  volume={},
  number={},
  pages={1-6},
  }

@article{zhang2025research,
  title={Research and Experimental Validation for {3GPP ISAC} Channel Modeling Standardization},
  author={Zhang, Yuxiang and Zhang, Jianhua and Zhang, Jiwei and Pei, Yuanpeng and Liu, Yameng and Tian, Lei and Jiang, Tao and Liu, Guangyi},
  journal={arXiv preprint arXiv:2504.09799},
  year={2025}
}

@article{liu2025coupling,
  author={Liu, Yameng and Zhang, Jianhua and Zhang, Yuxiang and Xing, Hongbo and Xiong, Yifeng and Yuan, Zhiqiang and Liu, Guangyi},
  journal={IEEE Trans. Cogn. Commun. Netw.}, 
  title={The Coupling Effect of Sensing Targets on the Environment for {3GPP} {ISAC} Channels: Observation, Modeling, and Validation}, 
  year={2025},
  volume={},
  number={},
  pages={1-1}
}

@article{xiong2023fundamental,
  title={On the fundamental tradeoff of integrated sensing and communications under Gaussian channels},
  author={Xiong, Yifeng and Liu, Fan and Cui, Yuanhao and Yuan, Weijie and Han, Tony Xiao and Caire, Giuseppe},
  journal={IEEE Trans. Inf. Theory},
  volume={69},
  number={9},
  pages={5723--5751},
  year={2023},
  publisher={IEEE}
}

@article{xiong2024torch,
  title={From torch to projector: Fundamental tradeoff of integrated sensing and communications},
  author={Xiong, Yifeng and Liu, Fan and Wan, Kai and Yuan, Weijie and Cui, Yuanhao and Caire, Giuseppe},
  journal={IEEE BITS Inf. Theory Mag.},
  year={2024},
  volume={4},
  number={1},
  pages={73-90},
  publisher={IEEE}
}

@article{yu2022location,
  title={Location sensing and beamforming design for {IRS}-enabled multi-user {ISAC} systems},
  author={Yu, Zhouyuan and Hu, Xiaoling and Liu, Chenxi and Peng, Mugen and Zhong, Caijun},
  journal={IEEE Trans. Signal Process.},
  volume={70},
  pages={5178--5193},
  year={2022},
  publisher={IEEE}
}

@article{chen2024empirical,
  title={An empirical study of {ISAC} channel characteristics with human target impact at 105 {GHz}},
  author={Chen, Wenjun and Zhang, Yuxiang and Liu, Yameng and Zhang, Jianhua and Gong, Huiwen and Jiang, Tao and Xia, Liang},
  journal={Electron. Lett.},
  volume={60},
  number={17},
  pages={e70017},
  year={2024},
  publisher={Wiley Online Library}
}

@article{barneto2022millimeter,
  title={Millimeter-wave mobile sensing and environment mapping: Models, algorithms and validation},
  author={Barneto, Carlos Baquero and Rastorgueva-Foi, Elizaveta and Keskin, Musa Furkan and Riihonen, Taneli and Turunen, Matias and Talvitie, Jukka and Wymeersch, Henk and Valkama, Mikko},
  journal={IEEE Trans. Veh. Technol.},
  volume={71},
  number={4},
  pages={3900--3916},
  year={2022},
  publisher={IEEE}
}

@inproceedings{jiang2024novel,
  title={A Novel Approach to Model the Scattering Environment in Channel Modeling for Integrated Sensing and Communications},
  author={Jiang, Chuangxin and Liu, Junchen and Lou, Junpeng and Liu, Ruiqi and Yang, Qi and Wang, Xinhui},
  booktitle={Proc. Int. Wireless Commun. Mobile Comput. Conf. (IWCMC)},
  pages={1809--1813},
  year={2024},
  organization={IEEE}
}

@inproceedings{bader2022nlos,
  title={{NLoS detection for enhanced 5G mmWave-based positioning for vehicular iot applications}},
  author={Bader, Qamar and Saleh, Sharief and Elhabiby, Mohamed and Noureldin, Aboelmagd},
  booktitle={Proc. IEEE GLOBECOM},
  pages={5643--5648},
  year={2022},
  organization={IEEE}
}

@article{zhang2016measurement,
  title={{Measurement and modeling of angular spreads of three-dimensional urban street radio channels}},
  author={Zhang, Ruonan and Lu, Xiaofeng and Zhao, Jianping and Cai, Lin and Wang, Jiao},
  journal={IEEE Trans. Veh. Technol.},
  volume={66},
  number={5},
  pages={3555--3570},
  year={2016},
  publisher={IEEE}
}

@ARTICLE{wen2025survey,
  author={Wen, Dingzhu and Zhou, Yong and Li, Xiaoyang and Shi, Yuanming and Huang, Kaibin and Letaief, Khaled B.},
  journal={IEEE Commun. Surv. Tutorials}, 
  title={A Survey on Integrated Sensing, Communication, and Computation}, 
  year={2025},
  volume={27},
  number={5},
  pages={3058-3098}
}

@article{blaunstein2006signal,
  title={{Signal power distribution in the azimuth, elevation and time delay domains in urban environments for various elevations of base station antenna}},
  author={Blaunstein, Nathan and Toeltsch, Martin and Laurila, Juha and Bonek, Ernst and Katz, Dmitry and Vainikainen, Pertti and Tsouri, Nissim and Kalliola, Kimmo and Laitinen, Heikki},
  journal={IEEE Trans. Antennas Propag.},
  volume={54},
  number={10},
  pages={2902--2916},
  year={2006},
  publisher={IEEE}
}

@article{zhong2019outdoor,
  title={Outdoor-to-Indoor Channel Measurement and Coverage Analysis for 5G Typical Spectrums},
  author={Zhong, Zhimeng and Zhao, Jianyao and Li, Chao},
  journal={Int. J. Antennas Propag.},
  volume={2019},
  number={1},
  pages={3981678},
  year={2019},
  publisher={Wiley Online Library}
}

@article{kuutti2018survey,
	title={{A survey of the state-of-the-art localization techniques and their potentials for autonomous vehicle applications}},
	author={Kuutti, Sampo and Fallah, Saber and Katsaros, Konstantinos and Dianati, Mehrdad and Mccullough, Francis and Mouzakitis, Alexandros},
	journal={IEEE Internet Things J.},
	volume={5},
	number={2},
	pages={829--846},
	year={2018},
	publisher={IEEE}
}

@inproceedings{al2002ml,
	title={{ML and Bayesian TOA location estimators for NLOS environments}},
	author={Al-Jazzar, S and Caffery, Jr},
	booktitle={Proc. IEEE 56th Veh. Technol. Conf. (VTC)},
	volume={2},
	pages={1178--1181},
	year={2002},
	organization={IEEE}
}

@article{liu2023shared,
  title={A shared cluster-based stochastic channel model for integrated sensing and communication systems},
  author={Liu, Yameng and Zhang, Jianhua and Zhang, Yuxiang and Yuan, Zhiqiang and Liu, Guangyi},
  journal={IEEE Trans. Veh. Technol.},
  volume={73},
  number={5},
  pages={6032--6044},
  year={2023},
  publisher={IEEE}
}

@inproceedings{jiang2025novel,
  title={A Novel Environment Object Modeling Method for Vehicular ISAC Scenarios},
  author={Jiang, Hanyuan and Zhang, Yuxiang and Liu, Yameng and Zhang, Jianhua and Tian, Lei and Jiang, Tao},
  booktitle={Proc. IEEE Wireless Commun. Netw. Conf. (WCNC)},
  pages={1--6},
  year={2025},
  organization={IEEE}
}

@inproceedings{chen2024and,
	title={{LOS and NLOS Targets Localization in an L-Shaped Corner}},
	author={Chen, Jiahui and Zhu, Zhihao and Qiu, Chen and Wu, Peilun and Guo, Shisheng and Cui, Guolong and Yang, Xiaobo},
	booktitle={Proc. IEEE Int. Geosci. Remote Sens. Symp.},
	pages={10487--10490},
	year={2024},
	organization={IEEE}
}

@article{ali2020leveraging,
  title={Leveraging sensing at the infrastructure for mmWave communication},
  author={Ali, Anum and Gonzalez-Prelcic, Nuria and Heath, Robert W and Ghosh, Amitava},
  journal={IEEE Commun. Mag.},
  volume={58},
  number={7},
  pages={84--89},
  year={2020},
  publisher={IEEE}
}

@book{molisch2012wireless,
  title={Wireless communications},
  author={Molisch, Andreas F},
  volume={34},
  year={2012},
  publisher={John Wiley \& Sons}
}

@article{tang2021channel,
	title        = {{Channel measurement and path loss modeling from 220 GHz to 330 GHz for 6G wireless communications}},
	author       = {Tang, Pan and Zhang, Jianhua and Tian, Haoyu and others},
	year         = 2021,
	month        = may,
	journal      = {{China Commun.}},
	publisher    = {IEEE},
	volume       = 18,
	number       = 5,
	pages        = {19--32}
}

@article{liu2025novel,
  title={A Novel Multi-Reference-Point Modeling Framework for Monostatic Background Channel: Toward {3GPP ISAC} Standardization},
  author={Liu, Yameng and Zhang, Jianhua and Zhang, Yuxiang and Yuan, Zhiqiang and Jiang, Chuangxin and Liu, Junchen and Hong, Wei and Li, Yingyang and Li, Yan and Liu, Guangyi},
  journal={arXiv preprint arXiv:2511.03487},
  year={2025}
}

@ARTICLE{yu2022implementation,
  author={Yu, Li and Zhang, Yuxiang and Zhang, Jianhua and Yuan, Zhiqiang},
  journal={China Commun.}, 
  title={Implementation framework and validation of cluster-nuclei based channel model using environmental mapping for 6G communication systems}, 
  year={2022},
  volume={19},
  number={4},
  pages={1-13},
  doi={10.23919/JCC.2022.04.001}}

@ARTICLE{zhang2016interdisciplinary,
  author={Zhang, Jianhua},
  journal={China Commun.}, 
  title={The interdisciplinary research of big data and wireless channel: A cluster-nuclei based channel model}, 
  year={2016},
  volume={13},
  number={Supplement2},
  pages={14-26},
  doi={10.1109/CC.2016.7833457}}

@article{wang2020small,
  title={Small target tracking in satellite videos using background compensation},
  author={Wang, Yunming and Wang, Taoyang and Zhang, Guo and Cheng, Qian and Wu, Jia-qi},
  journal={IEEE Trans. Geosci. Remote Sens.},
  volume={58},
  number={10},
  pages={7010--7021},
  year={2020},
  publisher={IEEE}
}

@article{xu2023spatial,
  title={Spatial Structure Matching Based {DoA} Estimation and Tracking for Integrated Sensing and Communication Massive {MIMO OFDM} System},
  author={Xu, Kui and Xia, Xiaochen and Li, Chunguo and Hu, Guojie and Su, Qiao and Xie, Wei},
  journal={IEEE Trans. Cogn. Commun. Netw.},
  year={2023},
  volume={10},
  number={2},
  pages={526-540},
  publisher={IEEE}
}

@misc{3gppRan116,
  note= {3GPP TSG RAN WG1 \#116, Chair Notes, Feb. 26–Mar. 1, 2024. Available: https://www.3gpp.org/}
}

@ARTICLE{gong2024new,
  author={Gong, Huiwen and Zhang, Jianhua and Zhang, Yuxiang and Liu, Guangyi},
  journal={IEEE Veh. Technol. Mag.}, 
  title={New Characteristics and Modeling of {6G} Channels: Toward a Unified Channel Model for Standardization}, 
  year={2025},
  volume={20},
  number={3},
  pages={68-77}
}

@ARTICLE{Cakir2014FDTD,
  author={Çakir, Gonca and Çakir, Mustafa and Sevgi, Levent},
  journal={IEEE Antennas and Propagation Magazine}, 
  title={An FDTD-Based Parallel Virtual Tool for RCS Calculations of Complex Targets}, 
  year={2014},
  volume={56},
  number={5},
  pages={74-90},
  doi={10.1109/MAP.2014.6971919}}

@ARTICLE{Lucente2008Iteration,
  author={Lucente, Eugenio and Monorchio, Agostino and Mittra, Raj},
  journal={IEEE Trans. Antennas Propag.}, 
  title={An Iteration-Free MoM Approach Based on Excitation Independent Characteristic Basis Functions for Solving Large Multiscale Electromagnetic Scattering Problems}, 
  year={2008},
  volume={56},
  number={4},
  pages={999-1007},
  doi={10.1109/TAP.2008.919166}}

@ARTICLE{Dunn2006Numerical,
  author={Dunn, E.A. and Jin-Kyu Byun and Branch, E.D. and Jian-Ming Jin},
  journal={IEEE Trans. Antennas Propag.}, 
  title={Numerical Simulation of BOR scattering and radiation using a higher order FEM}, 
  year={2006},
  volume={54},
  number={3},
  pages={945-952},
  doi={10.1109/TAP.2006.869936}}

@ARTICLE{Youssef1989Radar,
  author={Youssef, N.N.},
  journal={Proc. IEEE}, 
  title={Radar cross section of complex targets}, 
  year={1989},
  volume={77},
  number={5},
  pages={722-734},
  doi={10.1109/5.32062}}

@misc{3gppRan121FL,
  note= {Moderator (Xiaomi), “Summary \#6 on ISAC channel modelling,” 3GPP, Malta, Tech. Rep. R1-2504945, May 2025, [Online]. Available: https://www.3gpp.org/}
}

@article{liu2024cooperative,
  title={Cooperative sensing for {6G} mobile cellular networks: feasibility, performance and field trial},
  author={Liu, Guangyi and Xi, Rongyan and Han, Zixiang and Han, Lincong and Zhang, Xiaozhou and Ma, Liang and Wang, Yajuan and Lou, Mengting and Jin, Jing and Wang, Qixing and others},
  journal={IEEE J. Sel. Areas Commun.},
  year={2024},
  volume={42},
  number={10},
  pages={2863-2876},
  publisher={IEEE}
}

@misc{3gppRan117FL,
  note= {Moderator (Xiaomi), “Summary\#3 on ISAC channel modelling,” 3GPP, Japan, Tech. Rep. R1-2404636, May. 2024, [Online]. Available: https://www.3gpp.org/}
}

@ARTICLE{wang2023doppler,
  author={Wang, Yirun and Wang, Gongpu and He, Ruisi and Ai, Bo and Tellambura, Chintha},
  journal={IEEE Trans. Commun.}, 
  title={Doppler Shift and Channel Estimation for Intelligent Transparent Surface Assisted Communication Systems on High-Speed Railways}, 
  year={2023},
  volume={71},
  number={7},
  pages={4204-4215},
  doi={10.1109/TCOMM.2023.3275590}}

@misc{itu6G,
  note = {International Telecommunication Union (ITU), ``Framework and Overall Objectives of the Future Development of IMT for 2030 and Beyond,'' ITU-R M.2160-0, Tech. Rep., Dec. 2023. [Online]. Available: https://www.itu.int}
}

@misc{itu2412,
  note = {ITU, ``Guidelines for Evaluation of Radio Interface Technologies for IMT-2020,'' ITU-R M.2412, Tech. Rep., Oct. 2017. [Online]. Available: https://www.itu.int}
}

@ARTICLE{liu2023seventy,
  author={Liu, Fan and Zheng, Le and Cui, Yuanhao and Masouros, Christos and Petropulu, Athina P. and Griffiths, Hugh and Eldar, Yonina C.},
  journal={IEEE Signal Process. Mag.}, 
  title={Seventy Years of Radar and Communications: The road from separation to integration}, 
  year={2023},
  volume={40},
  number={5},
  pages={106-121},
  doi={10.1109/MSP.2023.3272881}}

@article{NYUSIM,
	title        = {{A tutorial on NYUSIM: Sub-terahertz and millimeter-wave channel simulator for 5G, 6G, and beyond}},
	author       = {Poddar, Hitesh and Ju, Shihao and Shakya, Dipankar and others},
	year         = 2023,
	month        = {2nd Quart.},
	journal      = {{IEEE Commun. Surveys Tuts.}},
	publisher    = {IEEE},
	volume       = 26,
	number       = 2,
	pages        = {824--857}
}

@article{CMG_Ref2,
	title        = {{QuaDRiGa: A 3-D multi-cell channel model with time evolution for enabling virtual field trials}},
	author       = {Jaeckel, Stephan and Raschkowski, Leszek and B\"{o}rner, Kai and others},
	year         = 2014,
	month        = jun,
	journal      = {{IEEE Trans. Antennas Propag.}},
	publisher    = {IEEE},
	volume       = 62,
	number       = 6,
	pages        = {3242--3256},
	fjournal     = {IEEE Trans. Antennas Propag.}
}

@article{thor,
	title        = {{THz communications and the demonstration in the ThoR--Backhaul link}},
	author       = {K{\"u}rner, Thomas and Braun, Ralf-Peter and Ducournau, Guillaume and others},
	year         = 2024,
	month        = jun,
	journal      = {{IEEE Trans. THz Sci. Technol.}},
	publisher    = {IEEE},
	volume       = 14,
	number       = 5,
	pages        = {554--567}
}

@article{kucg,
	title        = {{User guide for sub-THz link-level evaluations based on the Kyoto University channel generator}},
	author       = {Koda, Yusuke and Harada, Hiroshi},
	year         = 2025,
	journal      = {{Authorea Preprints}},
	publisher    = {Authorea}
}

@inproceedings{CMG_SimRIS,
	title        = {{SimRIS channel simulator for reconfigurable intelligent surface-empowered communication systems}},
	author       = {Basar, Ertugrul and Yildirim, Ibrahim},
	year         = 2020,
	month        = nov,
	booktitle    = {Proc. IEEE Latin-Amer. Conf. Commun. (LATINCOM)},
	pages        = {1--6},
	doi          = {10.1109/LATINCOM50620.2020.9282349}
}

@inproceedings{nirvawave,
	title        = {{NirvaWave: An accurate and efficient near field wave propagation simulator for 6G and beyond}},
	author       = {Yazdnian, Vahid and Ghasempour, Yasaman},
	year         = 2025,
	month        = mar,
	booktitle    = {Proc. IEEE Wireless Commun. Netw. Conf. (WCNC)},
	pages        = {1--7},
	organization = {IEEE}
}

@article{zhang2025unifiedRCS,
  title={A Unified {RCS} Modeling of Typical Targets for {3GPP ISAC} Channel Standardization and Experimental Analysis},
  author={Y. Zhang and J. Zhang and H. Gong and X. Hu and J. Zhang and S. Luo and Y. Xiong and L. Yu and Z. Yuan and G. Liu and T. Jiang},
  journal={IEEE J. Sel. Areas Commun.},
  year={2025},
  note={early access},
  doi={10.1109/JSAC.2025.3608732}}

\appendices
\section{}

Table \ref{tab:calibration_LSP} lists the simulation assumptions for large scale calibration for UAV sensing targets.

\begin{table*}[t]
    \centering
    \caption{Simulation assumptions for large scale calibration for UAV sensing targets}
    \begin{tabular}{m{4cm}<{\raggedright}| m{13cm}<{\raggedright}} 
    \toprule
    \hline
        \textbf{Parameters} & \textbf{Values} \\
        \hline
        Scenario & UMa-AV \\
        \hline
        Sensing mode & TRP monostatic, TRP-TRP bistatic, TRP-UT bistatic, UT-UT bistatic\\
        \hline
        Sectorization & Single 360-degree sector can be assumed \\
        \hline
        Carrier Frequency & FR1: 6 GHz \newline FR2: 30 GHz \\
        \hline
        BS antenna configurations & Single dual-pol isotropic antenna \\
        \hline
        BS Tx power & FR1: 56 dBm \newline FR2: 41 dBm \\
        \hline
        Bandwidth & FR1: 100 MHz \newline FR2: 400 MHz \\
        \hline
        BS noise figure & FR1: 5 dB \newline FR2: 7 dB \\
        \hline
        UT antenna configurations & $(M,N,P,M_g,N_g;M_p,N_p) = (1,1,2,1,1;1,1)$ \\
        \hline
        UT Tx power & 23dBm \\ \hline
        UT noise figure & FR1: 9 dB \newline FR2: 10 dB \\
        \hline
        UT Distribution & - The overall number of UTs is 30 uniformly distributed in the center cell. \newline - All of the UTs are either terrestrial UTs or aerial UTs, all outdoors. \newline - Vertical distribution of aerial UT: Fixed height value of 200 m. \newline - FR1 is assumed for aerial UT. \\ 
        \hline
        Sensing target distribution & 1 target uniformly distributed (across multiple drops) within the center cell. \newline Vertical distribution: Fixed height value of 200 m. \\
        \hline
        Component A of the RCS for each scattering point & -12.81 dBsm \\
        \hline
        Minimum 3D distances between pairs of Tx/Rx and sensing target & 10 m \\
        \hline
        Wrapping Method & No wrapping method is used if interference is not modelled, \newline otherwise \textit{geographical distance based wrapping} \\

        \hline
        Coupling loss for target channel & Power scaling factor (pathloss, shadow fading, and RCS component A included):
        $\begin{array}{r}
        L_{Tx-SPST-Rx} = PL_{dB}(d_1) + PL_{dB}(d_2) + 10 \log_{10}\left( \frac{c^2}{4 \pi f^2} \right) \\
        \hspace{9mm} - 10 \log_{10}(\sigma_{RCS,A}) + SF_{dB,1} + SF_{dB,2}
        \end{array}$
        \\ 
        
        \hline
        STx/SRx selection & Best N = 4 STx-SRx pairs to be selected for the target. \newline
        see note 1 \\
        
        \hline
        Metrics & Coupling loss for target channel  \\
        \hline
        \bottomrule
        \multicolumn{2}{p{0.8\linewidth}}{Note 1: Based on the STx-SRx pairs with the smallest power scaling factor of the target channel.}\\
        \multicolumn{2}{p{0.8\linewidth}}{Note 2: CDFs can be separately generated for target channel, background channel.}
    \end{tabular}
    \label{tab:calibration_LSP}
\end{table*}

\vfill

\end{document}